\documentclass[12pt,journal,onecolumn,draftcls]{IEEEtran}

\usepackage{epsfig}
\usepackage{times}
\usepackage{float}
\usepackage{afterpage}
\usepackage{amstext}
\usepackage{amssymb,bm}
\usepackage{latexsym}
\usepackage{color}
\usepackage{amsmath}
\usepackage{theorem}
\usepackage{stfloats}
\usepackage{pstricks}
\usepackage{subfigure}
\usepackage{enumerate}
\newcommand{\gap}{3.16}%{7.3}

\newcommand{\gapwholecapstrong}{1.8}%\log(12)/2=1.8
\newcommand{\gapwholecapZ}{2}

\newcommand{\Rp}{R_\mathsf{p}}
\newcommand{\Rc}{R_\mathsf{c}}
\newcommand{\Xc}{X_{\mathsf{c}}}%{\mathsf{X}_{\mathsf{c}}}
\newcommand{\Xp}{X_{\mathsf{p}}}%{\mathsf{X}_{\mathsf{p}}}
\newcommand{\Yf}{Y_{\mathsf{f}}}%{\mathsf{Y}_{\mathsf{f}}}
\newcommand{\Yc}{Y_{\mathsf{c}}}%{\mathsf{Y}_{\mathsf{c}}}
\newcommand{\Yp}{Y_{\mathsf{p}}}%{\mathsf{Y}_{\mathsf{p}}}
\newcommand{\Zf}{Z_{\mathsf{f}}}%{\mathsf{Z}_{\mathsf{f}}}
\newcommand{\Zc}{Z_{\mathsf{c}}}%{\mathsf{Z}_{\mathsf{c}}}
\newcommand{\Zp}{Z_{\mathsf{p}}}%{\mathsf{Z}_{\mathsf{p}}}
\newcommand{\Uc}{U_{\mathsf{c}}}
\newcommand{\Up}{U_{\mathsf{p}}}
\newcommand{\snr}{\mathsf{S}}
\newcommand{\inr}{\mathsf{I}}
\newcommand{\Sp}{{\snr}_{\mathsf{p}}}
\newcommand{\Sc}{{\snr}_{\mathsf{c}}}
\newcommand{\Ip}{{\inr}_{\mathsf{p}}}
\newcommand{\Ic}{{\inr}_{\mathsf{c}}}
\newcommand{\Cc}{\mathsf{C}}
\newcommand{\eac}{{\rm e}^{{\rm j}\theta_{\mathsf{c}}}}
\newcommand{\eap}{{\rm e}^{{\rm j}\theta_{\mathsf{p}}}}
\newcommand{\ap}{\alpha_{\mathsf{p}}}
\newcommand{\ac}{\alpha_{\mathsf{c}}}
\newcommand{\gp}{\gamma_{\mathsf{p}}}
\newcommand{\gc}{\gamma_{\mathsf{c}}}
\newcommand{\gdof}{\mathsf{d}}

\newtheorem{theorem}{Theorem}

\begin{document}

%\title{On the Gaussian Interference Channel with Causal Cognition, or  with Unilateral Source Cooperation}
\title{On the Capacity of the Two-user Gaussian Causal Cognitive Interference Channel}

%\author{\hspace{-10mm} \parbox{3 in}{ \centering Martina Cardone, Raymond Knopp\\
%        %Mobile Communications Dept.\\
%        Institut Eurecom\\
%        Sophia Antipolis, 06560, France\\
%        {\tt\small \{cardone,knopp\}@eurecom.fr}}
%        \hspace{-16mm}
%        \parbox{3 in}{ \centering Daniela Tuninetti\\
%        %Electrical and Computer Engineering Dept.\\
%        University of Illinois at Chicago\\
%        Chicago, IL 60607, USA\\
%        {\tt\small danielat@uic.edu}}
%        \hspace{-16mm}
%        \parbox{3 in}{ \centering Umer Salim\\
%        Intel Mobile Communications\\
%        Sophia Antipolis, 06560, France\\
%        {\tt\small umer.salim@intel.com}}
%}

\author{Martina~Cardone, Daniela~Tuninetti, Raymond~Knopp and Umer~Salim %
\thanks{M. Cardone and R. Knopp are with the Mobile Communications Department at Eurecom, Biot, 06410, France (e-mail: cardone@eurecom.fr; knopp@eurecom.fr). D. Tuninetti is with the Electrical and Computer Engineering Department of the University of Illinois at Chicago, Chicago, IL 60607 USA (e-mail: danielat@uic.edu). U. Salim is with Algorithm Design Group of Intel Mobile Communications, Sophia Antipolis, 06560, France (e-mail: umer.salim@intel.com).

The work of D.~Tuninetti was partially funded by NSF under award number 0643954;
the contents of this article are solely the responsibility of the author and
do not necessarily represent the official views of the NSF.
The work of D.~Tuninetti was possible thanks to the generous support of Telecom-ParisTech, Paris France,
while the author was on a sabbatical leave at the same institution.
Eurecom's research is partially supported by its industrial partners:
BMW Group Research \& Technology, IABG, Monaco Telecom, Orange, SAP, SFR, ST Microelectronics, Swisscom and Symantec.
The research work carried out at Intel by U. Salim has received funding from the
European Community's Seventh Framework Program (FP7/2007-2013) SACRA project (grant agreement number 249060)

The results in this paper have been presented in part to the 2013 IEEE International Conference on Communications (ICC 2013).}
}
%
%\IEEEauthorblockN{Martina Cardone$^{\dagger}$, Raymond Knopp$^{\dagger}$, Daniela Tuninetti$^*$, Umer Salim$^{\ddagger}$}\\
%$^{\dagger}$Institut Eurecom,
%Sophia Antipolis, 06560, France
%Email: \{cardone, knopp\}@eurecom.fr\\
%$^*$ University of Illinois at Chicago,
%Chicago, IL 60607, USA
%Email: danielat@uic.edu\\
%$^{\ddagger}$ Intel Mobile Communications
%Sophia Antipolis, 06560, France
%Email: umer.salim@intel.com}

%\author{Martina Cardone*, Daniela Tuninetti$^{\star}$ and Raymond Knopp*}% <-this % stops a space
%\thanks{*This work was not supported by any organization}% <-this % stops a space
%\thanks{$^{\star}$H. Kwakernaak is with Faculty of Electrical Engineering, Mathematics and Computer Science,}
%%        University of Twente, 7500 AE Enschede, The Netherlands
%        {\tt\small h.kwakernaak at papercept.net}}%
%\thanks{$^{2}$P. Misra is with the Department of Electrical Engineering, Wright State University,
%        Dayton, OH 45435, USA
%        {\tt\small p.misra at ieee.org}}%

\maketitle

%\thispagestyle{empty}
%\pagestyle{empty}

%%%%%%%%%%%%%%%%%%%%%%%%%%%%%%%%%%%%%%%%%%%%%%%%%%%%%%%%%%%%%%%%%%%%%%%%%%%%%%%%
\begin{abstract}
This paper considers the two-user Gaussian Causal Cognitive Interference Channel (GCCIC), which consists of two source-destination pairs that share the same channel and where one full-duplex cognitive source can causally learn the message of the primary source through a noisy link. The GCCIC is an interference channel with unilateral source cooperation that better models practical cognitive radio networks than the commonly used model which assumes that one source has perfect non-causal knowledge of the other source's message.

First the sum-capacity of the symmetric GCCIC is determined to within a constant gap.
Then, the insights gained from the derivation of the symmetric sum-capacity are extended to characterize the whole capacity region to within a constant gap for more general cases. In particular, the capacity is determined (a) to within 2~bits for the fully connected GCCIC when, roughly speaking, the interference is not weak at both receivers, (b) to within 2~bits for the Z-channel, i.e., when there is no interference from the primary user, and (c) to within 2~bits for the S-channel, i.e., when there is no interference from the secondary user. 
%The sum-capacity of the GCCIC is studied in both the {\em interference symmetric} and {\em  interference asymmetric} cases. In the former case the two interfering links and the two direct links are of the same strength, in the latter case one of the interfering links is absent. { These scenarios reflect different possible topologies of cognitive networks.} It is shown through evaluation of various achievable schemes that known sum-rate upper bounds are achievable to within a constant gap regardless of the actual value of the channel parameters. Interestingly, except for a small set of parameters, the achievable schemes are quite simple in the sense that only superposition coding is used. More complex schemes based on binning and superposition coding are shown to achieve a smaller gap than superposition coding alone.

The parameter regimes where the GCCIC is equivalent, in terms of generalized degrees-of-freedom, to the noncooperative interference channel (i.e., unilateral causal cooperation is not useful), to the non-causal cognitive interference channel (i.e., causal cooperation attains the ultimate limit of cognitive radio technology), and to bilateral source cooperation are identified. These comparisons shed lights into the parameter regimes and network topologies that in practice might provide an unbounded throughput gain compared to currently available (non cognitive) technologies.
%
%Finally, it is shown through numerical evaluations that the developed insights based on the study on the sum-capacity can be extended to the whole capacity region. The complete characterization of the two-dimensional capacity region to within a constant gap for any possible choice of the five parameters that specify the GCCIC model is outside the scope of this work. It is conjectured that the inner and outer bounds used in the work would suffice, except in the regime of very weak interference and very weak cooperation where novel outer bounds might be necessary akin to a similar parameter regime in the noncooperative case.
\end{abstract}


%%%%%%%%%%%%%%%%%%%%%%%%%%%%%%%%%%%%%%%%%%%%%%%%%%%%%%%%%%%%%%%%%%%%%%%%%%%%%%%%
\section{Introduction}
\label{sec:intro}
%{\bf \Large A shorter version of this paper has been submitted to the 50th Annual Allerton Conference on Communication, Control, and Computing (Allerton 2012).}

This work considers the cognitive radio overlay paradigm~\cite{goldsmith:spectrymgridlock} that consists of two source-destination pairs sharing the same channel in which the pair with cognitive abilities attains its communication goals while helping the other (non cognitive) pair. The sources are indicated as PTx and CTx, and the destinations as PRx and CRx. PTx and PRx are referred to as the {\em primary} pair, while CTx and CRx as the {\em cognitive} pair. The prime features of overlay cognitive radio are to firstly allow the cognitive nodes to communicate without hindering the communication of the primary nodes, and secondly to enhance the communication reliability of the primary nodes. To this end, the CTx is assumed to operate in a full-duplex mode on the same channel as the PTx.  Due to the broadcast property of the wireless media, the CTx overhears the PTx through a lossy communication link. Contrary to the commonly studied cognitive radio model that assumes perfect non-causal primary message knowledge available at the CTx~\cite{Devroye}, in this work we treat the causal case, that is, the CTx has access only to primary information it receives over the air. We refer to this system as the Causal Cognitive Interference Channel (CCIC).

From an application standpoint, the CCIC fits future 4G networks with heterogeneous deployments \cite{3GPPRel10doc} where the CTx corresponds to the so-called {\em small-cell} base-station, or eNB. In this scenario, the CTx would listen to the PTx transmission but not make use of a dedicated point-to-point backhaul link (i.e., on either another channel or through a wired link). We consider deployment scenarios where the CTx$\rightarrow$CRx link is on the same carrier frequency as PTx$\rightarrow$PRx link and the CTx operates in a full-duplex mode. This implies that the CTx can listen to the PTx's transmission while transmitting.  Full-duplex communication is possible thanks to sophisticated self-interference cancellation techniques at the CTx~\cite{RICEpaper}.
%and sufficient antenna-spacing between the PTx$\rightarrow$CTx receive antenna and CTx$\rightarrow$(PRx,CRx) transmit antenna. 
Moreover, we assume that the PRx and CRx can implement sophisticated interference-mitigation techniques which exploit knowledge of the codebooks used at both PTx and CTx. These codebooks are conceived for the interference scenario (e.g. superposition-coding~\cite{HK} or Dirty Paper Coding (DPC)~\cite{costaDPC}).  It should be noted that, since 4G air-interfaces already specify up to 8-level superposition coding for point-to-point MIMO or point-to-multipoint MIMO transmission~\cite{3GPPRel10doc}, it is feasible to assume that extensions for distributed superposition coding could also be envisaged.

Different interference scenarios are considered and can correspond to the choice of appropriate deployment configurations in cognitive radio networks. The first class is the fully connected CCIC where both destinations suffer from interference, i.e., in this case both destinations are in the coverage area of both sources.
%{\em symmetric} Gaussian CCIC where the two interfering links are of the same strength and the two direct links are of the same strength.  
The second class is the interference-asymmetric Gaussian CCIC where either the link PTx$\rightarrow$CRx is non-existent (referred to as the Z-channel) or the link CTx$\rightarrow$PRx is non-existent (referred to as the S-channel).
%while keeping the strength of the direct links the same. 
In the noncooperative IC these two asymmetric scenarios %with direct links of the same strength, the Z-channel and the S-channel have the same capacity region, 
are the same, up to a relabeling of the nodes. In the CCIC case, due to the asymmetry in the cooperation, the two scenarios are different and must be treated separately.
The Z-channel models a situation such as an indoor CTx$\rightarrow$CRx with another receiver (PRx) connected to an outdoor base station (PTx) in the vicinity of CTx.  The S-channel models the case where PRx is out-of-range of CTx and the base station (PTx) schedules traffic to both PRx and CTx/CRx concurrently. Both scenarios are relevant for practical cognitive radio deployments and their ultimate performance is investigated in this work.

\subsection{Related Past Work}
The presence of a lossy communication link between PTx and CTx enables CTx to cooperate with PTx.
% in order to send the PTx's message. 
CTx, in fact, through this noisy channel overhears the signal sent by the PTx and gathers information about PTx's message, which serves as the basis for unilateral cooperation between the two sources. Unilateral source cooperation is a special case of the {\em IC with generalized feedback}, or bilateral source cooperation~\cite{HostMadsenIT06,YANG-TUNINETTI,PVIT11,TuninettiITA10,TndonUlukusIT11}.

\subsubsection{IC with Bilateral Source Cooperation}
Bilateral source cooperation has been actively investigated recently. 
Host-Madsen~\cite{HostMadsenIT06} first studied outer and inner bounds for the Gaussian IC with either source or destination bilateral cooperation. For outer bounds, the author in~\cite{HostMadsenIT06} evaluated the different cut-set upper bounds and then tightened the sum-rate upper bound by extending the sum-rate outer bounds originally developed by Kramer~\cite{Kramer} for the Gaussian noncooperative IC in weak and strong interference to the cooperative case.
Tuninetti~\cite{TuninettiITA10} derived a general outer bound for the IC with bilateral source cooperation by extending Kramer's Gaussian noise sum-rate upper bounds in~\cite[Theorem 1]{Kramer} to any memoryless IC with source cooperation, and more recently to any form of source and destination cooperation~\cite{TuninettiITW12}.
Prabhakaran and Viswanath~\cite{PVIT11} extended the idea of~\cite[Theorem 1]{etw} to derive a sum-rate outer bound for a class of injective semi-deterministic IC with bilateral source cooperation in the spirit of the work by Telatar and Tse~\cite{TelatarTse}, and evaluated it for the Gaussian channel with independent noises (this assumption is not without loss of generality when cooperation and feedback are involved).
Tandon and Ulukus \cite{TndonUlukusIT11} developed an outer bound for the IC with bilateral source cooperation based on the dependence-balance idea of Hekstra and Willems~\cite{kekstrawillems} and proposed a novel method to evaluate it for the Gaussian channel with independent noises.

The largest known achievable region for general bilateral source cooperation, to the best of our knowledge, is the one presented in \cite[Section V]{YANG-TUNINETTI}. In \cite[Section V]{YANG-TUNINETTI} each source splits its message into two parts, i.e., a {\em common} and a {\em private} message, as in the Han-Kobayashi's scheme for the noncooperative IC~\cite{HK}; these two messages are further sub-divided into a {\em noncooperative} and a {\em cooperative} part. The noncooperative messages are transmitted as in the noncooperative IC~\cite{HK}, while the cooperative messages are delivered to the destinations by exploiting the cooperation among the two sources. In \cite[Section V]{YANG-TUNINETTI} each source, e.g. source~1, after learning the cooperative messages of source~2, sends the common cooperative message of source~2 and uses Gelfand-Pinsker's binning~\cite{gelfandpinsker}, or Dirty Paper Coding (DPC) \cite{costaDPC} in the Gaussian noise case, against the private cooperative message of source~2 in an attempt to rid its own receiver of this interference. The achievable scheme in \cite[Section V]{YANG-TUNINETTI} only uses partial-decode-and-forward for cooperation. A possibly larger achievable region could be obtained by also including compress-and-forward as cooperation mechanism in the spirit of~\cite{coveElGamal} for the relay channel. 

For the two-user Gaussian noise IC with bilateral source cooperation, {\em under the assumption that the cooperation links have same strength}, the scheme of~\cite[Section V]{YANG-TUNINETTI} was sufficient to match the sum-capacity upper bounds of~\cite{TuninettiITA10,PVIT11} to within a constant gap~\cite{PVIT11,YangHighCoop}.~\cite{PVIT11} characterized the sum-capacity to within 20/2~bits (in this work we consider the gap per user) of the IC with bilateral source cooperation under the condition that the cooperation links have the same strength, but otherwise arbitrary direct and interfering links. The gap was reduced to 2~bits in the `strong cooperation regime' in~\cite{YangHighCoop} with symmetric direct links, symmetric interfering links and symmetric cooperation links.
In this work we seek extensions of these results to the case where the cooperation links have different strengths. In particular, motivated by the cognitive radio technology, we focus on the case of unilateral source cooperation where one of the cooperation links is absent.  Moreover, we seek to determine the whole capacity region to within a constant gap, not simply the sum-capacity.
To the best of our knowledge, the case of asymmetric cooperation links, of which unilateral cooperation is a special case, has not been considered in the literature. Moreover, the whole capacity region with source cooperation, to the best of our knowledge, has never been characterized to within a constant gap in the literature, which is a major contribution of this work. %for the case of unilateral cooperation.

\subsubsection{IC with Unilateral Source Cooperation}
Unilateral source cooperation is clearly a special case of the general bilateral cooperation case where the cooperation capabilities of the two sources are not restricted to be the same. This case has been specifically considered in \cite{DBLP:journals/corr/abs-1111-3966} where the cooperating transmitter works either in full-duplex or in half-duplex mode.  For full-duplex unilateral cooperation, the authors of \cite{DBLP:journals/corr/abs-1111-3966} evaluated the performance of two achievable schemes: one that exploits partial-decode-and-forward and binning and a second one that extends the first by adding rate splitting. It was observed, through numerical evaluations, that the proposed inner bounds are not too far from the outer bound of \cite{TndonUlukusIT11} for certain Gaussian noise channels. In this work we formally prove that the outer bound region obtained from~\cite{HostMadsenIT06,PVIT11,TuninettiITA10} is achievable to within a constant gap, for the different network scenarios considered. Moreover, we use as unifying framework the achievable scheme of \cite[Section V]{YANG-TUNINETTI}, of which the schemes of \cite{DBLP:journals/corr/abs-1111-3966} are special cases.

An extension of the IC with unilateral source cooperation was studied in \cite{MirmohseniIT2012}, where it was assumed 
% for the case of ``look ahead'' at the cognitive source, meaning
that at any given time instant the cognitive source has a non-causal access to $L\geq 0$ future channel outputs. The case $L=0$ corresponds to the strictly causal case considered in this paper, while the case $L\to\infty$ to the limiting non-causal cognitive IC \cite{Devroye}.
The authors of \cite{MirmohseniIT2012} derived potentially tighter outer bounds for the CCIC channel (i.e., case $L=0$) than those of \cite{PVIT11,TuninettiITA10} specialized to unilateral source cooperation; unfortunately it is not clear how to evaluate these bounds in Gaussian noise because they are expressed as a function of auxiliary random variables jointly distributed with the inputs and for which no cardinality bounds on the corresponding alphabets are known. 
The achievable region in \cite[Corollary 1]{MirmohseniIT2012} is also no smaller than the region in \cite[Section V]{YANG-TUNINETTI} specialized to the case of unilateral source cooperation (see \cite[Remark 2, point 6]{MirmohseniIT2012}). Although \cite[Corollary 1]{MirmohseniIT2012} is, to the best of our knowledge, the largest known achievable region for the general memoryless CCIC with unilateral cooperation, its evaluation in general is quite involved as the rate region is specified by 9 jointly distributed auxiliary random variables and by 30 rate constraints.
In \cite{MirmohseniIT2012} inner bounds were compared numerically to the $2\times2$ MIMO outer bound for the Gaussian CCIC; the $2\times2$ MIMO outer bound is loose in general compared to the bounds in \cite{HostMadsenIT06,PVIT11,TuninettiITA10}. Although it was noted in \cite{MirmohseniIT2012} that, for the simulated set of channel gains, the proposed bounds are not far away from one another, a performance guarantee in terms of (sum-)capacity to within a constant gap was not given. In this work we characterize the capacity to within a constant gap for several channel configurations.

\subsubsection{Non-Causal Cognitive Radio Channel}
The cognitive radio channel is commonly modeled following the pioneering work of Devroye {\em et al}~\cite{Devroye} in which the superior capabilities of the cognitive source are modeled as perfect non-causal knowledge of PTx's message at CTx. For this non-causal model the capacity region in Gaussian noise is known exactly for some parameter regimes and to within 1~bit otherwise \cite{riniJ1}.
%; for an in-depth review of known results on this model we refer the interested reader to~\cite{riniJ1,riniJ2} and references therein. 
In this work we remove the ideal non-causal message knowledge assumption by considering a more realistic scenario where CTx causally learns the PTx's message through a noisy link. The study of the causal model stems from the question of whether cognitive radio can offer a substantial rate gain over the noncooperative IC. Since the answer was in the positive for the non-causal model~\cite{riniJ1}, the next question is whether such gains can be attained in practical channels where message knowledge must be obtained through a noisy channel. This work answers this question in the positive. In particular, we identify the set of the channel parameters sufficient to attain, to within a constant gap, the ultimate performance limits of cognitive radio as predicted by the non-causal model \cite{riniJ1}.

\subsection{Contributions and Paper Organization}

The rest of the paper is organized as follows. 
Section \ref{sec:channel model} describes the channel model, defines the concept of capacity to within a constant gap and of generalized degrees of freedom (gDoF), and summarizes known inner and outer bounds. 
%Section \ref{sec:dof} shows that known sum-capacity upper bounds can be achieved to within a constant gap irrespectively of the actual value of the channel parameters for the {interference-symmetric} and the two {interference-asymmetric} cases with symmetric direct links.
%Section \ref{sec:achievable scheme} reports two simple different achievable schemes for the Gaussian CCIC based on time-sharing. 
Section~\ref{sec:gap sumcapacity Symmetric Channel} characterizes the capacity region of the symmetric GCCIC to within 1~bit for almost all parameter regimes, and the sum-capacity to within \gap~bits otherwise (see Theorem~\ref{thm:gap symmetric}).
%a constant gap in strong interference and in weak interference when the cooperation link is sufficiently strong.  For the set of parameters outside this range, Section \ref{sec:gap sumcapacity Symmetric Channel} characterizes the sum-capacity to within a constant gap. 
Section~\ref{sec:general extension in strong interference} considers the general GCCIC and characterizes its capacity region to within 2~bits for a large set of channel parameters that, roughly speaking, excludes the case of weak interference at both receivers (see Theorem~\ref{thm:gap full}). In order to better understand the weak interference regime, we analyze the `interference asymmetric' GCCIC  in which one of the interfering links is absent which models different network topologies; we determine the capacity region to within 2~bits for the Z-channel in Section~\ref{sec:general extension Z} (see Theorem~\ref{thm:gap Z}), and to within 2~bits for the S-channel in Section~\ref{sec:general extension S} (see Theorem~\ref{thm:gap S}). Section~\ref{sec:Conclusion} concludes the paper. Most of the proofs are reported in the Appendix. In particular, the Appendix contains the details of the relatively simple proposed achievable schemes, which can be used to provide design insights into practical schemes for future cognitive networks.  
For all system models considered, we compare the gDoF attained with causal unilateral cooperation with that of other known forms of cooperation to quantify when causal cognitive radio might be worth implementing in practice.

\section{System Model and Background}
\label{sec:channel model}

Throughout the paper we adopt the notation convention of~\cite{ElGamalKimBook}. In particular,
$[n_1:n_2]$ denotes the set of integers from $n_1$ to $n_2 \geq n_1$;
$[x]^+ := \max\{0,x\}$ for $x\in\mathbb{R}$;
$\log^+(x) := \max\{0,\log(x)\}$ for $x\in\mathbb{R}$;
$Y^{j}$ is a vector of length $j$ with components $(Y_1,\ldots,Y_j)$.
%for an index set $\mathcal{A}$ we let $Y_{\mathcal{A}} = \{ Y_j : j\in \mathcal{A} \}$;
%$\mathbf{0}_j$ is the all zero row vector of length $j$;
%$\mathbf{1}_j$ is the all one column vector of length $j$;
%$\mathbf{I}_j$ is the identity matrix of dimension $j$;
%$f_1(x)\doteq f_2(x)$ means that $\lim_{x\to+\infty}f_1(x)/f_2(x)=1$.
The subscript $\mathsf{c}$ (in sans serif font) is used for quantities related to the cognitive pair, while the subscript $\mathsf{p}$ (in sans serif font) for those related to the primary pair. The subscript $\mathsf{f}$ (in sans serif font) is used to refer to generalized feedback information received at CTx.  The subscript $c$ (in roman font) is used to denote common messages, while the subscript $p$ (in roman font) to denote private messages. The notation ${\rm eq}(n)$ is used to denote the rightmost side of the equation number $n$.

\subsection{The Gaussian noise channel}

\begin{figure}%[h]
\centering
\includegraphics[width=0.45\textwidth]{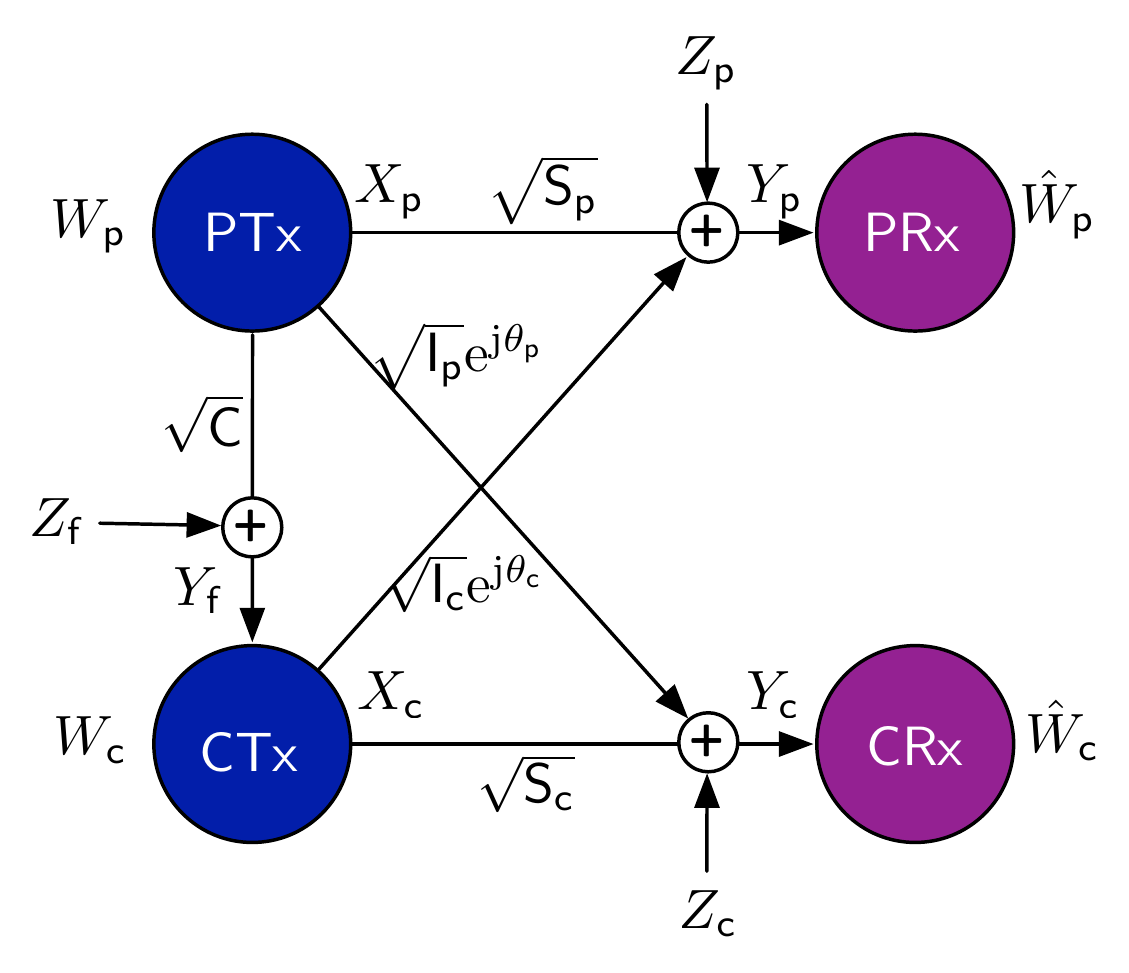}%{figure1.eps}
\caption{The two-user Gaussian Causal Cognitive Interference Channel (GCCIC).}
\label{fig:channelmodel}
\end{figure}

A single-antenna full-duplex GCCIC, shown in Fig.~\ref{fig:channelmodel}, is described by the input/output relationship
\begin{align}
\begin{bmatrix}
\Yf \\ \Yp \\ \Yc \\
\end{bmatrix}
=
\begin{bmatrix}
\sqrt{\Cc} & \star  \\ 
\sqrt{\Sp} & \sqrt{\Ic} \eac \\
\sqrt{\Ip} \eap & \sqrt{\Sc} \\
\end{bmatrix}
\begin{bmatrix}
\Xp \\ \Xc \\
\end{bmatrix}
+
\begin{bmatrix}
\Zf \\ \Zp \\ \Zc \\ 
\end{bmatrix}
\label{eq:awgn full}
\end{align}
where $\star$ indicates the channel gain that does not affect the capacity region (because CTx can remove its transmit signal $\Xc$ from its channel output $\Yf$).
%}
The channel gains are constant, and therefore known to all nodes. Without loss of generality certain channel gains can be taken to be real-valued and non-negative because a node can compensate for the phase of one of its channel gains.
The channel inputs are subject to a unitary average power constraint  without loss of generality, i.e., $\mathbb{E} \left [ |X_i|^2 \right ] \leq 1, i \in \{ \mathsf{p},\mathsf{c} \}$.
The noises are independent circularly symmetric Gaussian random variables with, without loss of generality, zero mean and unit variance.

%A general memoryless CCIC consists of two input alphabets $\left (\mathcal{X}_{\mathsf{p}},\mathcal{X}_{\mathsf{c}} \right )$, three output alphabets $\left (\mathcal{Y}_{\mathsf{f}},\mathcal{Y}_{\mathsf{p}},\mathcal{Y}_{\mathsf{c}} \right )$ and a memoryless transition probability $\mathbb{P}_{\mathsf{{Y}_{f},{Y}_p,{Y}_{c}}|\mathsf{{X}_p,{X}_c}}[a,b,c|d,e] : \mathcal{Y}_{\mathsf{f}} \times \mathcal{Y}_{\mathsf{p}}  \times \mathcal{Y}_{\mathsf{c}} \to [0,1], \ \forall (d,e)\in \mathcal{X}_{\mathsf{p}} \times \mathcal{X}_{\mathsf{c}}$.
%
PTx has a message $W_\mathsf{p}\in [1:2^{N \Rp}]$ for PRx and CTx has a message $W_\mathsf{c}\in [1:2^{N \Rc}]$ for CRx, where $N\in\mathbb{N}$ denotes the codeword length and $\Rp\in\mathbb{R}_+$ and $\Rc\in\mathbb{R}_+$ the transmission rates for PTx and CTx, respectively.
%, measured in bits per channel use. %(logarithms are in base 2).
The messages $W_\mathsf{p}$ and $W_\mathsf{c}$ are independent and uniformly distributed on their respective domains. At time $i, \ i\in [1:N],$ PTx maps its message $W_\mathsf{p}$ into a channel input symbol $X_{\mathsf{p},i}(W_\mathsf{p})$ and CTx maps its message $W_\mathsf{c}$ and its past channel observations into a channel input symbol $X_{\mathsf{c},i}(W_\mathsf{c},\Yf^{i-1})$.
At time $N$, PRx makes an estimate of its intended message based on all its channel observations as $\widehat{W}_{\mathsf{p}}(\Yp^N)$, and similarly CRx outputs $\widehat{W}_{\mathsf{c}}(\Yc^N)$.
The capacity region is the convex closure of all non-negative rate pairs $\left ( \Rp, \Rc \right )$ such that $\max_{u\in\{\mathsf{c},\mathsf{p}\}}\mathbb{P}[\widehat{W}_u \neq W_u] \to 0$ as $N\to+\infty$.

The noncooperative IC is obtained as a special case of the CCIC by setting $\Cc=0$ %$\mathsf{{Y}_{f}}=\emptyset$,
and the non-causal cognitive IC in the limit for  $\Cc\to+\infty$. %by setting $\mathsf{{Y}_{f}}$ to be a noiseless bit-pipe of infinite capacity.

A GCCIC is said to be a Z-channel if $\Ip=0$, i.e., the CRx does not experience interference from PTx, and an S-channel if $\Ic=0$, i.e., the PRx does not experience interference from CTx.

\bigskip
{\bf Capacity region to within a constant gap.}
The capacity region of the GCCIC is said to be known to within $\mathsf{GAP}$~bits if we can show an inner bound region $\mathcal{I}$ and an outer bound region $\mathcal{O}$ such that 
\[
(\Rp,\Rc)\in\mathcal{O} \Longrightarrow ([\Rp-\mathsf{GAP}]^+,[\Rc-\mathsf{GAP}]^+)\in\mathcal{I}.
\]

\bigskip
{\bf Generalized Degrees of Freedom (gDoF).}
Knowledge of the capacity region to within a constant gap implies an exact capacity characterization at high SNR. The gDoF is a performance measure introduced in \cite{etw} for the noncooperative IC to capture the high SNR behavior of the sum-capacity as a function of the relative strengths of direct and interference links. The gDoF represents a more refined characterization of the sum-capacity at high SNR compared to the classical DoF.
%The gDoF in \cite{etw} was introduced for a noncooperative IC with symmetric direct and interfering links. 
In order to quantify the gain of causal unilateral source cooperation compared to the noncooperative IC, we shall use the gDoF as a performance measure.
% for a {\em symmetric} setting.
Let ${\snr}>1$ and parameterize
\begin{subequations}
\begin{align}
     \Sp  &:= {\snr}^1, \ \text{primary direct link},
\\   \Sc  &:= {\snr}^1, \ \text{cognitive direct link}, 
\\   \Ip  &:= {\snr}^{\ap},   \ \ap\geq 0,   \ \text{interference at CRx from PTx},
\\   \Ic  &:= {\snr}^{\ac},   \ \ac\geq 0,   \ \text{interference at PRx from CTx},
\\   \Cc  &:= {\snr}^\beta, \ \beta\geq 0, \ \text{cooperation link},
\end{align}
\label{eq:paramet}
\end{subequations}
where $\ap$ and $\ac$ measure the strength of the interference links compared to the direct link, while $\beta$  the strength of the cooperation link compared to the direct link.  We remark that the parameterization in \eqref{eq:paramet}, with direct links of the same strength, is used only for evaluation of the gDoF. Moreover, in order to capture different network topologies, we focus on
\begin{enumerate}
\item interference-symmetric channel:
$%\begin{align*}
\ap= %\alpha, \
\ac= \alpha%, \
%\beta = \beta
%\ \beta_{p} = 0,
$;%\end{align*}
\item Z-channel:
$%\begin{align*}
\ap= 0, \
\ac= \alpha%, \
%\beta = \beta
%\ \beta_{p} = 0,
$;%\end{align*}
\item S-channel: 
$%\begin{align*}
\ap= \alpha, \
\ac= 0%, \
%\beta = \beta.
%\ \beta_{p} = 0.
$.%\end{align*}
\end{enumerate}
The case $\ap=\ac=0$ is not interesting since in this case the GCCIC reduces to two parallel point-point links for which cooperation is useless.  For the above three cases, the system is parameterized by the triplet $({\snr},\alpha,\beta)$, where ${\snr}$ is referred to as the (direct link) SNR, $\alpha$ as the interference exponent and $\beta$ as the cooperation exponent.\footnote{In principle the system performance also depends on the phases of the interfering links $(\theta_\mathsf{c},\theta_\mathsf{p})$. However, as far as gDoF and sum-capacity to within a constant gap are concerned, the phases $(\theta_\mathsf{c},\theta_\mathsf{p})$ only matter if the IC channel matrix
${\small \begin{bmatrix}
\sqrt{\Sp} & \sqrt{\Ic} \eac \\
\sqrt{\Ip} \eap & \sqrt{\Sc} \\
\end{bmatrix}}$
is rank deficient, in which case one received signal is a noisier version of the other and the overall channels behave, sum-capacity-wise, as a Multiple Access Channel (MAC).}
The gDoF is defined as
\begin{align}
{\gdof}(\alpha,\beta) 
&:= \lim_{{\snr}\to+\infty} \frac{\max\{\Rp + \Rc\}}{2\log(1+{\snr})}
\label{eq:gDoF definition}
\end{align}
where the maximization is intended over all possible achievable rate pairs $(\Rc,\Rp)$. 
Without cooperation, the gDoF ${\gdof}(\alpha,0)$ reduces to the gDoF characterized in~\cite{etw} while for $\beta\to+\infty$ to the gDoF that can be evaluated from the capacity characterization to within 1 bit of~\cite{riniJ1}.
Here we are interested in determining under which condition on the cooperation exponent $\beta$ we have ${\gdof}(\alpha,\beta)  > {\gdof}(\alpha,0)$ since a strict improvement in gDoF implies an unbounded gain in terms of sum-capacity as the SNR grows to infinity.

%The gDoF region is an asymptotically exact characterization of capacity at infinite ${\snr}$. At finite ${\snr}$ the capacity is said to be known to within $\mathsf{b}$~bits if we can show an inner bound region  $\mathcal{I}$ and an outer bound region $\mathcal{O}$ such that $(\Rc,\Rp)\in{\rm ConvexClosure}[\mathcal{I}] \Longrightarrow (\Rc+\mathsf{b},\Rp+\mathsf{b})\not\in\mathcal{O}$.

\subsection{Known outer bounds for the GCCIC}
\label{sec:outer}

\begin{figure*}
\begin{subequations}
\begin{align}
%%%\hline %\nonumber
   \Rc &\leq \log\left ( 1 + \Sc \right )
\label{eq:CutSetRc}
\\ \Rp &\leq \min\Big\{
   \log\left ( 1+(\sqrt{\Sp} +  \sqrt{\Ic})^2\right ), \ \log\left ( 1+\Cc+\Sp \right )   
\Big\}
\label{eq:CutSetRp}
\\ \Rp+\Rc &\leq \min \left \{r^{(\rm{CS})},r^{(\rm{DT})},r^{(\rm{PV})}  \right \}
\label{eq:ourSumRate}
\\ r^{(\rm{CS})} &\leq  \log\left ( 1 + \Sc \right ) + \min\Big\{
   \log\left ( 1+(\sqrt{\Sp} +  \sqrt{\Ic})^2\right ), \ \log\left ( 1+\Cc+\Sp \right )   
\Big\}
\label{eq:CutSet}
\\
  r^{(\rm{DT})} &\leq \min \Big\{ 
     \log  \left( \frac{1 +      \max\{\Ic,\Sc\}}{1+ \Ic}  \right)
   + \log  \left( 1+ (\sqrt{\Sp} + \sqrt{\Ic})^2  \right), \nonumber
\\& \qquad \qquad
     \log  \left( \frac{1 + \Cc +  \max\{\Sp,\Ip\}}{1+ \Ip}  \right) 
   + \log  \left( 1+ (\sqrt{\Sc} + \sqrt{\Ip})^2 \right)   \Big\}
\label{eq:Tuninetti}
\\
 r^{(\rm{PV})}& \leq
    \log \left( \left( 1 + \left ( \frac{\sqrt{\Sp}}{\sqrt{\max\{1,\Ip\}}}  +  \sqrt{\Ic} \right)^2 \right) 
         \left( 1 + \left ( \frac{\sqrt{\Sc}}{\sqrt{\max\{1,\Ic\}}}  +  \sqrt{\Ip} \right)^2 \right) \right )
   +  \Delta
\label{eq:PV}
\\
\Delta&:=  
  \log \left( \left( 1+\Cc \right)
         \frac{1+\left( \frac{\sqrt{\Sc}}{\sqrt{\max\{1,\Ic\}}} + \frac{\sqrt{\Ip}}{\sqrt{\max\{1,\Cc\}}} \right)^2 }
              {1+\left( \frac{\sqrt{\Sc}}{\sqrt{\max\{1,\Ic\}}} + \sqrt{\Ip} \right)^2 }
         \right)
\label{eq:deltaPV}
%%%\\\hline \nonumber
\end{align}
\label{eq:ourRate}
\end{subequations}
\end{figure*}

In the literature several outer bounds are known for bilateral source cooperation~\cite{HostMadsenIT06,PVIT11,TuninettiITA10,TndonUlukusIT11}. Here we specialize some of them for the GCCIC in~\eqref{eq:awgn full}. We let $\mathbb{E} \left [\Xp \Xc^* \right ]=\rho$, for some $\rho\in\mathbb{C}$ such that $|\rho|\leq 1$.
An outer bound region for the GCCIC is reported in~\eqref{eq:ourRate} at the top of next page and is obtained by upper bounding over $(\rho,\theta_{\mathsf{c}},\theta_{\mathsf{p}})$ each mutual information term in the bounds in~\cite{HostMadsenIT06,PVIT11,TuninettiITA10} (the details can be found in Appendix~\ref{app:outer}). In particular,
the bounds on the individual rates in~\eqref{eq:CutSetRc} and~\eqref{eq:CutSetRp} are cut-set bounds, and
the sum-rate upper bound in~\eqref{eq:ourSumRate} is the minimum of three quantities obtained as follows:
from the cut-set bounds on the individual rates we obtain~\eqref{eq:CutSet}, 
from~\cite{TuninettiITA10} we obtain~\eqref{eq:Tuninetti}, and
from~\cite{PVIT11} we obtain~\eqref{eq:PV}.

\bigskip
The upper bound in~\eqref{eq:ourRate} for $\Cc\to+\infty$ reduces to the upper bound for the non-causal cognitive IC in~\cite[Theorem III.1]{riniJ1}, which unifies previously known outer bounds for the weak ($\Sc>\Ic$) and strong ($\Sc\leq\Ic$) interference regimes. The region in~\cite[Theorem III.1]{riniJ1} is known to be achievable to within 1~bit in all parameter regimes. However, in weak interference ($\Sc>\Ic$), the capacity region of the non-causal cognitive IC is known exactly and is given by%
\begin{subequations}
\begin{align}
   \Rp &\leq \log \left( 1+\frac{\Sp +|\gc|^2 \Ic + 2|\gc|  \ \sqrt{\Sp \Ic} }{1+ (1-|\gc|^2)\Ic} \right)
\label{eq:the noncausalcognitive capacity weak rp init}
\\ \Rc &\leq  \left( 1 +(1-|\gc|^2)\Sc \right)
\label{eq:the noncausalcognitive capacity weak rc init}
\end{align}
\label{eq:the noncausalcognitive capacity weak}
\end{subequations}
union over all $|\gc|\leq 1$. Therefore, the region in~\eqref{eq:the noncausalcognitive capacity weak} is an outer bound for the GCCIC for $\Sc>\Ic$.

\medskip
%We aim to show that the sum-rate upper bound in \eqref{eq:ourRate} is achievable to within a constant gap for the interference-symmetric and the interference-asymmetric cases (recall that here we assume that the direct links have the same strength, see the parameterization in~\eqref{eq:paramet}). As a consequence, the gDoF of the channel is completely characterized in these cases. 
From the sum-rate upper bound in~\eqref{eq:ourSumRate}, with the parameterization in~\eqref{eq:paramet}, we can immediately obtain the following gDoF upper bound
\begin{subequations}
\begin{align}
 {\gdof} %(\alpha_p,\alpha_c,\beta) \nonumber
  \leq \frac{1}{2}\min\Big\{
  & {{\gdof}^{\rm(CS)}(\ac,\beta)+{\gdof}^{\rm(CS)}(\ap,0)},\label{eq:Dof upper asym:CS}
\\& {\min\{{\gdof}^{\rm(DT)}(\ac,0),{\gdof}^{\rm(DT)}(\ap,\beta)\}},\label{eq:Dof upper asym:DT}
\\& {{\gdof}^{\rm(PV)}(\ap,\ac,\beta)}\label{eq:Dof upper asym:PV}
\Big\}
\end{align}
where
\begin{align}
&
{\gdof}^{\rm(CS)}(\alpha,\beta) := \max\{1,\min\{\alpha,\beta\}\} \label{eq:Dof upper asym:CS gen}
\\&
{\gdof}^{\rm(DT)}(\alpha,\beta) := \max\{\beta,\alpha,1\}-\alpha+\max\{\alpha,1\} \label{eq:Dof upper asym:DT gen}
\\&
{\gdof}^{\rm(PV)}(\ap,\ac,\beta) := 
 \max \left \{ 1-\ap      ,\ac \right \} \nonumber\\&\qquad
+\max \left \{ 1-\ac+\beta,\ap \right \}. \label{eq:Dof upper asym:PV gen}
\end{align}
\label{eq:Dof upper asym}
\end{subequations}
The proof follows by using the upper bound in~\eqref{eq:ourSumRate} in the gDoF definition in~\eqref{eq:gDoF definition} (the details can be found in Appendix~\ref{app:outer}). The achievability for the interference-symmetric ($\ap= \ac= \alpha$) and the interference-asymmetric cases (either $\ap= 0, \ \ac= \alpha$ or $\ap= \alpha, \ \ac= 0$) will follow from the constant gap results in the next sections.

\subsection{Known inner bounds for the general memoryless CCIC}
\label{sec:knowninner}
To the best of our knowledge, the largest known achievable region for the general memoryless IC with generalized feedback, or bilateral source cooperation, is the {\em superposition+binning} region from~\cite[Section V]{YANG-TUNINETTI}. In this scheme, adapted to the case of unilateral source cooperation, the PTx's message is split into four parts:
the {\em noncooperative common message} and 
the {\em noncooperative private message} 
are sent as in the Han-Kobayashi's scheme for the noncooperative IC~\cite{HK};
the {\em cooperative common message} and 
the {\em cooperative private message} are
decoded at CTx in a given slot and retransmitted in the next slot by using a decode-and-forward based block-Markov scheme. 
The CTx's message is split into two parts:
the {\em noncooperative common message} and 
the {\em noncooperative private message} 
that are sent as in the Han-Kobayashi's scheme for the noncooperative IC~\cite{HK}.
The common messages are decoded at both destinations while non-intended private messages are treated as noise. 
For cooperation, the two sources `beam form' the PTx's cooperative common message to the destinations as in a distributed MIMO system, and the CTx precodes its private messages against the interference created by the PTx's cooperative private message as in a MIMO broadcast channel.
The achievable region in~\cite[Section V]{YANG-TUNINETTI} is quite complex to evaluate because it is a function of 11 auxiliary random variables and is described by about 30 rate constraints per source-destination pair. In this work we will use a small subset of these 11 auxiliary random variables in each parameter regime (see Appendices~\ref{sec:allachschemsappSUP} and~\ref{sec:allachschemsappDPC}) and show that the corresponding schemes are to within a constant gap from the outer bound in~\eqref{eq:ourRate}.

As noted in the Introduction, the largest known achievable region for the IC with unilateral source cooperation is, to the best of our knowledge, the region in \cite[Corollary 1]{MirmohseniIT2012}. The difference between \cite[Corollary 1]{MirmohseniIT2012} and the region in~\cite[Section V]{YANG-TUNINETTI} adapted to the case of unilateral source cooperation is, see \cite[Remark 2, point 6]{MirmohseniIT2012}: ``in \cite[Section V]{YANG-TUNINETTI} binning is done sequentially and conditionally, while \cite[Corollary 1]{MirmohseniIT2012} utilizes joint binning technique. [...] In \cite[Corollary 1]{MirmohseniIT2012} uses joint backward decoding at the receivers, while two-step decoding is used in \cite[Section V]{YANG-TUNINETTI}.'' As far as capacity to within a constant gap is concerned, the results in this paper show that these differences are not fundamental for approximate optimality.

\medskip
Next, in Section~\ref{sec:gap sumcapacity Symmetric Channel} we characterize to within a constant gap the capacity of the {\em symmetric} GCCIC, where the direct links have the same strength and the interfering links have the same strength. This will allow us to identify the key features of the proposed achievable schemes in the strong and weak interference regimes, and set the stage for the gap derivation for the general GCCIC in Section~\ref{sec:general extension in strong interference}, for the general Z-channel in Section~\ref{sec:general extension Z}, and for the general S-channel in Section~\ref{sec:general extension S}.

\section{The capacity region to within a constant gap for the symmetric GCCIC}
\label{sec:gap sumcapacity Symmetric Channel}

The symmetric GCCIC is defined by $\Sp=\Sc={\snr}$ and $\Ip=\Ic={\inr}={\snr}^\alpha$. % in~\eqref{eq:paramet}.
Following the naming convention of the noncooperative IC, we say that the symmetric GCCIC has strong interference if ${\snr} \leq {\inr}$, that is $1\leq \alpha$, and weak interference otherwise. 
Our main result for the symmetric GCCIC is as follows:
\begin{theorem}
\label{thm:gap symmetric} 
For the symmetric GCCIC we have:
\begin{enumerate}%[A)]
\item %(Strong interference) 
${\snr} \leq {\inr}$:  capacity region to within 1~bit with a cooperative scheme based on superposition coding,
\item %(Weak interference) 
${\snr} > {\inr}$ when $\Cc \geq \left({\snr} + {\inr} + 2 \sqrt{{\inr}{\snr} \ \frac{{\inr}}{1+{\inr}}} \right)(1+{\inr})$: capacity region to within 1~bit with a cooperative scheme based on DPC and superposition coding,
\item %(Weak interference) 
${\snr} > {\inr}$ when $\Cc < \left({\snr} + {\inr} + 2 \sqrt{{\inr}{\snr} \ \frac {{\inr}}{1+{\inr}}} \right)(1+{\inr})$:  sum-capacity to within \gap~bits.
\end{enumerate}
\end{theorem}
%\begin{IEEEproof}
%The proof can be found Appendix~\ref{app:gap sym}. 

The rest of the section is devoted to the proof of Theorem~\ref{thm:gap symmetric}.
In order to highlight the key steps in the proof, we use the gDoF as starting point for our discussion. 
The gDoF upper bound for the symmetric GCCIC is obtained by setting $\ap = \ac = \alpha$ in~\eqref{eq:Dof upper asym}.  
%{\red IN  Fig.~\ref{fig:fig3}, SHOULD THE GAPS BE DIVIDED BY 2 (PER USER GAP)?}
Fig.~\ref{fig:fig3} shows the gDoF and the gap (per user) for the symmetric GCCIC for the different regions in the $(\alpha,\beta)$ plane, where the whole set of parameters has been partitioned into multiple sub-regions depending upon different levels of cooperation ($\beta$) and interference ($\alpha$) strengths.  
%These regimes are numbered from 1 to 6 and the details of the gap proof for the $i$-th regime appear in $i$-th item in Appendix~\ref{app:gap sym}, $i\in[1:6]$. We provide next an overview of the achievable schemes used to prove the constant gap result in Theorem~\ref{thm:gap symmetric} for the different parameter regimes.
In regimes~1,~3,~4 and~5 of Fig.~\ref{fig:fig3} the gDoF attained by the symmetric GCCIC is the same as that achieved by the noncooperative IC given by~\cite{etw}
\begin{align*}
{\gdof}^{\rm IC}(\alpha) = \min\{\max\{1-\alpha,\alpha\}, \ \max\{1-\alpha/2,\alpha/2\}, \ 1\}.
\end{align*}
%In these regimes we shall use as an achievable scheme the optimal (to within a constant gap) strategy of the corresponding noncooperative IC and show it is to within a constant number of bits from the outer bound in \eqref{eq:ourRate}. 
Unilateral cooperation therefore provides strict gDoF gain over the noncooperative IC in regimes~2 and~6 of Fig.~\ref{fig:fig3}. 
%In the following we adhere to the naming convention of the noncooperative IC and we say that a link is {\em strong} if its SNR exponent is larger than that of the direct link (which we set to~1 without loss of generality) and {\em weak} otherwise.   For the different parameter regimes in Fig.~\ref{fig:fig3} we have the following.
For reference, the gDoF on the non-causal cognitive IC can be evaluated from~\cite{riniJ1} as 
\begin{align*}
{\gdof}^{\rm CIC}(\alpha) = \max\{1-\alpha/2,\alpha/2\}.
\end{align*}
In general we have
\[
{\gdof}(\alpha,0) = {\gdof}^{\rm IC}(\alpha) 
\leq {\gdof}(\alpha,\beta)
\leq {\gdof}^{\rm CIC}(\alpha) = \lim_{\beta\to+\infty} {\gdof}(\alpha,\beta).
\]
From Fig.~\ref{fig:fig3}, in regime~2 with $\beta \geq \alpha-1$, in regimes 3 and 4, and in regime~6 with $\beta \geq \min \{\alpha,1-\alpha \}$, causal unilateral source cooperation attains the ultimate gDoF limit of the non-causal cognitive IC.
%a sufficient condition for ${\gdof}(\alpha,\beta)={\gdof}^{\rm CIC}(\alpha)$, i.e., causal cooperation attains the ultimate limit of non-causal cognitive radio, is that $\beta \geq \min\{\alpha, |1-\alpha|\}$.

At a high level, the approximately optimal coding schemes are as follows.
In the strong interference and weak cooperation regime both users employ a noncooperative common message.
In the strong interference and strong cooperation regime, PTx's common message becomes cooperative and is forwarded to PRx by CTx. 
In the weak interference regime, each user splits its message into a common and a private part; 
for CTx the two message parts are noncooperative while for PTx are cooperative;
PTx's cooperative common message is the `cloud center' of a superposition coding scheme, and 
PTx's cooperative private message is the `known interference' against which CTx's message is precoded in a DPC-based scheme. 
Binning/DPC is used in the weak interference and strong cooperation regime where CTx can easily decode the signal from PTx because of strong cooperation, but CRx cannot because of weak interference; therefore in this regime it makes sense that the best use of CTx's knowledge of PTx's message is to treat it as a `known state' to precode its message against it.

We shall now discuss each regime of Fig.~\ref{fig:fig3} separately.

%\begin{itemize}

%\item
\subsection{Regime~1 (strong interference): same gDoF as in the noncooperative IC, and capacity region to within 1~bit with a noncooperative scheme} \label{subsec:sum regime 1}
Regime~1 corresponds to very strong interference ($\alpha\geq 2$) and weak cooperation ($\beta\leq 1$).
In the noncooperative IC with very strong interference it is exactly optimal to use only (noncooperative) common messages in order to achieve the whole capacity region; since the interference is very strong, it can be decoded by treating the intended signal as noise, after which each receiver is left with an interference-free point-to-point channel from its transmitter; this noncooperative strategy achieves
\begin{subequations}
\begin{align}
\mathcal{I}^{\ref{subsec:sum regime 1}} : \quad
\Rc&\leq \log(1+{\snr}),
\label{eq:sum ac regime 1 rc}
\\
\Rp&\leq \log(1+{\snr}),
\label{eq:sum ac regime 1 rp}
\end{align}
\label{eq:sum ac regime 1}
\end{subequations}
or ${\gdof}\geq(1+1)/2=1$.
Since the cooperation link is weak in regime~1, the amount of data PTx could communicate to CTx for cooperation is very limited. As a result in this regime unilateral cooperation does not improve performance compared to the noncooperative case. In other words, in regime~1, cooperation provides a `beam forming gain' but not a gDoF gain. To see this, the cut-set upper bounds on individual rates in~\eqref{eq:CutSetRc} and~\eqref{eq:CutSetRp}, in the symmetric case for $\beta\leq 1 \Longleftrightarrow \Cc \leq {\snr}$, give the following upper bounds on the individual rates
\begin{subequations}
\begin{align}
\mathcal{O}^{\ref{subsec:sum regime 1}} : \quad
   \Rc &\leq \log(1+{\snr}),
\label{eq:sum up regime 1 rc}
\\ \Rp &\leq \log(1+{\snr}+\Cc) \leq \log(1+2{\snr})\leq \log(1+{\snr})+\log(2).
\label{eq:sum up regime 1 rp}
\end{align}
\label{eq:sum up regime 1}
\end{subequations}
From the upper bound on $\Rp$ in~\eqref{eq:sum up regime 1 rp}, we see that unilateral cooperation can at most double the SNR on the primary direct link, which can at most increase the rate by 1~bit compared to the noncooperative case. As a result, the gDoF with unilateral cooperation is ${\gdof}=1$ and the rate pair in~\eqref{eq:sum ac regime 1} is optimal to within 1~bit, i.e., $\max\{{\rm eq}\eqref{eq:sum up regime 1 rc}-{\rm eq}\eqref{eq:sum ac regime 1 rc}, {\rm eq}\eqref{eq:sum up regime 1 rp}-{\rm eq}\eqref{eq:sum ac regime 1 rp}\}\leq \max\{0,\log(2)\}=1 \ \rm{bit}$. 

%Note that in regime~1, the 1~bit gap result holds for the whole capacity region and not just for the sum-capacity as claimed in Theorem~\ref{thm:gap symmetric}.  Moreover, the optimal noncooperative scheme is optimal to within 1~bit for the whole region.

%\item 
\subsection{Regime~2 (strong interference): improved gDoF compared to the noncooperative IC, and capacity region to within 1~bit with a cooperative scheme} \label{subsec:sum regime 2}
In regime~2 the interference is very strong ($\alpha \geq 2$) and the cooperation is strong ($\beta > 1$). 
Similarly to the noncooperative very strong interference regime, the transmitters send a common message only.
As opposed to regime~1, where both messages were sent noncooperatively, here the PTx takes advantage of the strong cooperation link and sends its message to PRx with the help of the CTx. In order to enable cooperation, a block Markov coding scheme is used as follows. Transmission is over a frame of $B\gg 1$ slots. In slot $t\in[1:B]$, the PTx sends its old (cooperative common) message $W_{\mathsf{p},t-1}$ and superposes to it the new (cooperative common) message $W_{\mathsf{p},t}$, while the CTx forwards the primary old (cooperative common) message $W_{\mathsf{p},t-1}$ and superposes to it its (noncooperative common) message $W_{\mathsf{c},t}$. At the end of slot $t$, CTx decodes the new message $W_{\mathsf{p},t}$ after subtracting the contribution of the old message $W_{\mathsf{p},t-1}$. The destinations wait until the whole frame has been received and then proceed to jointly backward decode all messages. The details can be found in Appendix~\ref{sec:scheme 1} and the achievable region is given in~\eqref{eq:sup.only ach.region with common messages only in unilateral cooperation V1=X1 gaussian}, which in the symmetric GCCIC in very strong interference reduces to
\begin{subequations}
\begin{align}
\mathcal{I}^{\ref{subsec:sum regime 2}} : \quad
   \Rc &\leq \log(1+{\snr}),
\label{eq:sum ac regime 2 rc}
\\ \Rp &\leq \log(1+\Cc),
\label{eq:sum ac regime 2 rp}
\\ \Rp+\Rc &\leq \log(1+{\snr}+{\inr}).
\label{eq:sum ac regime 2 rsum}
\end{align}
\label{eq:sum ac regime 2}
\end{subequations}
The region in~\eqref{eq:sum ac regime 2} is strictly larger than the noncooperative capacity region in very strong interference given by~\eqref{eq:sum ac regime 1} for ${\snr}(1+{\snr})\leq {\inr}$, or $\alpha \geq 2$, and $\Cc > {\snr}$, or $\beta>1$, which is precisely the definition of regime~2.
The sum-capacity from~\eqref{eq:sum ac regime 2} can take two possible values, depending on which one among the MAC sum-rate bound in~\eqref{eq:sum ac regime 2 rsum} and the sum of the bounds on the individual rates in~\eqref{eq:sum ac regime 2 rc}-\eqref{eq:sum ac regime 2 rp} is the most stringent. In particular, the following sum-rate is achievable
%In the symmetric case $\Sc=\Sp={\snr}, \ \Ic=\Ip={\inr}$ the following sum-rate is therefore achievable based on~\eqref{eq:sup.only ach.region with common messages only in unilateral cooperation V1=X1 gaussian}
\begin{align*}
\Rp + \Rc &\leq
\left\{\begin{array}{ll}
\log(1+\Cc)+\log(1+{\snr}) & \text{if $\Cc(1+{\snr})\leq {\inr}$}\\
\log\left(1+{\snr}+{\inr}\right)& \text{if $\Cc(1+{\snr})> {\inr}$} \\
\end{array}\right.,
\end{align*}
that is,
${\gdof} \geq (\beta+1)/2$ if $\beta+1 \leq \alpha$ and
${\gdof} \geq \alpha/2$ otherwise (in either case the gDoF is larger than ${\gdof}^{\rm IC}=1$).

%To prove the approximate optimality of the region in~\eqref{eq:sum ac regime 2},
From the outer bound region obtained from the cut-set upper bounds on the individual rates in~\eqref{eq:CutSetRc} and~\eqref{eq:CutSetRp} and the sum-rate upper bound in~\eqref{eq:Tuninetti}, under the condition $\beta> 1 \Longleftrightarrow \Cc > {\snr}$, we have that any achievable rate pair must satisfy
\begin{subequations}
\begin{align}
\mathcal{O}^{\ref{subsec:sum regime 2}} : \quad
   \Rc &\leq \log(1+{\snr}),
\label{eq:sum up regime 2 rc}
\\ \Rp &\leq \log(1+{\snr}+\Cc) \leq  \log(1+\Cc)+\log(2)
\label{eq:sum up regime 2 rp},
\\ \Rp + \Rc &\leq \log\left(1+(\sqrt{\snr}+\sqrt{\inr})^2\right)\leq\log(1+{\snr}+{\inr})+\log(2),
\label{eq:sum up regime 2 rsum}
\end{align}
\label{eq:sum up regime 2}
\end{subequations}
since $(\sqrt{x}+\sqrt{y})^2 \leq 2(x+y), \ \forall(x,y)\in\mathbb{R}^2_+$,
The upper bound in~\eqref{eq:sum up regime 2} and the achievable region in~\eqref{eq:sum ac regime 2} are to within 1~bit of one another since
\[
\mathsf{GAP} \leq \max\left\{
{\rm eq}\eqref{eq:sum up regime 2 rc}-{\rm eq}\eqref{eq:sum ac regime 2 rc},
{\rm eq}\eqref{eq:sum up regime 2 rp}-{\rm eq}\eqref{eq:sum ac regime 2 rp},
\frac{{\rm eq}\eqref{eq:sum up regime 2 rsum}-{\rm eq}\eqref{eq:sum ac regime 2 rsum}}{2}
\right\} 
%\leq \max\{0, \log(2), 1/2 \ \log(2)\} 
\leq \log(2).
\]
This shows that the whole capacity region, and therefore  the gDoF ${\gdof}=\min\{\beta+1, \ \alpha\}/2$ too, is achievable to within 1~bit in regime~2. 

%As for regime~1, also in regime~2 the 1~bit gap result holds for the whole capacity region and not just for the sum-capacity as claimed in Theorem~\ref{thm:gap symmetric}. In regime~2 the optimal scheme to within 1~bit requires cooperation.

%Due to the interference regime, the scheme used here involves only common messages, i.e., messages that are decoded at both destinations, which are superimposed one over the other (power split at the sources in Gaussian). Moreover, the CTx (cooperative source), at the end of each time slot, fully decodes the message of the PTx and cooperates in sending it in the next slot. In other words, in this regime, we exploit a block Markov superposition encoding scheme that involves three auxiliary random variables: $Q$ which conveys the cooperative common message of the PTx from the previous time slot, $V_1$ which conveys the cooperative common message of PTx of the current time slot and $U_2$ which conveys the noncooperative common message of CTx of the current time slot. Notice that, since PTx does not assist the transmission of CTx, the message of the cognitive source is noncooperative. Finally, both the receivers apply backward decoding after the last slot of the frame has been received.

%\item 
\subsection{Regime~3 (strong interference): same gDoF as in the noncooperative IC, and capacity region to within 1~bit with a cooperative scheme} \label{subsec:sum regime 3}
Regime~3 corresponds to strong but not very strong interference ($\alpha\in[1,2)$). Note that there are no restrictions on the cooperation exponent $\beta$ in this regime.
Similarly to regimes~1 and~2, here we use only 
%noncooperative 
common messages -- a strategy that is capacity achieving in the corresponding noncooperative IC. The difference between regime~1 and regime~3 is that stripping decoding is no longer optimal and the receivers must instead jointly decode the intended and non-intended messages as in a MAC. By taking the largest between the achievable region developed for regime~2 in~\eqref{eq:sum ac regime 2} and the noncooperative achievable region for this regime (i.e., common messages only, which has $\Rp \leq \log(1+{\snr})$ as a bound on the primary rate rather than $\Rp \leq \log(1+\Cc)$) we obtain the following achievable region
\begin{subequations}
\begin{align}
\mathcal{I}^{\ref{subsec:sum regime 3}} : \quad
   \Rc &\leq \log(1+{\snr}),
\label{eq:sum ac regime 3 rc}
\\ \Rp &\leq \log(1+\max\{\Cc,{\snr}\}),
\label{eq:sum ac regime 3 rp}
\\ \Rp+\Rc &\leq \log(1+{\snr}+{\inr}),
\label{eq:sum ac regime 3 rsum}
\end{align}
\label{eq:sum ac regime 3}
\end{subequations}
which implies ${\gdof} \geq \min\{1+\max\{1,\beta\}, \ \max\{1,\alpha\} \}/2 = \alpha/2$, i.e., the sum-rate bound in~\eqref{eq:sum ac regime 3} is the tightest.
In regime~3, no matter how strong the cooperation link is, cooperation does not improve the noncooperative gDoF.
%An intuitive reason for this is that when the interfering signal is not received at a significantly different power level compared to the intended signal (i.e., $\alpha$ is not so much larger than~1), then there is no better strategy than decoding the interference along with the intended signal as far as gDoF is concerned.

From the outer bound region obtained from the cut-set upper bounds on the individual rates in~\eqref{eq:CutSetRc} and~\eqref{eq:CutSetRp} and the sum-rate upper bound in~\eqref{eq:Tuninetti}, we have that any achievable rate pair must satisfy
\begin{subequations}
\begin{align}
\mathcal{O}^{\ref{subsec:sum regime 3}} : \quad
   \Rc &\leq \log(1+{\snr}),
\label{eq:sum up regime 3 rc}
\\ \Rp &\leq \log(1+{\snr}+\Cc) \leq  \log(1+\max\{\Cc,{\snr}\})+\log(2),
\label{eq:sum up regime 3 rp}
\\ \Rp + \Rc &\leq \log\left(1+(\sqrt{\snr}+\sqrt{\inr})^2\right)\leq\log(1+{\snr}+{\inr})+\log(2).
\label{eq:sum up regime 3 rsum}
\end{align}
\label{eq:sum up regime 3}
\end{subequations}
%the cooperative sum-rate upper bound in~\eqref{eq:Tuninetti}, we have that with unilateral cooperation the sum-capacity must satisfy
%\[
%\Rp+\Rc 
%\leq \log\left(1+(\sqrt{\snr}+\sqrt{\inr})^2\right)
%\leq \log(1+{\snr}+{\inr})+\log(2).
%\]
It is easy to see that the regions in~\eqref{eq:sum up regime 3} and~\eqref{eq:sum ac regime 3} are to within 1~bit of one another.
%, which implies ${\gdof} = \alpha/2$.
% since
%\[
%\max\left\{
%{\rm eq}\eqref{eq:sum up regime 3 rc}-{\rm eq}\eqref{eq:sum ac regime 3 rc},
%{\rm eq}\eqref{eq:sum up regime 3 rp}-{\rm eq}\eqref{eq:sum ac regime 3 rp},
%\frac{{\rm eq}\eqref{eq:sum up regime 3 rsum}-{\rm eq}\eqref{eq:sum ac regime 3 rsum}}{2}
%\right\} \leq \max\{0, \log(2), 1/2 \ \log(2)\} = \log(2).
%\]

%As for regimes~1 and~2, also in regime~3 the 1~bit gap result holds for the whole capacity region and not just for the sum-capacity as claimed in Theorem~\ref{thm:gap symmetric}.  In regime~3 the optimal scheme to within 1~bit requires cooperation.

%\item 
\subsection{Regime~4 (weak interference): same gDoF as in the noncooperative IC} \label{subsec:sum regime 4}
Regime~4 corresponds to moderately weak interference ($\alpha\in[2/3, 1)$). In this regime, rate splitting is needed to achieve the capacity to within 1 bit in the noncooperative IC~\cite{etw}. Therefore we propose to use here the noncooperative scheme that consists of two messages for each user: the noncooperative common and the noncooperative private. The power of the noncooperative private message (which is treated as noise at the non-intended receiver) is such that it is received at or below the receiver noise floor~\cite{etw}. As shown in~\cite{etw}, in the moderately weak interference regime the sum-rate upper bound of~\cite[Theorem 1]{Kramer} can be achieved to within 1~bit per user, that is, the following sum-rate is achievable
\begin{align}
\Rp+\Rc &\leq \log\left(1+{\snr}+{\inr}\right) 
          + \log(1+{\snr})-\log(1+{\inr}) -2\log(2),
\label{eq:sum ac regime 4}
\end{align}
or ${\gdof} \geq \frac{\max\{1,\alpha\}+(1-\alpha)}{2} = 1-\alpha/2$.
The cooperative sum-rate upper bound in~\eqref{eq:Tuninetti} can be further upper bounded as 
\begin{align}
\Rp+\Rc & \leq
%\leq \log\left(1+(\sqrt{\snr}+\sqrt{\inr})^2\right) 
%         + \log(1+{\snr})-\log(1+{\inr})\nonumber
%      \\&\leq 
      \log\left(1+{\snr}+{\inr}\right)
         + \log(1+{\snr})-\log(1+{\inr})+\log(2).
\label{eq:sum up regime 4}
\end{align}
Therefore, the gap is at most $\mathsf{GAP} \leq\frac{{\rm eq}\eqref{eq:sum up regime 4}-{\rm eq}\eqref{eq:sum ac regime 4}}{2}\leq 3/2 \ \log(2)$ and is achieved by the noncooperative scheme with rate splitting as in~\cite{etw}.

\bigskip
In order to claim capacity to within a constant gap in the weak interference regime, we must derive an upper bound that reduces to, or is to within a constant gap of, the capacity outer bound in~\cite[Theorem 3]{etw} when $\Cc=0$. The outer bound region \cite[Theorem 3]{etw} is characterized by bounds on the individual rates, bounds on the sum-rate, and by bounds on $2\Rc+\Rp$ and $\Rc+2\Rp$.
%(see for example the gDoF region in~\cite[Figure 23]{etw}). 
Therefore, unless outer bounds on $2\Rc+\Rp$ and $\Rc+2\Rp$ for the cooperative case are developed, it is not possible to claim optimality to within a finite gap of the upper bound in~\eqref{eq:ourRate} for small $\Cc$. 
Developing outer bounds on $2\Rc+\Rp$ and $\Rc+2\Rp$ for the general IC with source cooperation is an important open problem, which is outside the scope of this work. An interesting question that could be answered by such a line of research is as follows. In~\cite{suhtse:ICwithfeedback}, the authors interpreted the bounds on $2\Rc+\Rp$ and $\Rc+2\Rp$ as a measure of the amount of `resource holes', or inefficiency, due to the distributed nature of the noncooperative IC. In~\cite{suhtse:ICwithfeedback}, the authors showed that with output feedback from a destination to its source, such `resource holes' are no longer present; in other words, feedback enables coordination among the sources which results in a full utilization of the channel resources.
An interesting open question is whether unilateral cooperation enables sufficient coordination among the sources for full utilization of the channel resources.
In the limiting case where unilateral cooperation equals non-causal cognition, we know from~\cite{riniJ1} that the capacity region does not have bounds on $2\Rc+\Rp$ and $\Rc+2\Rp$, i.e., there are no `resource holes'. Therefore the question can be rephrased as: is there a minimum strength of the cooperation link $\Cc$ above which unilateral causal cooperation results in no `resource holes' in weak interference, i.e., bounds on $2\Rc+\Rp$ and $\Rc+2\Rp$ are not needed to (approximately) characterize the capacity region?
%With the achievable schemes presented in this paper, we have not been able to answer this question in the positive.

%\item 
\subsection{Regime~5 (weak interference): same gDoF as in the noncooperative IC}  \label{subsec:sum regime 5}
In regime~5 the interference is moderately weak ($\alpha\in[1/2, 2/3)$) and the cooperation is fairly weak ($0\leq \beta < 2\alpha-1$).
The gDoF upper bound gives ${\gdof}=\alpha$ as for the noncooperative IC. Hence in this regime we use the scheme that is approximately optimal for the sum-capacity of the noncooperative IC, with noncooperative common and private messages and with power splits as in~\cite{etw}. The noncooperative scheme achieves 
\begin{align}
\Rp+\Rc &\leq 2\log\left(1+{\inr}+\frac{{\snr}}{\max\{1,{\inr}\}}\right) -2\log(2).
\label{eq:sum ac regime 5}
\end{align}
The cooperative sum-rate upper bound in~\eqref{eq:PV} can be further upper bounded as 
\begin{align}
\Rp+\Rc &\leq 2\log\left(1+{\inr}+\frac{{\snr}}{\max\{1,{\inr}\}}\right)+2\log(2)+\Delta^\prime,
\label{eq:sum up regime 5}
\end{align}
where $\Delta^\prime$ is the latest $\Delta$ in~\eqref{eq:deltaPV} in the regime 
$\beta < 2\alpha-1 \Longleftrightarrow \Cc < {\inr}^2/{\snr}  \Longleftrightarrow \frac{{\snr}}{{\inr}} < \frac{{\inr}}{\Cc}$
within the weak interference regime $1 \leq \frac{{\snr}}{{\inr}}$, that is,
\begin{align*}
\Delta^\prime 
&= 
 \max_{ 1\leq \frac{{\snr}}{{\inr}} < \frac{{\inr}}{\Cc}}%,  \ 1 \leq {\inr}    
\log\frac{(1 + \Cc)
\left( 1 +  \left( \sqrt{\frac{{\snr}}{{\inr}}}  +  \sqrt{\frac{{\inr}}{\Cc}} \right)^2 \right)}
      {1 +  \left( \sqrt{\frac{{\snr}}{{\inr}}}  +  \sqrt{{\inr}}       \right)^2} 
\\&\leq 
 \max_{ 1\leq \frac{{\snr}}{{\inr}} < \frac{{\inr}}{\Cc}}%,  \ 1 \leq {\inr}    
\log\frac{(1 + \Cc)
\left( 1 +  2\frac{{\snr}}{{\inr}}+2\frac{{\inr}}{\Cc} \right)}
      {1 +  \frac{{\snr}}{{\inr}}+{\inr}} 
%\\
%\\&=\max_{1 \leq {\inr}\leq \Cc}
%\log\frac{(1 + \Cc)
%\left( 1 +  4 \ {\inr}/{\Cc} \right)}
%      {1 +  \left( \sqrt{\frac{{\inr}}{\Cc}}  +  \sqrt{{\inr}}       \right)^2} 
\\&=\max_{1 \leq \frac{{\inr}}{\Cc}}
\log\frac{(1 + \Cc)
\left( 1 +  4 \ \frac{{\inr}}{\Cc} \right)}
      {1 +  \frac{{\inr}}{\Cc}(1+\Cc)}
\\&=
\log
\max\left\{
\frac{(1 + \Cc)
\ 5 }{2+\Cc},
\
\frac{(1 + \Cc)
\ 4 }{1+\Cc}
\right\}
\leq  \log(5),
\end{align*}
where in the derivation we used $1\leq \Cc$ (note that for $\Cc<1$ the outer bounds in~\eqref{eq:ourRate} are to within a constant gap of the corresponding bounds for $\Cc=0$).
Therefore, the gap (per user) is at most $\mathsf{GAP} \leq\frac{{\rm eq}\eqref{eq:sum up regime 5}-{\rm eq}\eqref{eq:sum ac regime 5}}{2}\leq \frac{(2+2)\log(2)+\log(5)}{2} \approx 3.16 \log(2)$ and is achieved by the noncooperative scheme.

The observations we made for regime~4, regarding possible extensions to the whole capacity region in the general case, apply to regime~5 as well.

%Note that the largest gap occurs in this regime. A possible way to reduce the gap would due to develop a tighter upper bound than the one used here from~\cite{PVIT11}, or more sophisticated achievable schemes than what we did so far.

%\item 
\subsection{Regime~6 (weak interference): improved gDoF compared to the noncooperative IC}  \label{subsec:sum regime 6}
In regime~6, the interference is quite weak ($\alpha < 2/3$) and the cooperation exponent satisfies $\beta \geq [2\alpha -1]^+$. Since the interference is weak, we split the messages into a common part and a private part, as for the noncooperative IC. For CTx the two messages are noncooperative, but for PTx the common message is cooperative and the private message is noncooperative, in other words, in regime~6 we extend the scheme used in regime~2 by adding a private message.
The cooperation mechanism is based on decode-and-forward: at any given time slot of a block Markov coding scheme CTx decodes the primary common message, which PTx and CTx `beam form' to the receivers in the next slot. The new common and private messages of each user are superposed to the old primary cooperative common message. The details of the achievable scheme are reported in Appendix~\ref{sec:scheme 2}, where we show that the sum-rate in~\eqref{eq:after long computations scheme 2 sym}, namely
%\begin{subequations}
\begin{align*}
\Rp + \Rc \leq  \min 
\Big \{ 
& \log \left( 1+\frac{{\snr}}{2{\inr}} \right) 
+ \log \left( \frac{{\snr}+{\inr}+1}{2} \right),
\\& 
\log \left( 1+\frac{{\snr}}{2{\inr}} \right)
 +\log \left( \frac{1+\Cc}{{\inr}+\Cc} \right)
 +\log \left( \frac{{\snr}+{\inr}^2+{\inr}}{2} \right)
\Big\},
\end{align*}
%\label{eq:after long computations scheme 2 sym again}
%\end{subequations}
is achievable. 
Depending on which expression attains the minimum, we obtain the four subregions, indicated as from 6a to 6d, into which regime~6 is subdivided. In particular, for subregions~6a and~6b the tightest outer bound is the one in~\eqref{eq:Tuninetti}, while for subregions~6c and~6d the tightest sum-rate outer bound is the one in~\eqref{eq:PV}. Note that the outer bound in~\eqref{eq:PV} reduces to the more involved part of the W-curve of~\cite{etw} for $\alpha <2/3$ when $\beta=0$.
In Appendix~\ref{app:gap sym regime 6} we show that this scheme is optimal to within 2.5~bits.

%Differently from a common message, a private message is just decoded at the intended destination and treated as noise at the other destination. These messages are then superimposed one over the other, which means that a power split is applied at the two sources in the Gaussian noise channel. Therefore, in this regime, we exploit a block Markov superposition encoding scheme that involves four auxiliary random variables: $V_1$ which conveys the cooperative common message of PTx of the current time slot (as for regime 2), $T_1$ which conveys the noncooperative private message of PTx of the current time slot, $U_2$ which conveys the noncooperative common message of CTx of the current time slot (as for regime 2) and $T_2$ which conveys the noncooperative private message of CTx of the current time slot. As for regime 2, at the end of the transmission, both the receivers apply backward decoding.

\medskip
The achievable scheme used for regime~6 (defined as $\alpha < 2/3$) is also optimal to within a constant gap for most of regime~4 (defined as $\alpha \in [2/3,1)$). 
In particular, as a consequence of the gap derivation in Appendix~\ref{app:gap sym regime 6}, the achievable scheme for regime~6 and the outer bound in~\eqref{eq:Tuninetti} are to within a constant gap of one another when the interference is weak ($\alpha \leq 1$) and the cooperation satisfies $\beta \geq \min\{\alpha, 1-\alpha\}$.

%The gap result for the ``Weak Interference~2'' regime (item~\ref{item: weak 2 a gap} in Appendix~\ref{app:gap sym}) also holds for a big part of the ``Moderately Weak Interference'' regime (item~\ref{item:mod weak} in Appendix~\ref{app:gap sym}). In particular, it only does not apply in the regime $\alpha \in [2/3,1]$ and $\beta < 1 - \alpha$ where the proposed scheme for the ``Weak Interference 2'' regime (item~\ref{item: weak 2 a gap} in Appendix~\ref{app:gap sym}) does not achieve the optimal gDoF.

\medskip
The largest gap in regime~6 is of 2.5~bits in sub-regimes 6c and 6d, where the tightest sum-rate outer bound is the one in~\eqref{eq:PV}. This gap may be decreased in several ways. For example, one can develop tighter bounds than the one in~\eqref{eq:PV}, or develop more involved coding schemes.  An example of the latter method can be found next, where we consider a DPC-based achievable scheme for the weak interference regime / regimes~4 and~6.

\subsection{Regimes~4 and~6 (weak interference) with strong cooperation: capacity to within 1~bit with a cooperative scheme} \label{subsec:sum regimes 4 and 6}
We return on an observation made earlier, namely, that when the cooperation link gain $\Cc$ is sufficiently large, we expect the performance of the GCCIC to approach that of the non-causal cognitive IC. We next show that a DPC-based scheme 
%from Appendix~\ref{sec:scheme 3}, which achieves the sum-capacity to within 1~bit for $\Cc \geq {\snr}$, as shown in Section~\ref{subsec:sum regime 6}, is actually 
is optimal to within 1~bit for the whole capacity region in the weak interference regime when the cooperation gain $\Cc $ is sufficiently strong, and we give a sufficient condition to quantify what `sufficiently strong $\Cc$' means.

In the DPC-based achievable scheme in Appendix~\ref{sec:scheme 3}, the primary private message is cooperative, while in the scheme used previously for regime~6 in Appendix~\ref{sec:scheme 2} it was noncooperative. Here we propose that CTx, with knowledge of PTx's primary private message, uses DPC to rid CRx of the interference due to the primary private message. In particular, PTx  sends $\Xp = \gp S+ \sqrt{1-|\gp |^2} \Up$, for some $|\gp |^2 \leq 1$, where $S$ carries the PTx's old private cooperative message and $\Up$ carries the PTx's new private cooperative message in a block Markov coding scheme. CTx sends $\Xc = \gc S+ \sqrt{1-|\gc |^2} \Uc$, for some $|\gc |^2 \leq 1$, where $\Uc$ carries the CTx's private noncooperative message. In a given time slot, CTx knows PTx's old private cooperative message $S$ and decodes PTx's new private cooperative message $\Up$ from its channel output. CTx then precodes its private noncooperative message against the `known interference' $S$; thanks to DPC, CRx decodes $\Uc$ as if the interference $S$ was not present~\cite{costaDPC}, while treating $\Up$ as noise. PRx does backward decoding in order to recover its message while treating $\Uc$ as noise. 
This DPC-based scheme %from Appendix~\ref{sec:scheme 3}
is similar to the capacity achieving scheme for the non-causal cognitive IC in weak interference~\cite{sriranCICweak,viswanathCICweak}, except for the fact that now CTx must decode PTx's message in $\Up$, and that CRx's equivalent noise variance includes the interference due to $\Up$. To overcome this last problem, inspired by~\cite{etw}, we choose the power split $\gp$ in such a way that the interference created by $\Up$ at CRx is at the same level of the noise. %, that is, $(1-|\gp |^2) = \frac{1}{1+{\inr}}$. 
With this choice of parameters the achievable region in~\eqref{eq:the simple DPC region}, specialized to the symmetric case, becomes 
\begin{subequations}
\begin{align}
\mathcal{I}^{\ref{subsec:sum regimes 4 and 6}} : \quad
   \Rp &\leq \log \left( 1+\frac{\Cc}{1+{\inr}} \right)
\label{eq:the simple DPC region sym+etw rp1}
\\ \Rp &\leq \log \left( 1+\frac{{\snr} +|\gc|^2 {\inr} + 2|\gc|  \sqrt{{\inr}{\snr} \ \frac{{\inr}}{1+{\inr}}} }{1+ (1-|\gc|^2){\inr}} \right)
\label{eq:the simple DPC region sym+etw rp2}
\\ \Rc &\leq  \left( 1 +  \frac{(1-|\gc|^2){\snr}}{1+\frac{{\inr}}{1+{\inr}}} \right)
\label{eq:the simple DPC region sym+etw rc}
\end{align}
\label{eq:the simple DPC region sym+etw}
\end{subequations}
for all $|\gc|\leq 1$. Under the condition
\begin{align}
&\frac{\Cc}{1+{\inr}} 
\geq \max_{|\gc|\leq 1}\frac{{\snr} +|\gc|^2 {\inr} + 2|\gc|   \sqrt{{\inr}{\snr} \ \frac{{\inr}}{1+{\inr}}} }{1+ (1-|\gc|^2){\inr}}
\Longleftrightarrow
\nonumber\\
&\Cc 
\geq \left({\snr} + {\inr} + 2 \sqrt{{\inr}{\snr} \ \frac{{\inr}}{1+{\inr}}} \right)(1+{\inr})
%= \frac{{\snr}}{1+ {\inr}}+{\inr}\left(\sqrt{\frac{{\snr}}{1+ {\inr}}}+1\right)^2
%\stackrel{\text{for ${\inr}\leq {\snr}$}}{\in}[1,\ 4]{\snr}(1+{\snr}),
\quad (\Longleftrightarrow \beta \geq 1+\alpha)
\label{eq:the simple DPC region sym+etw condition}
\end{align}
the constraint in~\eqref{eq:the simple DPC region sym+etw rp1} is redundant. 

The achievable region under the condition in~\eqref{eq:the simple DPC region sym+etw condition} must next be compared to an outer bound.
We use here as an outer bound the capacity region of the non-causal cognitive IC given in~\eqref{eq:the noncausalcognitive capacity weak}. By comparing~\eqref{eq:the noncausalcognitive capacity weak rp init} with~\eqref{eq:the simple DPC region sym+etw rp2}, and~\eqref{eq:the noncausalcognitive capacity weak rc init} with~\eqref{eq:the simple DPC region sym+etw rc}, it is easy to see that for every value of $|\gc|\leq 1$ the two regions are at most $\mathsf{GAP} \leq \log\left(1+\frac{{\inr}}{1+{\inr}}\right)\leq\log(2)=$~1~bit away. 
This capacity result to within a constant gap holds for sufficiently large $\Cc$ and it agrees with the intuition that the GCCIC should perform more and more as the non-causal cognitive IC as $\Cc$ increases.

\medskip
If we only consider the sum-capacity, in Appendix~\ref{sec:scheme 3gap} we show that the scheme in~\eqref{eq:the simple DPC region}, of which the scheme in~\eqref{eq:the simple DPC region sym+etw} is a special case, achieves the sum-capacity upper bound in~\eqref{eq:Tuninetti} to within 1~bit when the channel gains satisfy $\Cc \geq {\snr}$, that is, $\beta \geq 1$, which is smaller than the gap of~2.5~bits we found with the superposition-based scheme. Note that the condition $\Cc \geq {\snr}$ for sum-capacity approximate optimality is less restrictive than the one in~\eqref{eq:the simple DPC region sym+etw condition} (which is approximately $\Cc\geq 4{\snr}(1+\inr)$) needed for the approximate optimality of the whole rate region.
%In Appendix~\ref{sec:scheme 3gap} we use the DPC-based achievable scheme for the weak interference ($\alpha<1$) and strong cooperation ($\beta>1$) regime and show it achieves the sum-capacity within $2$~bits, rather than $3$~bits as the previous proposed scheme based on superposition coding only. 

%\end{itemize}
%\end{IEEEproof}

\medskip
We have now concluded the proof of Theorem~\ref{thm:gap symmetric}. Before concluding this Section, we compare the gDoF performance of the symmetric GCCIC with that of other channel models so as to determine when unilateral cooperation may be worth implementing in practical systems.

%\item 
\subsection{Comparisons}
When the gDoF, or high SNR throughput, is the desired performance metric, we can make the following observations:
\begin{itemize}%[1.]

\item
Causal unilateral source cooperation does not improve on the gDoF of the noncooperative IC when
\[
\alpha\in\left[\frac{2}{3}, 2 \right] \ \text{or} \ \beta \leq \min \left \{ 1,[2\alpha-1]^+ \right \}
\] 
as shown by the green and yellow-shaded regions in Fig.~\ref{fig:fig6}, that is, 
the regimes 1, 3, 4 and 5 in Fig.~\ref{fig:fig3}.
%In this case our baseline strategies with $\tau = 0$ achieve the gDoF upper bound.
%{OK \blue
%\begin{itemize}
%\item $\alpha \geq 2, \beta \leq 1$;
%\item $\frac{2}{3} \leq \alpha \leq 2$;
%\item $\frac{1}{2} \leq \alpha \leq \frac{2}{3}, \beta \leq 2 \alpha -1$.
%\end{itemize}
%}
%
For this set of parameters, unilateral cooperation might not be worth implementing in practical systems since the same gDoF is achieved without explicit cooperation, i.e., unilateral cooperation only provides a power gain.

\item
In the regime $1 \leq \alpha \leq \beta$, unilateral cooperation attains the gDoF of the classical relay channel given by ${\gdof}^{\rm RC} = \max\{1,\min\{\alpha,\beta\}\}=\alpha$,
as shown by the red and yellow-shaded regions in Fig.~\ref{fig:fig6},
i.e., parts of the regime 2 and regime 3 in Fig.~\ref{fig:fig3} where ${\gdof}=\alpha/2$,
which correspond to a subset of the strong interference where the cooperation link is greater than the interference link.
%In this case our baseline strategies with $\tau = 1$ achieve the gDoF upper bound.
%
For this set of parameters cognitive radio might not be worth implementing in practical systems since the rate $\Rc=0$ for the cognitive pair is approximately sum-capacity optimal. There are however other rate pairs $(\Rc,\Rp)$ attaining the optimal sum-rate with $\Rc>0$.

\item
The gDoF of the GCCIC is equal to that of the non-causal cognitive IC, given by ${\gdof}=\max\{1-\alpha/2,\alpha/2\}$, everywhere except in the regimes 5, 6c and 6d in Fig.~\ref{fig:fig3}, and for $\alpha \geq \max\{2, \beta + 1\}$,
as shown by the horizontal-line-shaded region in Fig.~\ref{fig:fig6}.
%OK
%\begin{itemize}
%\item $\alpha \geq 2, \beta \geq \alpha-1$;
%\item $\frac{2}{3} \leq \alpha \leq 2$;
%\item $0 \leq \alpha \leq \frac{2}{3}, \beta \geq \min \{1-\alpha,\alpha \}$.
%\end{itemize}
For this set of parameters unilateral cooperation attains the ultimate performance limits of non-causal cognitive radio and therefore represents the ideal channel condition for cognitive radio.

\item
The gDoF of unilateral cooperation equals that of bilateral cooperation, with cooperation links of the same strength as considered in~\cite{PVIT11}, when
%{\blue in regime 5 are the same}
\[
\beta \leq 1 \ \text{or} \ \beta \in \Big[[\alpha -1]^+, \alpha\Big] 
\ \text{except in the regimes 6c and 6d in Fig.~\ref{fig:fig3}}
\]
as shown by the vertical-line-shaded region in Fig.~\ref{fig:fig6}.
%OK
%{\blue
%\begin{itemize}
%\item $\alpha \geq 2, \beta \leq 1$;
%\item $\alpha \geq 2, \alpha -1 \leq \beta \leq \alpha$;
%\item $1 \leq \alpha \leq 2, \beta \leq \alpha$;
%\item $\frac{2}{3} \leq \alpha \leq 1, \beta \leq 1$;
%\item $\frac{1}{2} \leq \alpha \leq \frac{2}{3}, \beta \leq 2 \alpha -1$;
%\item $0 \leq \alpha \leq \frac{1}{2}, \alpha \leq \beta \leq 1-\alpha$;
%\item $0 \leq \alpha \leq \frac{2}{3}, 1-\alpha \leq \beta \leq 1$.
%\end{itemize}} 
For this set of parameters unilateral cooperation attains the same gDoF of bilateral cooperation but with less resources and therefore represents a better trade-off in practical systems.
%{\red WOULD BE INTERESTING TO SEE WHETHER THIS UNILATERAL SCHEME GIVES A SMALLER GAP FOR BILATERAL COOPERATION  THAN WHAT IN THE PV SCHEME/20 BITS.}

\item
For the symmetric case, our analysis suggests that superposition coding is approximately optimal if either the interference is strong or the cooperation is strong; when both interference and cooperation are weak, then cooperation based on DPC coding is approximately optimal. Even when superposition coding is approximately optimal in weak interference, DPC coding might lead to a smaller gap.
The DPC-based scheme is more complex to implement in practice than superposition coding; hence 
there might be an interesting practical trade-off between complexity and constant gap.

\end{itemize}

\section{The capacity region to within a constant gap for the general GCCIC}
\label{sec:general extension in strong interference}

We now focus on the general GCCIC, which is more complex to analyze due to the fact that one has to deal with 5~different channel parameters. Following the naming convention of the noncooperative IC, we say that the general GCCIC has strong interference if $\{ {\Sp} \leq {\Ip}, \ {\Sc} \leq {\Ic}\}$, weak interference if $\{ {\Sp} > {\Ip}, \ {\Sc} > {\Ic} \}$, and mixed interference otherwise. Moreover, we say that the general GCCIC has strong cooperation if $\Cc >\Sp$ and weak cooperation otherwise.
Our main result for the general GCCIC is as follows:
\begin{theorem}
\label{thm:gap full}
For the capacity region of the general GCCIC we have:
\begin{enumerate}[A)]
\item
%(Weak cooperation and not too weak interference)   
$\Cc \leq \Sp, \ \Sc\Sp \leq (1+\Ip)(1+\Ic)$: capacity region to within 2~bits with a noncooperative scheme,
\item 
%(Strong cooperation and strong interference at CRx) 
$\Sp < \Cc \leq \Ip$: capacity region to within 1~bit with a cooperative scheme based on superposition coding (cooperation on common message only),
\item
%(Strong cooperation~2)
$\max\{\Sp, \Ip\} < \Cc, \ \Sc \frac{1+ \Ip+\Sp}{1+ 2\Ip} \leq \Ic , \ \Sc \leq \Ic$: capacity region to within \gapwholecapstrong~bits with a cooperative scheme based on superposition coding (cooperation on both common and private messages),
%MULTIPLE CONDITIONS%{\red OLD $\Sp \left(1+\Sc \right) \leq \Ip \left(1+\Ic \right)$}
\item %(Weak interference) 
$\Sc > \Ic$ and $\Cc \geq \left(\Sp + \Ic + 2 \sqrt{\Sp  \Ic \ \frac{\Ip}{1+\Ip}} \right)(1+\Ip)$: capacity region to within 1~bit with a cooperative scheme based on DPC and superposition coding (private messages only).
\end{enumerate}
\end{theorem}

\begin{figure}%[h]
\centering
\includegraphics[width=0.8\textwidth]{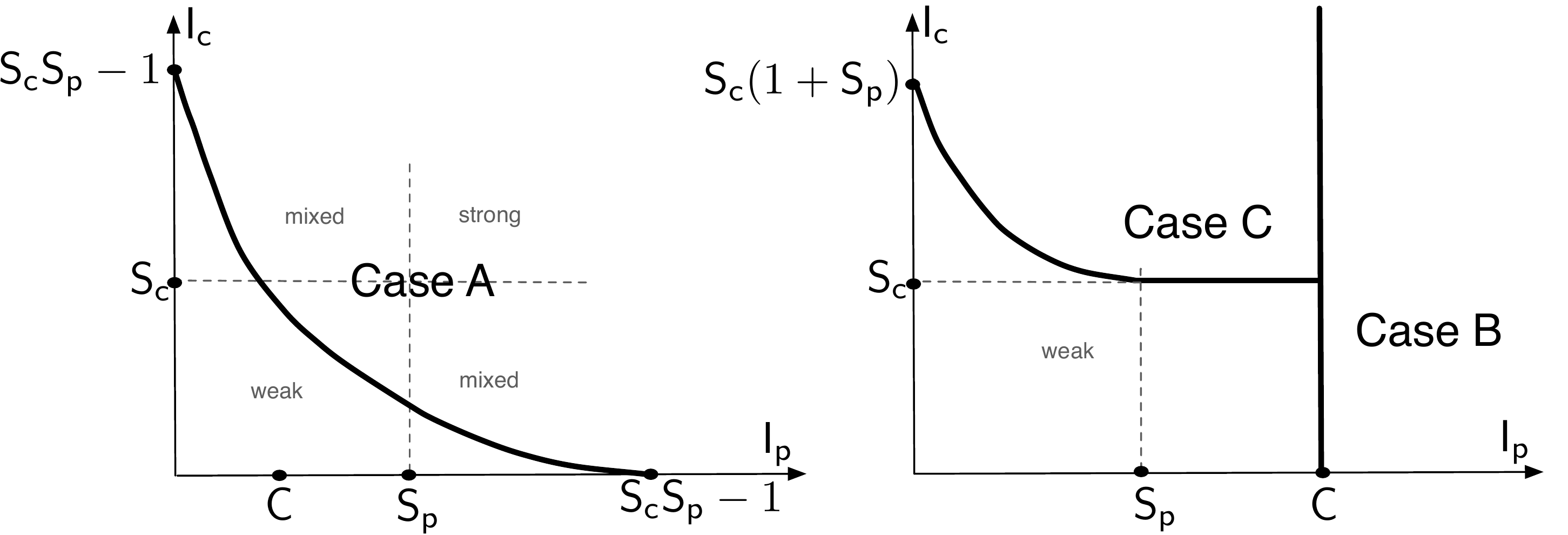}
\caption{The regimes identified by~Theorem~\ref{thm:gap full} where capacity is known to within a constant gap (indicated as ``Case A'', ``Case B'' and ``Case C'').}
\label{fig:gen case plot of where we have a gap}
\end{figure}

The rest of the section is devoted to the proof of Theorem~\ref{thm:gap full}. We divide the whole set of parameters depending on the strength of the cooperation link $\Cc$ compared to the direct link $\Sp$ and the interference link $\Ip$. Fig.~\ref{fig:gen case plot of where we have a gap} shows the regimes of Theorem~\ref{thm:gap full} for which we have an approximate capacity result (indicated as ``Case A'', ``Case B'' and ``Case C'' as in Theorem~\ref{thm:gap full}). As it can be noted from Fig.~\ref{fig:gen case plot of where we have a gap}, our capacity characterization to within a constant gap roughly excludes the weak interference regime. Case~D is a straightforward generalization of the condition in~\eqref{eq:the simple DPC region sym+etw condition} for the symmetric case studied in Section~\ref{subsec:sum regimes 4 and 6} and shall therefore not be further discussed.
We shall now discuss each case separately.

\subsection{The case $\Cc\leq\Sp$: when unilateral cooperation may not be useful}
\label{subsec:cap regime Cc<Sp}

We start our discussion with a simple observation.
Under the condition $\Cc\leq\Sp$ 
%all terms that depend on $\Cc$ in the outer bound in~\eqref{eq:ourRate} `disappear' in the sense that
we can further bound the region in~\eqref{eq:ourRate} as
\begin{subequations}
\begin{align}
\mathcal{O}^{\ref{subsec:cap regime Cc<Sp}} : \quad
   \Rc &\leq \log(1+\Sc),
\label{eq:when we drop C rc}
\\ \Rp &\leq 
%\log(1+\Sp+\Cc)  \nonumber
%\\& \leq  
\log(1+\Sp)+\log(2),
\label{eq:when we drop C rp}
\\ \Rp + \Rc &\leq    
%     \log  \left( \frac{1 +      \max\{\Ic,\Sc\}}{1+ \Ic}  \right)
%   + \log  \left( 1+ (\sqrt{\Sp} + \sqrt{\Ic})^2  \right)  \nonumber
%\\&\leq
     \log^+  \left( \frac{1 +\Sc}{1+ \Ic}  \right)
   + \log  \left( 1+ \Sp+\Ic \right)+\log(2),
\label{eq:when we drop C kra1}
\\ \Rp + \Rc &\leq   
%     \log  \left( \frac{1 + \Cc +  \max\{\Sp,\Ip\}}{1+ \Ip}  \right) 
%   + \log  \left( 1+ (\sqrt{\Sc} + \sqrt{\Ip})^2 \right)   \nonumber
%\\&\leq
     \log^+  \left( \frac{1 + \Sp}{1+ \Ip}  \right) 
   + \log  \left( 1+ \Sc+\Ip \right) +2\log(2).
\label{eq:when we drop C kra2}
\end{align}
\label{eq:when we drop C everywhere}
\end{subequations}
The bounds in~\eqref{eq:when we drop C everywhere} are to within 1~bit of the following rate region
\begin{subequations}
\begin{align}
\mathcal{I}^{\ref{subsec:cap regime Cc<Sp}} : \quad
   \Rc &\leq \log(1+\Sc),%-\log(2),
\\ \Rp &\leq \log(1+\Sp),%-\log(2),
\\ \Rp+\Rc &\leq \log(1+\Sp+\Ic)+\log^+\left(\frac{1+\Sc}{1+\Ic}\right),%-2\log(2),
\\ \Rp+\Rc &\leq \log(1+\Sc+\Ip)+\log^+\left(\frac{1+\Sp}{1+\Ip}\right),%-2\log(2),
\end{align}
\label{eq:non coop out without etw}
\end{subequations}
which is achievable to within 1 bit
%
%which is the well known upper bound of~\cite[Theorem 1]{Kramer} for the noncooperative IC.
%
%The region in~\eqref{eq:non coop out without etw} is tight for the noncooperative IC in the strong interference regime, defined as $\{\Sp \leq \Ip, \ \Sc \leq \Ic\}$. This last observation immediately implies that the noncooperative scheme with only common messages, which achieves the region in~\eqref{eq:non coop out without etw} in strong interference, is optimal to within 1~bit for the GCCIC in the strong interference regime when $\Cc\leq\Sp$.
%
%In the mixed and weak interference regimes, defined as either $\Sp > \Ip$  or $\Sc > \Ic$, the region in~\eqref{eq:non coop out without etw} is achievable to within 1~bit 
for the noncooperative IC when the `$R_1+2R_2, 2R_1+R_2$'-type of bounds in~\cite[Theorem 3]{etw} are redundant\footnote{By using the `worst noise covariance argument' as in~\cite{TuninettiITA10}, one can show that the upper bound in~\cite[Theorem 3]{etw}, which was derived for the noncooperative IC in weak interference, is actually valid for all channel parameters if one replaces $\log\left(\frac{1+\mathsf{SNR}_i}{1+\mathsf{INR}_j}\right)$ with $\log^+\left(\frac{1+\mathsf{SNR}_i}{1+\mathsf{INR}_j}\right)$, $i\not=j, \ i=1,2$. 
By using the notation of~\cite{etw}, the steps of the proof are as follows
\begin{align*}
  n(R_1 + 2R_2 -3\epsilon)
  &\leq I(X_1^n; Y_1^n)      +I(X_2^n; Y_2^n)            +I(X_2^n; Y_2^n)
\\&\leq I(X_1^n; Y_1^n,S_1^n)+I(X_2^n; Y_2^n,Y_1^n,X_1^n)+I(X_1^n,X_2^n; Y_2^n)-I(X_1^n; Y_2^n|X_2^n)
\\&= \underbrace{I(X_1^n; Y_1^n,S_1^n)+I(X_2^n; Y_1^n|X_1^n)-I(X_1^n; Y_2^n|X_2^n)}_{=h(Y_1^n|S_1^n)-h(Z_1^n)}
  + \underbrace{I(X_2^n; Y_2^n|X_1^n,Y_1^n)}_{\text{use worst noise covariance}}+I(X_1^n,X_2^n; Y_2^n)
%\\&\leq n\log(1+|h_{12}|^2+\frac{|h_{11}|^2}{1+|h_{21}|^2})
%\\&+n\log^+(\frac{1+|h_{22}|^2}{1+|h_{12}|^2})+n\log(1+|h_{21}|^2+|h_{22}|^2) 
\end{align*}
}; with the notation adopted in this paper, one can easily show that these bounds are redundant if
\begin{align}
%\Big\{\Sp \leq \Ip, \ \Sc > \Ic, \ \Sc\Sp \leq (1+\Ip)(1+\Ic)\Big\} \cup \Big\{\Sc \leq \Ic, \ \Sc \leq \Ic, \ \Sp\Sc \leq (1+\Ip)(1+\Ic)\Big\}.
\Sc\Sp \leq (1+\Ip)(1+\Ic).
\label{eq:non coop IC where etw-type bounds do not matter}
\end{align}
%\end{subequations}
Hence we can immediately conclude that the noncooperative scheme of~\cite{etw} is optimal to within 2 bits in the regime identified by~\eqref{eq:non coop IC where etw-type bounds do not matter} when the cooperation link gain satisfies $\Cc\leq\Sp$.
Notice that the regime in~\eqref{eq:non coop IC where etw-type bounds do not matter}, depicted in Fig.~\ref{fig:gen case plot of where we have a gap} on the left, includes the strong interference regime and most of the mixed interference regime; in other words, it roughly excludes the weak interference regime.
%and at high SNR it approximately corresponds to $\Sp \Sc \leq \Ip \Ic$. 

%To conclude we showed that: 
%\begin{theorem}\label{eq:cooperation is useless; strong and most of mixed}
%The noncooperative Han-Kobayashi scheme is to within 2 bits of the GCCIC outer bound in~\eqref{eq:when we drop C everywhere} if the interference links satisfy $\Sc\Sp \leq (1+\Ip)(1+\Ic)$ and the cooperation link gain satisfies $\Cc\leq\Sp$.
%\end{theorem}
The capacity result that we just proved is the generalization of the symmetric capacity result of Theorem~\ref{thm:gap symmetric} in Regime~1 and part of Regime~3 of Fig.~\ref{fig:fig3} (i.e., in the symmetric case the condition in~\eqref{eq:non coop IC where etw-type bounds do not matter} simplifies to ${\snr} \leq 1+{\inr}$, which at high SNR corresponds to $1 \leq \alpha$, and the condition $\Cc\leq {\snr}$ at high SNR corresponds to $\beta \leq 1$). As for Theorem~\ref{thm:gap symmetric} in the corresponding regime, a noncooperative scheme is approximately optimal.

\medskip
When $\Sc\Sp > (1+\Ip)(1+\Ic)$ and $\Cc\leq\Sp$ (which in the symmetric case corresponds to $1>\alpha$ and $\beta \leq 1$ and for which we could only show a sum-capacity result to within a constant gap in Theorem~\ref{thm:gap symmetric}) we expect that, in order to show an approximate capacity result, 
%, the issues raised in Section~\ref{subsec:sum regime 4} must be addressed, namely, determine 
upper bounds on $\Rp+2\Rc$ and $2\Rp+\Rc$ must be derived. 
%
%{\red In this case, a noncooperative scheme achieves the sum-capacity in OUTERBOUND when the etw-sum-rate in ETW is redundant; this is the case if we are OUTSIDE
%\begin{align*}
%  & \Ip \leq \Sp
%\\& \Ic \leq \Sc
%\\
%&\left((1+\Ip)(1+\Ic)+\Sc\right)\left((1+\Ip)(1+\Ic)+\Sp\right)
% < (1+\Sp+\Ic)(1+\Sc)(1+\Ip)
%\\
%&\left((1+\Ip)(1+\Ic)+\Sc\right)\left((1+\Ip)(1+\Ic)+\Sp\right)
% < (1+\Sc+\Ip)(1+\Sp)(1+\Ic)
%	%&(1+\Ip)^2(1+\Ic)^2-(1+\Ip)(1+\Ic)
%	%<\Sc(\Sp\Ic-\Ip-\Ip\Ic)
%	\\&(cp+\Sp)(cp+\Sc)<(cp+p\Sp)(1+\Sc)
%	\\&(cp+\Sp)(cp+\Sc)<(cp+c\Sc)(1+\Sp)
%	\\
%	\\&(cp)^2+cp(\Sp-1)+\Sp\Sc < p\Sp(1+\Sc)
%	\\&(cp)^2+cp(\Sc-1)+\Sp\Sc < c\Sc(1+\Sp)
%\end{align*}
%}

\subsection{The case $\Sp < \Cc \leq \Ip$: when unilateral cooperation is useful}
\label{subsec:cap regime Sp<Cc<Ip}

For $\Sp < \Cc \leq \Ip$ we further bound the capacity upper bound in~\eqref{eq:ourRate} as
\begin{subequations}
\begin{align}
\mathcal{O}^{\ref{subsec:cap regime Sp<Cc<Ip}} : \quad
   \Rc &\leq \log(1+\Sc),
\\ \Rp &\leq %\log(1+\Sp+\Cc)  \nonumber \\& \leq  
\log(1+\Cc)+\log(2),
%\\ \Rp &\leq \log(1+(\sqrt{\Sp}+\sqrt{\Ic})^2)  \nonumber
%\\& \leq  \log(1+\Sp+\Ic)+\log(2),
\\ \Rp + \Rc &\leq    
%     \log  \left( \frac{1 +      \max\{\Ic,\Sc\}}{1+ \Ic}  \right)
%   + \log  \left( 1+ (\sqrt{\Sp} + \sqrt{\Ic})^2  \right)  \nonumber
%\\&\leq
     \log^+  \left( \frac{1 +\Sc}{1+ \Ic}  \right)
   + \log  \left( 1+ \Sp+\Ic \right)+\log(2),
\label{eq:when we drop C in summate krap}
\\ \Rp + \Rc &\leq   
%     \log  \left( \frac{1 + \Cc +  \max\{\Sp,\Ip\}}{1+ \Ip}  \right) 
%   + \log  \left( 1+ (\sqrt{\Sc} + \sqrt{\Ip})^2 \right)   \nonumber
%%\\&\leq
%%     \log  \left( 1+\frac{\Cc}{1+ \Ip}  \right) 
%%   + \log  \left( 1+ \Sc+\Ip \right) +\log(2).
%\\&\leq
     \log  \left( 1+ \Sc+\Ip \right) + 2\log(2).
\label{eq:when we drop C in summate krac}
\end{align}
\label{eq:when we drop C in summate}
\end{subequations}
In this regime, unilateral cooperation helps increasing the rate of the primary user.
In the symmetric case, the upper bound in~\eqref{eq:when we drop C in summate} reduces to the part of Regime~2 and~3 of Fig.~\ref{fig:fig3} for $1<\beta\leq\alpha$; we therefore consider the generalization of the achievable scheme we used for Regime~2 of Fig.~\ref{fig:fig3} to the case of general channel gains. Here PTx takes advantage of the strong cooperation link and sends its message with the help of the CTx. The sum-rate upper bound in~\eqref{eq:when we drop C in summate krac} suggests that CRx should decode the PTx message in addition to its intended message, that is, PTx should use a (cooperative) common message only. The sum-rate upper bound in~\eqref{eq:when we drop C in summate krap}, suggests that PRx should decode CTx's message only when $\Ic>\Sc$, that is, CTx should use both a (noncooperative) common and a (noncooperative) private message. This is exactly the strategy described in Appendix~\ref{sec:scheme 1} and the resulting achievable region is given in~\eqref{eq:sup.only ach.region with common messages only in unilateral cooperation V1=X1 gaussian}, namely
\begin{subequations}
\begin{align}
\mathcal{I}^{\ref{subsec:cap regime Sp<Cc<Ip}} : \quad
   \Rc &\leq \log(1+\Sc),
\\ \Rp &\leq \log(1+\Cc),
\\ \Rp+\Rc &\leq \log(1+\Sp+\Ic)+\log^+\left(\frac{1+\Sc}{1+\Ic}\right),
\\ \Rp+\Rc &\leq \log(1+\Sc+\Ip).
\end{align}
\label{eq:sup.only ach.region with common messages only in unilateral cooperation V1=X1 gaussian again again}
\end{subequations}
By comparing the upper bound in~\eqref{eq:when we drop C in summate} with the achievable region in~\eqref{eq:sup.only ach.region with common messages only in unilateral cooperation V1=X1 gaussian again again} we conclude that
the capacity region is known to within 1~bit for a general GCCIC where the channel gains satisfy $\Sp < \Cc \leq \Ip$.
Notice that we did not impose any condition on the strength of $\Ic$ compared to $\Sc$, i.e., in other words the gap result holds regardless of whether the interference at PRx is strong ($\Ic\geq\Sc$) or weak ($\Ic<\Sc$).

\subsection{The case $\max\{\Sp,\Ip\} < \Cc$  and $\Sc \leq \Ic$: when unilateral cooperation is useful}
\label{subsec:cap regime Sp<Cc and Ip<C and Sc<Ic}

For this case we further bound the capacity upper bound in~\eqref{eq:ourRate} as
%%%\begin{subequations}
%%%\begin{align}
%%%\mathcal{O}^{\ref{subsec:cap regime Sp<Cc and Ip<C and Sc<Ic} \ general} : \quad
%%%   \Rc &\leq \log(1+\Sc),
%%%\\ \Rp &\leq \log(1+\Cc)+\log(2),
%%%\\ \Rp &\leq {\red \log(1+\Sp+\Ic)+\log(2), } 
%%%\\ \Rp + \Rc &\leq    
%%%{\red  \log^+  \left( \frac{1 + \Sc}{1+ \Ic}  \right)+}
%%%    \log  \left( 1+ \Sp+\Ic \right)+\log(2), \label{eq:when we drop C in neither sumrateP}
%%%\\ \Rp + \Rc &\leq   
%%%     \log  \left(\frac{1+2\Cc}{1+ \Ip}  \right) 
%%%   + \log  \left( 1+ \Sc+\Ip \right) +\log(2). \label{eq:when we drop C in neither sumrateC}
%%%\end{align}
%%%\label{eq:when we drop C in neither}
%%%\end{subequations}
%%%
%%%{\magenta
%%%Strong: $\Sc \leq \Ic$
\begin{subequations}
\begin{align}
\mathcal{O}^{\ref{subsec:cap regime Sp<Cc and Ip<C and Sc<Ic}} : \quad % \ strong
   \Rc &\leq \log(1+\Sc),
\\ \Rp &\leq \log(1+\Cc)+\log(2),
\\ \Rp + \Rc &\leq    
    \log  \left( 1+ \Sp+\Ic \right)+\log(2), \label{eq:when we drop C in neither sumrateP}
\\ \Rp + \Rc &\leq   
     \log  \left(\frac{1+2\Cc}{1+ \Ip}  \right) 
   + \log  \left( 1+ \Sc+\Ip \right) +\log(2). \label{eq:when we drop C in neither sumrateC}
\end{align}
\label{eq:when we drop C in neither strong}
\end{subequations}
%%%
%%%
%%%Weak: $\Sc > \Ic$
%%%\begin{subequations}
%%%\begin{align}
%%%\mathcal{O}^{\ref{subsec:cap regime Sp<Cc and Ip<C and Sc<Ic} } : \quad % \ weak
%%%   \Rc &\leq \log(1+\Sc),
%%%\\ \Rp &\leq \log(1+\Cc)+\log(2),
%%%\\ \Rp &\leq \log(1+\Sp+\Ic)+\log(2), 
%%%\\ \Rp + \Rc &\leq    
%%%    \log\left( \frac{1 + \Sc}{1+ \Ic}  \right)+
%%%    \log  \left( 1+ \Sp+\Ic \right)+\log(2), 
%%%\\ \Rp + \Rc &\leq   
%%%     \log  \left(\frac{1+2\Cc}{1+ \Ip}  \right) 
%%%   + \log  \left( 1+ \Sc+\Ip \right) +\log(2).
%%%\end{align}
%%%\label{eq:when we drop C in neither weak}
%%%\end{subequations}
%%%}
In this regime, unilateral cooperation helps increasing both the rate of the primary user and the sum-capacity.
In the symmetric case, the upper bound in~\eqref{eq:when we drop C in neither strong} reduces to the part of Regime~2 and~3 of Fig.~\ref{fig:fig3} for $1<\alpha<\beta$.
%In this regime we use the scheme in Appendix~\ref{sec:scheme 6}, which is based on superposition coding only. 
Here PTx takes advantage of the strong cooperation link and sends its message with the help of the CTx.
The sum-rate upper bound in~\eqref{eq:when we drop C in neither sumrateP} suggests that PRx should decode the CTx message in addition to its intended message, that is, CTx should use a (noncooperative) common message only; this is so because the condition $\Sc \leq \Ic$ corresponds to strong interference at the PRx.
The sum-rate upper bound in~\eqref{eq:when we drop C in neither sumrateC}, suggests that PTx should use both a (cooperative) common and a (cooperative) private message; this is so because here we do not specify which one among $\Sp$ and $\Ip$ is the largest, and therefore the interference at CRx could be either strong or weak.
This is exactly the strategy described in Appendix~\ref{sec:scheme 4},
which is based on superposition coding only (as the cognitive common message is not precoded against the interference of the primary private message); both the common and the private message of PTx are cooperative; this scheme can be thought of as the extension of the scheme used in Section~\ref{subsec:cap regime Sp<Cc<Ip} so as to include a private message for PTx in case the interference at CRx is weak.

The achievable region is given in~\eqref{eq:scheme fully connected strong interference Gaussian}. With the possible suboptimal choice
$|\gp|^2 = \frac{1}{1+ \Ip}, \ |\gc|^2 = \frac{1}{1+ \Sc}$
inspired by~\cite{etw},
the achievable region in~\eqref{eq:scheme fully connected strong interference Gaussian} becomes
\begin{subequations}
\begin{align}
\mathcal{I}^{\ref{subsec:cap regime Sp<Cc and Ip<C and Sc<Ic}} : \quad
   \Rc & \leq 
   \log \left( \frac{1+ \frac{\Ip}{1+\Ip} + \Sc}{1+ \frac{\Ip}{1+\Ip} +  \frac{\Sc}{1+\Sc}} \right),
%   \log \left( 1+\frac{\Sc \frac{\Sc}{1+\Sc}}{1+ \frac{\Ip}{1+\Ip} +  \frac{\Sc}{1+\Sc}} \right),
%
\\ \Rp &\leq \log \left( 1+\Cc \right),
\\ \Rp+\Rc & \leq \log \left( 1+ \Sp + \Ic   \right),
\label{eq:scheme fully connected strong interference Gaussian again again first sumrate}
\\ \Rp + \Rc &  \leq  \log \left( 1 + \frac{\Cc}{1+\Ip} \right ) 
 +  \log \left( \frac{1+\Sc+\Ip}{1 +  \frac{\Ip}{1+\Ip}  +  \frac{\Sc}{1+\Sc} } \right),
\label{eq:scheme fully connected strong interference Gaussian again again middle sumrate}
\\ \Rp + \Rc & \leq \log \left(  1+ \frac{\Sp}{1+\Ip} + \frac{\Ic}{1+\Sc} \right)
 +  \log \left( \frac{1+\Sc+\Ip}{1 +  \frac{\Ip}{1+\Ip}  +  \frac{\Sc}{1+\Sc} } \right),
%  +  \log \left( 1 + \frac{\Sc \frac{\Sc}{1+\Sc} + \Ip \frac{\Ip}{1+\Ip}}{1 +  \frac{\Ip}{1+\Ip}  +  \frac{\Sc}{1+\Sc} } \right).
\label{eq:scheme fully connected strong interference Gaussian again again last sumrate}
\\ \Rp + 2\Rc & \leq \log \left(  1+ \frac{\Sp}{1+\Ip} + \Ic   \right)
 +  \log \left( \frac{1+\Sc+\Ip}{1 +  \frac{\Ip}{1+\Ip}  +  \frac{\Sc}{1+\Sc} } \right).
\end{align}
\label{eq:scheme fully connected strong interference Gaussian again again}
\end{subequations}
By comparing the upper bounds in~\eqref{eq:when we drop C in neither strong} with the inner bounds in~\eqref{eq:scheme fully connected strong interference Gaussian again again} it can be shown that they are at most
\[
\mathsf{GAP} \leq \max\left\{\log(3),\log(2),\frac{\log(2)}{2}, \frac{\log(12)}{2},\frac{\log(6)}{2} \right\} = \frac{\log(12)}{2} \approx \gapwholecapstrong~\text{bits},
\]
bits away when the condition in~\eqref{eq:scheme fully connected strong interference Gaussian drop r1+2rc condition 2} holds for the considered choice of parameters, namely
\begin{align}
\Sc \frac{1+ \Ip+\Sp}{1+ 2\Ip} \leq \Ic
%%%\quad {\red OLD \ \Sp \left(1+\Sc \right) \leq \Ip \left(1+\Ic \right) }
\label{eq:akin to "non coop IC where etw-type bounds do not matter}
\end{align}
so that the bound on $\Rp+2\Rc$ in~\eqref{eq:scheme fully connected strong interference Gaussian again again} can be dropped.
Notice that the sum-rate bound in~\eqref{eq:when we drop C in neither sumrateP} and the one in~\eqref{eq:scheme fully connected strong interference Gaussian again again last sumrate} are the same up to a constant gap, that is,
\begin{align*}
& \log \left( 1+ \Sp + \Ic   \right) + \log(2)
- \log \left(  1+ \frac{\Sp}{1+\Ip} + \frac{\Ic}{1+\Sc} \right)
- \log \left( \frac{1+\Sc+\Ip}{1 +  \frac{\Ip}{1+\Ip}  +  \frac{\Sc}{1+\Sc} } \right)
\\
&\leq
  \log \left( 1+ \Sp + \Ic   \right)
- \log \left(  1+ \frac{\Sp+\Ic}{1+\max\{\Ip,\Sc\}} \right)
- \log \left( 1+\Sc+\Ip \right)+\log(6)
\\&=
  \log \left( \frac{1+ \Sp + \Ic}{1+\max\{\Ip,\Sc\}+\Sp+\Ic}  \
              \frac{1+\max\{\Ip,\Sc\}}{1+\max\{\Ip,\Sc\}+\min\{\Ip,\Sc\}} \right)
+ \log(6) \leq \log(6).
\end{align*}

\medskip
The condition in~\eqref{eq:akin to "non coop IC where etw-type bounds do not matter} is similar to the condition in~\eqref{eq:non coop IC where etw-type bounds do not matter}, which we derived in order to claim that bounds of the form $\Rp+2\Rc / 2\Rp+\Rc$ were redundant in the noncooperative achievable region in the weak interference regime.
In general, as can be noticed from the analysis so far, the weak interference regime is more challenging than the other regimes. 
In the next sections we concentrate on two special GCCIC
where one of the interfering links is absent: %. Due to the asymmetry in the cooperation, we shall consider
the case where CRx does not experience interference (i..e., the so-called Z-channel for which $\Ip=0$), and
the case where PRx does not experience interference (i..e., the so-called S-channel for which $\Ic=0$),
for which we shall prove a constant gap result also in the weak interference regime.
As we shall see, DPC-based schemes appear to be needed for approximate optimality in weak interference.

\section{The capacity region to within a constant gap for the Z-channel}
\label{sec:general extension Z}

Our main result for the Z-channel is as follows:
\begin{theorem}
\label{thm:gap Z}
The capacity region of the Z-channel (i.e., the link PTx$\rightarrow$CRx is non-existent) is known to within \gapwholecapZ~bits.
\end{theorem}
The rest of the section is devoted to the proof of Theorem~\ref{thm:gap Z}, that is, the upper bound 
\begin{subequations}
\begin{align}
%O^{\rm(Z)}=
%\left\{\begin{array}{rl}
   \Rc &\leq \log \left(1+ \Sc \right),
\label{eq:outerBoundZ rc}
\\ \Rp &\leq \log \left(1+ \left(\sqrt{\Sp}+\sqrt{\Ic} \right)^2 \right),
\label{eq:outerBoundZ rp}
\\ \Rp &\leq \log \left( 1+\Cc+\Sp \right),
\label{eq:outerBoundZ rp with c}
%\\ \Rp+\Rc &\leq \log \left(1+C+\Sp \right) + \log \left(1+\Sc \right) %redundant
\\ \Rp+\Rc &\leq \log^+ \left( \displaystyle\frac{1+\Sc}{1+\Ic} \right)+\log \left(1+ \left(\sqrt{\Sp}+\sqrt{\Ic} \right)^2 \right),
\label{eq:outerBoundZ rsum}
%\end{array}\right.
\end{align}
\label{eq:outerBoundZ}
\end{subequations}
from~\eqref{eq:ourRate} by setting $\Ip=0$, can be achieved to within a constant gap.
The region in~\eqref{eq:outerBoundZ} without the bound in~\eqref{eq:outerBoundZ rp with c} (i.e., the only one that depends on $\Cc$) is the capacity upper bound for the non-causal cognitive IC in~\cite[Theorem III.1]{riniJ1}, which unifies previously known outer bounds for the weak ($\Sc>\Ic$) and strong ($\Sc\leq\Ic$) interference regimes and is achievable to within 1~bit. Hence, we interpret the bound  in~\eqref{eq:outerBoundZ rp with c} as the `cost' of causal cooperation on the Z-channel.

For the proof of Theorem~\ref{thm:gap Z}, we consider separately different parameter regimes. 
Given the result in Theorem~\ref{thm:gap full}, we only need to consider the case $\Ic \leq \Sc(1+\Sp)$ (since $\Sc\Sp - 1 < \Sc(1+\Sp)$).
In the symmetric case, the regime $\Ic \leq \Sc(1+\Sp)$ is equivalent to ${\inr} \leq {\snr}(1+{\snr})$, or $\alpha \leq 2$ at high SNR, that is, we need to focus on the case where the Z-channel does not exhibit very strong interference.

\subsection{Case $\Cc \leq \Sp$: when unilateral cooperation might not be useful}
\label{subsec:Z useless} 
%We start by those case where we can `drop' the dependance on $C$ in~\eqref{eq:outerBoundZ}.
For the case $\Cc \leq \Sp$ we further outer bound the capacity upper bound in~\eqref{eq:outerBoundZ} as
\begin{subequations}
\begin{align}
%O^{\rm(Z)}\subseteq
%O^{\rm(Z-1)}=
%\left\{\begin{array}{rl}
\mathcal{O}^{\ref{subsec:Z useless}} : \quad
   \Rc &\leq \log \left(1+ \Sc \right),
\\ \Rp &\leq \log \left(1+ \Sp \right) +\log(2),
\\ \Rp+\Rc &\leq \log^+ \left( \frac{1+\Sc}{1+\Ic} \right)+\log \left(1+ \Sp+\Ic\right)+\log(2).
%\end{array}\right.
\end{align}
\label{eq:outerBoundZ case 1}
\end{subequations}
The region in~\eqref{eq:outerBoundZ case 1} is at most 1~bit away from
\begin{subequations}
\begin{align}
%I^{\rm(Z-1)}=
%\left\{\begin{array}{rl}
\mathcal{I}^{\ref{subsec:Z useless}} : \quad
   \Rc &\leq \log \left(1+ \Sc \right),
\\ \Rp &\leq \log \left(1+ \Sp \right),
\\ \Rp+\Rc &\leq \log^+ \left( \frac{1+\Sc}{1+\Ic} \right)+\log \left(1+ \Sp+\Ic\right),
%\end{array}\right.
\end{align}
\label{eq:outerBoundZ noncoop}
\end{subequations}
which is achievable to within 1~bit by a noncooperative scheme \cite{etw}. Therefore, for this set of parameters we have that the outer bound in \eqref{eq:outerBoundZ case 1} is achievable to within 2~bits.

The difference between the case $\Cc \leq \Sp$ for the Z-channel and the corresponding case for the general channel in Theorem~\ref{thm:gap full} in Section~\ref{subsec:cap regime Cc<Sp} is that here we do not need to impose the condition in~\eqref{eq:non coop IC where etw-type bounds do not matter} to claim the redundancy of the bounds on $\Rp+2\Rc / 2\Rp+\Rc$ in the noncooperative achievable region. This is so because those bounds do not matter, up to a constant gap of 1~bit, in the corresponding noncooperative IC~\cite{etw}.

\subsection{Case $\Cc > \Sp, \ \Sc \leq \Ic$ (i.e., strong interference at PRx): when unilateral cooperation is useful}
\label{subsec:Z useful strong} 

%OLD {Case $\Cc > \max \left \{\Sp,\Ic \right \}$ or $\Sp < \Cc < \Ic$ and $\Ic \geq \Sc$: when unilateral cooperation is useful and strong interference}
%OLD For the case $\Cc > \max \left \{\Sp,\Ic \right \}$ or $\Sp < \Cc < \Ic$ and $\Ic \geq \Sc$
%we distinguish two cases depending on the strength of $\Ic$ compared to $\Sc$. 
%For
%\begin{align}
%%\Sp < \min\{\Ic,C\}, 
%\Sp < \Ic, \
%\Sp < \Cc,   \
%\Sc \leq \Ic 
%\end{align}
In this case, we further outer bound the region in~\eqref{eq:outerBoundZ} as
\begin{subequations}
\begin{align}
%O^{\rm(Z)}\subseteq
%O^{\rm(Z-2)}=
%\left\{\begin{array}{rl}
\mathcal{O}^{\ref{subsec:Z useful strong} } : \quad
   \Rc &\leq \log \left(1+\Sc \right),
\\ \Rp &\leq \log \left(1+\Cc \right) +\log(2),
\\ \Rp+\Rc &\leq \log \left(1+ \Sp+\Ic\right)+\log(2). 
%\end{array}\right.
\end{align}
\label{eq:outerBoundZ case 2}
\end{subequations}
In this regime, we use the same strategy employed for the general GCCIC in the same regime, i.e., for $\Cc > \Sp$ and $\Ic \geq \Sc$ in Fig.~\ref{fig:gen case plot of where we have a gap} Case C, by setting $\Ip=0$. Here PTx takes advantage of the strong cooperation link and sends its message with the help of the CTx.
Moreover, since the PTx does not create interference at the CRx ($\Ip=0$), it sends a (cooperative) private message only. On the other hand, since the interference at the PRx is strong, the CTx sends a (noncooperative) common message only.
This is exactly the strategy described in Appendix~\ref{sec:scheme 4} and the resulting achievable region is given by~\eqref{eq:scheme fully connected strong interference Gaussian} (this is the same achievable region we used in Section~\ref{subsec:cap regime Sp<Cc and Ip<C and Sc<Ic}). In~\eqref{eq:scheme fully connected strong interference Gaussian}, we further set $\Ip=0$ and $|\gp|=1$ so that the PTx sends a private message only. With the possible suboptimal choice $|\gc|^2 = \frac{1}{1+\Sc}$, the achievable region in~\eqref{eq:scheme fully connected strong interference Gaussian} becomes
\begin{subequations}
\begin{align}
\mathcal{I}^{\ref{subsec:Z useful strong} } : \quad
  \Rc & \leq \log \left( \frac{1+\Sc}{1 +  \frac{\Sc}{1+\Sc}} \right),
\\ \Rp &\leq \log \left( 1+\Cc \right),
\\ \Rp+\Rc & \leq \log \left( 1+ \Sp + \Ic   \right),
%
%\\ \Rp + \Rc &  \leq  \log \left( 1 + \frac{\Cc}{1+\Ip} \right )  +  \log \left( 1 + \frac{\Sc \frac{\Sc}{1+\Sc} + \Ip \frac{\Ip}{1+\Ip}}{1 +  \frac{\Ip}{1+\Ip}  +  \frac{\Sc}{1+\Sc} } \right) 
%
\\ \Rp + \Rc & \leq \log \left(  1+ \Sp + \frac{\Ic}{1+\Sc} \right)   
+  \log \left( \frac{1 + \Sc}{1 +  \frac{\Sc}{1+\Sc} } \right).
%
%\\ \Rp +\Rc & \leq \log \left(  1 + \frac{\Sp}{1+\Ip} + \Ic  \right)   +   \log \left( 1 + \frac{ \Ip \frac{\Ip}{1+\Ip}}{1 +  \frac{\Ip}{1+\Ip}  +  \Sc} \right)
\end{align}
\label{eq:scheme Z channel strong interference Gaussian again again}
\end{subequations}
Notice that the bound on $\Rp+2\Rc$ in~\eqref{eq:scheme fully connected strong interference Gaussian r1+2r2} is always redundant because of the condition in~\eqref{eq:scheme fully connected strong interference Gaussian drop r1+2rc condition 1} since here we set $|\gp|=1$; this implies that 
the difference between this case for the Z-channel and the corresponding case for the general channel in Theorem~\ref{thm:gap full} in Section~\ref{subsec:cap regime Sp<Cc and Ip<C and Sc<Ic} is that here we do not need to impose the condition in~\eqref{eq:akin to "non coop IC where etw-type bounds do not matter} to claim the redundancy of the bound on $\Rp+2\Rc$ in the achievable region.

It is not difficult to see that the outer bound in~\eqref{eq:outerBoundZ case 2} and the inner bound in~\eqref{eq:scheme Z channel strong interference Gaussian again again} are at most 2~bits away.

\subsection{Case $\Cc > \Sp, \ \Sc > \Ic$ (i.e., weak interference at PRx): when unilateral cooperation is useful}
\label{subsec:Z useful weak} 

For this case, an outer bound for the Z-channel is given by the capacity of the non-causal cognitive IC in weak interference in~\eqref{eq:the noncausalcognitive capacity weak} together with the cut-set bound in~\eqref{eq:CutSetRp}, i.e., 
\begin{subequations}
\begin{align}
\mathcal{O}^{\ref{subsec:Z useful weak} } : \quad
\Rc &\leq  \left( 1 +(1-|\gc|^2)\Sc \right),
\label{eq:the noncausalcognitive capacity weak rc}
\\ \Rp &\leq \log \left( 1+\frac{\Sp +|\gc|^2 \Ic + 2|\gc|  \ \sqrt{\Sp \Ic} }{1+ (1-|\gc|^2)\Ic} \right),
\label{eq:the noncausalcognitive capacity weak rp}
\\ \Rp &\leq \log \left(1+\Cc \right) +\log(2),
\end{align}
\label{eq:outerBoundZ2}
\end{subequations}
union over all $|\gc|\leq 1$.
Since $\Cc > \Sp$, PTx takes advantage of the strong cooperation link and sends its message with the help of the CTx.
Moreover, since the PTx does not create interference at the CRx ($\Ip=0$), it sends a (cooperative) private message only. The outer bound in \eqref{eq:the noncausalcognitive capacity weak rp} suggests that the PRx should treat as noise the message of the CTx, while the bound in \eqref{eq:the noncausalcognitive capacity weak rc} tells us that the CRx should decode its own message without experiencing interference. In order to model this last observation, we use a DPC-based scheme. In this strategy the CTx precodes its message against the `known interference' so that the CRx decodes its own message as if the interference was not present~\cite{costaDPC}. This is exactly the strategy described in Appendix~\ref{sec:scheme 3} and the resulting achievable region is given by~\eqref{eq:the simple DPC region} with $\Ip=0$. We further set $|\gp|=0$ in~\eqref{eq:the simple DPC region} and we obtain
\begin{subequations}
\begin{align}
\mathcal{I}^{\ref{subsec:Z useful weak} } : \quad
   \Rc &\leq  \left( 1 +  (1-|\gc|^2)\Sc \right),
\\ \Rp &\leq \log \left( 1+\frac{\Sp +|\gc|^2 \Ic}{1+ (1-|\gc|^2)\Ic} \right),
\\ \Rp &\leq \log \left( 1+\Cc \right),
\end{align}
\label{eq:achievable coope strongissimo}
\end{subequations}
for all $|\gc|\leq 1$. By simple computations, the achievable region in \eqref{eq:achievable coope strongissimo} can be shown to be at most 1~bit away from the upper bound in~\eqref{eq:outerBoundZ2}.

Note that here we used a DPC-based scheme in order to determine the capacity to within a constant gap in weak interference, while in Section~\ref{subsec:cap regime Sp<Cc and Ip<C and Sc<Ic} for the general GCCIC we only used superposition coding. 

\subsection{Comparisons}
We conclude the section by comparing the performance of unilateral cooperation on the Z-channel with other forms of cooperation. Moreover, we also consider whether the absence of an interfering link is beneficial in the GCCIC. We shall use as performance metric the gDoF, or high SNR throughput. In order to reduce the number of parameters, we restrict our attention to the case where the direct links have the same strength.
For future reference, the gDoF of the noncooperative Z-channel is given by~\cite{SASON}
\[
{\gdof}^{\rm IC-Z} = \min\{\max\{1-\alpha/2,\alpha/2\}, \ 1\}
\]
and that of the non-causal cognitive Z-channel, which can be evaluated from~\cite{riniJ1}, is
\[
{\gdof}^{\rm CIC-Z} = \max\{1-\alpha/2,\alpha/2\}.
\]
Fig.~\ref{fig:fig4} shows the gDoF and the gap for the Z-channel for different regions in the $(\alpha,\beta)$ plane. The whole set of parameters has been partitioned into multiple sub-regions depending upon different level of cooperation ($\beta$) and interference ($\alpha$) strengths.
%two sub-regions: regime~1 corresponds to strong interference $(\alpha\geq1)$ and regime~2  to weak interference $(\alpha< 1)$.
%%\begin{itemize}
%%
%%\item  {\bf Regime~1: Strong interference.}
%From Fig.~\ref{fig:fig4}, we immediately see that unilateral cooperation improves on the gDoF of the noncooperative Z-channel only when the interference is very strong $(\alpha\geq 2)$ and the cooperation is strong $(\beta\geq 1)$.
%%
%%\item {\bf Region 2: Weak interference.}
%In region 2 the gDoF attained by the Z-channel with unilateral cooperation is the same as that achieved by the noncooperative Z-channel. 
%%\end{itemize}

When comparing unilateral cooperation with other channel models in terms of gDoF we observe:
\begin{itemize}

\item
For the noncooperative IC, it is well known that removing an interference link cannot degrade the performance and the sum-capacity is known exactly for all channel parameters~\cite{SASON}. The same cannot be said in full generality for the cooperative channel because ``useful cooperative information'' can flow through the interference link.  
%For the Z-channel: By ignoring cooperation we immediately have ${\gdof} \geq \min \left \{ 1,\max \left \{ \frac{\alpha}{2},1-\frac{\alpha}{2} \right \} \right \}$ from~\cite[Th.2]{SASON}. 
Thus for the Z-channel, cooperation only improves the gDoF with respect to the noncooperative case in the regime $\alpha \geq 2$ and $\beta \geq 1$, i.e., in very strong interference and strong cooperation (the gDoF achieved with and without cooperation is the same in the green  and yellow regions in Fig.~\ref{fig:fig7}).

\item
For the Z-channel, unilateral cooperation attains the gDoF of the classical relay channel when
$1 \leq \alpha \leq \beta$, 
%{\red 1 WAS MISSING; UPDATE FIG \ref{fig:fig6}}
as shown by the red and yellow-shaded regions in Fig.~\ref{fig:fig7}.
%For this set of parameters the Z-channel with unilateral cooperation might not be worth implementing in practical systems since rate $\Rc=0$ for the cognitive pair is gDoF optimal.

\item
The Z-channel achieves the same gDoF of the non-causal cognitive channel everywhere except in $\alpha>\max\{2,\beta+1\}$ (region with horizontal lines in Fig.~\ref{fig:fig7}).

\item
The gDoF of unilateral cooperation equals the gDoF upper bound of bilateral cooperation~\cite{PVIT11} when $\beta \leq \max\{1,\alpha\}$
%{\blue
%\begin{itemize}
%\item $\alpha \geq 2, \beta \leq 1$;
%\item $\alpha \geq 2, \beta \leq \alpha -1$;
%\item $\alpha \geq 2, \alpha-1 \leq \beta \leq \alpha$;
%\item $1 \leq \alpha \leq 2, \beta \leq \alpha$;
%\item $\alpha \leq 1, \beta \leq 1$.
%\end{itemize}}
(region with vertical lines in Fig.~\ref{fig:fig7}) that corresponds to the case where the cooperation link is weaker than the best between the direct link and the interference link. In this case bilateral cooperation might not be worth implementing in practice.
Notice that here we compare the (provably achievable) gDoF for the case of unilateral cooperation to an upper bound for bilateral cooperation. To the best of our knowledge, it has not been shown that the gDoF upper bound for the Z-channel with bilateral source cooperation is achievable, which we expect to be.

\item
By comparing Fig.~\ref{fig:fig3} and Fig.~\ref{fig:fig4} we observe that the gDoF of the Z-channel is always greater or equal than that of the interference-symmetric channel. This is due to the fact that the PTx does not cooperate in sending the cognitive signal. Therefore by removing the link between PTx and CRx we rid CRx of only an interfering signal and this leads to an improvement in gDoF. 

The regimes where the Z-channel strictly outperforms the interference-symmetric channel are when $0 \leq \alpha \leq \frac{2}{3}$ and $\beta \leq \min\{\alpha, 1-\alpha\}$
% as a result of not having interference from the PTx since $\sqrt{I_p} {\rm e}^{{\rm j}\theta_p} = 0$ 
(region with vertical lines in Fig.~\ref{fig:fig9}),  i.e., weak interference and fairly weak cooperation.
This regime can be thought of as the one where interference is the most harmful for the  interference-symmetric channel.

\end{itemize}

\section{The capacity region to within a constant gap for the S-channel}
\label{sec:general extension S}

Our main result for the S-channel is as follows:
\begin{theorem}
\label{thm:gap S}
The capacity region of the S-channel (i.e., the link CTx$\rightarrow$PRx is non-existent) is known to within 2~bits.
\end{theorem}
The rest of the section is devoted to the proof of Theorem~\ref{thm:gap S}. We distinguish two cases, depending on whether the following upper bound
\begin{subequations}
\begin{align}
%O^{\rm(S)}=
%\left\{\begin{array}{rl}
    \Rc &\leq \log\left ( 1 + \Sc \right ),
\\  \Rp &\leq \log\left ( 1 + \Sp \right ),
\\  \Rp+\Rc &\leq \log\left ( 1 + (\sqrt{\Sc} + \sqrt{\Ip})^2\right ) 
                + \log \left( \displaystyle\frac{1+ \Cc + \max\{\Ip,\Sp\} }{1+\Ip} \right),
%\end{array}\right.
\end{align}
\label{eq:outerBoundS}
\end{subequations}
from~\eqref{eq:ourRate} with $\Ic=0$,
can be achieved with a noncooperative scheme or not.
Note that the bounds on $\Rp$ and $\Rc$ in~\eqref{eq:outerBoundS} are the capacity region of the corresponding non-causal cognitive IC; therefore we interpret the sum-rate bound in~\eqref{eq:outerBoundS} as the `cost' for causally learning the primary message at the CTx through a noisy channel. 

For the proof of Theorem~\ref{thm:gap S}, we consider separately different parameter regimes. 
Given the result in Theorem~\ref{thm:gap full}, we should only consider the case $\Ip \leq \Sc\Sp - 1$ when $\Cc \leq \Sp$, and $\Ip \leq \Cc$ when $\Cc > \Sp$. However, here we will use a DPC-based scheme for the case $\max\{\Sp,\Ip\} < \Cc$ when we only used superposition coding in Section~\ref{subsec:cap regime Sp<Cc and Ip<C and Sc<Ic}.
%In the symmetric case, the regime ??? is equivalent to ???.

\subsection{Case $\Cc \leq \max\{\Ip,\Sp\}$: when unilateral cooperation might not be uselful}
\label{subsec:S useless} 

For the case $\Cc \leq \max\{\Ip,\Sp\}$ we can further outer bound the region in~\eqref{eq:outerBoundS} as
\begin{subequations}
\begin{align}
%O^{\rm(S)}\subseteq
%O^{\rm(S-1)}=
%\left\{\begin{array}{rl}
\mathcal{O}^{\ref{subsec:S useless}} : \quad
   \Rc &\leq \log\left ( 1 + \Sc \right ),
\\ \Rp &\leq \log\left ( 1 + \Sp \right ),
\\ \Rp+\Rc &\leq \log\left ( 1 + \Sc + \Ip \right ) 
               + \log^+ \left( \frac{1+ \Sp }{1+\Ip} \right)+2\log(2).
%\end{array}\right.
\end{align}
\label{eq:outerBoundS case 1}
\end{subequations}
The region in~\eqref{eq:outerBoundS case 1} is at most 1~bit away from
\begin{subequations}
\begin{align}
%I^{\rm(S-1)}=
%\left\{\begin{array}{rl}
\mathcal{I}^{\ref{subsec:S useless}} : \quad
   \Rc &\leq \log \left(1+ \Sc \right),
\\ \Rp &\leq \log \left(1+ \Sp \right),
\\ \Rp+\Rc &\leq \log\left ( 1 + \Sc + \Ip \right ) 
              + \log^+ \left( \frac{1+ \Sp }{1+\Ip} \right),
%\end{array}\right.
\end{align}
\label{eq:outerBoundS noncoop}
\end{subequations}
which is achievable to within 1~bit by a noncooperative scheme \cite{etw}.
Therefore we conclude that for $\Cc \leq \max\{\Ip,\Sp\}$ a noncooperative scheme is optimal to with 2~bits.

As for the Z-channel, the difference between this case and the corresponding case for the general channel in Theorem~\ref{thm:gap full} is that here we do not need to impose extra conditions to claim the redundancy of the bounds on $\Rp+2\Rc / 2\Rp+\Rc$ in the noncooperative achievable region since those bounds do not matter, up to a constant gap, in the noncooperative IC~\cite{etw}.

\subsection{Case $\Cc > \max\{\Ip,\Sp \}$: when unilateral cooperation is useful}
\label{subsec:S useful} 

When $\Cc > \max\{\Ip,\Sp \}$, a sufficient condition for the sum-rate upper bound in~\eqref{eq:outerBoundS} to be redundant is that 
\begin{align}
1+ \Sp \leq \frac{1+ \Cc+\max\{\Ip,\Sp \} }{1+\Ip}
\Longleftrightarrow
\Cc\geq \min\{\Ip,\Sp \} (1+\max\{\Ip,\Sp \}).
\label{eq:condition S NOT case 1: C very strong}
\end{align}
For the set of parameters in~\eqref{eq:condition S NOT case 1: C very strong}, we use the achievable region in~\eqref{eq:the simple DPC region} from Appendix~\ref{sec:scheme 3}, adapted to the S-channel case by setting $\Ic=0$, and with $|\gamma_{\mathsf{c}}|=0, \ \Cc(1-|\gamma_{\mathsf{p}}|^2) = \Sp$, to obtain the following achievable region
\begin{subequations}
\begin{align}
%   \Rp &\leq \log \left( 1+C(1-|\gp|^2) \right)
\mathcal{I}^{\ref{subsec:S useful}} : \quad
     \Rc &\leq \left( 1 +  \frac{\Sc}{1+\frac{\Sp \Ip}{\Cc}} \right)
 \\  \Rp &\leq \log \left( 1+\Sp\right).
\end{align}
\label{eq:the simple DPC region Ic=0}
\end{subequations}
By comparing the rate bounds in~\eqref{eq:the simple DPC region Ic=0} with those in~\eqref{eq:outerBoundS}, we see that when~\eqref{eq:condition S NOT case 1: C very strong} holds the gap is at most 1~bit since
\begin{align*}
  &\log\left ( 1 + \Sc \right ) - \log \left( 1 +  \frac{\Sc}{1+\frac{\Sp \Ip}{\Cc}} \right)
   \leq \log\left( 1 +  \frac{\Sp \Ip}{\Cc} \right)
\\&\leq \log\left( 1 +  \frac{\min\{\Ip,\Sp \} \ \max\{\Ip,\Sp \}}{\min\{\Ip,\Sp \} (1+\max\{\Ip,\Sp \})} \right)
   \leq \log(2).
\end{align*}
This shows that, when the condition in~\eqref{eq:condition S NOT case 1: C very strong} holds, not only the upper bound is achievable to within 1~bit but we can also achieve to within 1~bit the ultimate capacity of the corresponding non-causal cognitive channel. This results agrees with the intuition that, as the strength of the cooperation link increases, the performance of the causal cognitive channel should approach that of the corresponding non-causal model. The condition in~\eqref{eq:condition S NOT case 1: C very strong} can thus be interpreted as a sufficient condition on the strength of the cooperation link to achieve the capacity region of the corresponding non-causal model to within a constant gap.

\bigskip
We are now left with the case
\begin{align}
&\Big\{\max\{\Ip,\Sp \} < \Cc, \ \Cc < \min\{\Ip,\Sp \} (1+\max\{\Ip,\Sp \})\Big\}\subseteq
%\nonumber\\& 
\left\{ \Sp < \Cc < \Sp(1+\Ip) \right\}.
\label{eq:condition S NOT case 1: C strong}
\end{align}
%where we assume that $\min\{\Ip,\Sp \} \geq 1$ (if this is not the case, then we reason as follows: 
%if $\Sp\leq 1$ then $\Rp\leq\log(1+\Sp)\leq \log(2)$ and by silencing PTx the upper bound region is achieved to within 1~bit; 
%if $\Ip\leq 1$ then treating interference as noise looses at most a factor 2, or 1 bit). 
In the regime $\Sp < \Cc < \Sp(1+\Ip)$ we use the DPC-based in Appendix~\ref{app:gap for the general S-channel}. In this scheme CTx sends a private message only since $\Xc$ is not received at PRx; PTx sends a private and a common message, both with the help of CTx. The PTx's common message is forwarded by CTx to facilitate decoding at both receivers. The PTx's private message is decoded at CTx and its effect is `pre-canceled' at CRx thanks to DPC. The achievable region is given by~\eqref{eq:final dpc ach region Q=0 Ic=0 2} in Appendix~\ref{app:gap for the general S-channel}, namely 
\begin{subequations}
\begin{align}
\mathcal{I}^{\ref{subsec:S useful}} : \quad
   \Rp &\leq \log(1+\Sp),
\\ \Rc &\leq \log\left(1+\frac{\Sc}{1+\frac{\Ip}{1+\Ip}}\right),
\\ \Rp+\Rc &\leq  
  \log\left(\frac{1+\Sc+\Ip}{1+\Sc+\frac{\Ip}{1+\Ip} \ \frac{\Cc}{\Sp}}\right)
 +\log\left(1+\frac{\Sc}{1+\frac{\Ip}{1+\Ip}}\right)
 +\log\left(1+\frac{\Cc}{1+\Ip}\right).
\end{align}
\label{eq:final dpc ach region Q=0 Ic=0 2 again}
\end{subequations}
In Appendix~\ref{app:gap for the general S-channel} we show that the achievable region in~\eqref{eq:final dpc ach region Q=0 Ic=0 2 again} is optimal to within 2~bits when $\Sp < \Cc < \Sp(1+\Ip)$.
%, in strong interference ($\Ip \geq \Sp$) can be shown to be at most 1.5~bits away of the upper bound in~\eqref{eq:outerBoundS}, while in weak interference ($\Sp > \Ip$) to be at most 2~bits away of the upper bound in~\eqref{eq:outerBoundS}. The details of the gap derivation can be found 

Note that here we used a DPC-based scheme in order to determine the capacity to within a constant gap in weak interference, while  for the general GCCIC we only used superposition coding. Also, the choice of parameters in Appendix~\ref{app:gap for the general S-channel}  is unconventional, i.e., not inspired by~\cite{etw}, and might be necessary to show an approximate capacity result in weak interference for the general GCCIC.

\subsection{Comparisons}
We conclude the section by comparing the performance of unilateral cooperation on the S-channel with other forms of cooperation.
In order to reduce the number of parameters, we restrict our attention to the case where the direct links have the same strength.
For future reference, the gDoF of the noncooperative S-channel is given by~\cite{SASON}
\[
{\gdof}^{\rm IC-S} = \min\{\max\{1-\alpha/2,\alpha/2\}, \ 1\}
\]
and that of the non-causal cognitive  S-channel is given by~\cite{riniJ1}
\[
{\gdof}^{\rm CIC-S} = 1.
\]

Fig.~\ref{fig:fig5} shows the gDoF and the gap for the S-channel in the $(\alpha,\beta)$ plane. The whole set of parameters has been partitioned into multiple sub-regions depending upon different levels of cooperation ($\beta$) and interference ($\alpha$) strengths. We observe:
\begin{itemize}
%\item 
%In order to claim capacity for the S-channel, we make use of DPC technique \cite{costaDPC}. We believe that the use of this more involved encoding technique \cite{costaDPC} is necessary because in this scenario the CTx can not enhance the rate performance of the primary system since the PRx is out of the range of the CTx, but it can exploit the information overheard through the cooperation link in order to enhance the rate of its system. This can be attained by using a smarter encoding technique at the CTx that aims to pre-cancel some ``known'' interference at the CRx \cite{costaDPC}.

\item
Unilateral cooperation achieves the same gDoF of the noncooperative IC when $\alpha \geq 2$ or $\beta \leq \max \{ 1, \alpha \}$ (green region in Fig.~\ref{fig:fig8}).
In other words, unilateral cooperation is worth implementing in practice when the interference is not very strong and the cooperation link is the strongest among all links.

\item
The gDoF of unilateral cooperation never equals the gDoF of the relay channel. Actually when the link CTx$\rightarrow$CRx is not present, the channel achieves ${\gdof}=\frac{1}{2}$ (since $\Rc = 0$) that is always smaller than the gDoF achieved when the link CTx$\rightarrow$CRx exists, i.e. $\Rc \neq 0$. 

\item
The S-channel achieves the same gDoF of the non-causal cognitive IC everywhere except in $\alpha\leq 2$ and $\beta \leq \min\{2,\alpha+1\}$ (region with horizontal lines in Fig.~\ref{fig:fig8}).

\item
The gDoF of unilateral cooperation equals the gDoF upper bound of bilateral cooperation when $\alpha \geq 2$ and $ \beta \leq 1$ or when $\alpha\leq 2$  and $\beta \leq \min\{2,\alpha+1\}$
%{\blue
%\begin{itemize}
%\item $\alpha \geq 2, \beta \leq 1$;
%\item $1 \leq \alpha \leq 2, \beta \leq 2$;
%\item $0 \leq \alpha \leq 1, \beta \leq \max \{1,\alpha+1 \}$.
%\end{itemize}}
(region with vertical lines in Fig.~\ref{fig:fig8}).

\item
The S-channel outperforms the interference-symmetric CCIC when either $0 \leq \alpha\leq \frac{2}{3}$ and $\beta \leq \min\{\alpha, 1 - \alpha\}$ or when $\alpha \leq 2$ and $\beta \geq \max \{ 1, \alpha \}$ (green region in Fig.~\ref{fig:fig9}).

On the other hand, the interference-symmetric GCCIC outperforms the S-channel in very strong interference and strong cooperation, i.e., $\alpha \geq 2$ and $\beta \geq 1$. This is due to the fact that the information for the PRx can no longer be routed through the CTx since $\sqrt{\Ic} {\rm e}^{{\rm j}\theta_c}=0$ (red region in Fig.~\ref{fig:fig9}). 

%\item
%When $\alpha \geq 2$, i.e. very strong interference regime, we have an exact sum-capacity result, i.e., the gap between the outer bound and inner bound is equal to zero.

\end{itemize}

\section{Conclusions}
\label{sec:Conclusion}

In this work we considered the CCIC, a network with two source-destination pairs sharing the same channel. In contrast to the noncooperative IC, in the CCIC the CTx exploits information about the PTx from its own channel observations. This scenario represents a more practically relevant model for cognitive radio than the non-causal cognitive IC, where the CTx is assumed to have a priori knowledge of the PTx's message.  In particular, we believe that it is applicable in some practical heterogeneous deployments for 4G cellular networks.

%We developed and evaluated a baseline simple achievable scheme based on time-sharing between the achievable regions of the classical interference and relay channel which we showed to be gDoF optimal for some parameter regimes.
%% by numerical evaluations that this simple strategy may be not too far from an outer bound in certain parameter regimes.
%
%We exploited known outer bounds for bilateral source cooperation adapting them to the case of unilateral cooperation. Our main contribution consisted in showing
We proposed achievable schemes that match known outer bounds to within a constant gap if, roughly speaking, the channel does not exhibit weak interference at both destinations. We characterized the capacity region to within a constant gap for the case where one interfering link is absent, which includes cases of weak interference. From our analysis a practical guideline for system design is that superposition coding is approximately optimal when the interference at the primary receiver is strong and that binning / dirty paper coding is approximately optimal when the interference at the primary receiver is weak.
%that both for the interference-symmetric and interference-asymmetric cases the sum-capacity of the two-user Gaussian CCIC with independent noises can be achieved to within a constant gap. 
%{\blue Interestingly, the achievable schemes use superposition coding in strong interference and dirty paper coding otherwise.}
%Moreover, we gave an example where a more complex scheme employing binning/dirty-paper-coding achieves a smaller gap than superposition at the cost of an increase in the coding complexity.
We identified the set of parameters where causal cooperation achieves the same gDoF of the noncooperative IC and of the relay channel.
We also highlighted under which channel conditions the gDoF achieved with bilateral cooperation and with non-causal cognition equals that achieved with only unilateral causal cooperation.

\appendices

\section{Capacity region upper bound and gDoF upper bound}
\label{app:outer}
%In order  to prove~\eqref{eq:Dof upper asym} 
In this work we use known outer bounds from~\cite{HostMadsenIT06,TuninettiITA10,PVIT11}.
These bounds were developed for the case of bilateral source cooperation.
Here we adapt them to the case of unilateral source cooperation.

%{\red RECHECK ALL SUBSCRIPTS AS I WROTE THE BOUNDS FOR THE GENERAL CASE.}
%\begin{itemize}
%
%\item
\subsection{Cut-set upper bounds}
The cut-set upper bound for a relay channel with 
gain ${\snr}$ on the link from the source to the destination,
gain $\Cc$ on the link from the source to the relay, and
gain ${\inr}$ on the link from the relay to the destination is
upper bounded by~\cite{ElGamalKimBook}
\begin{align}
 &\max_{|\rho|\leq1}
\min \Big\{
    \log\left ( 1 + {\snr} + {\inr} + 2|\rho|\sqrt{ {\snr} {\inr}}\right ), 
    \log\left ( 1  +  \left ( 1 - |\rho|^2 \right ) \left ( \Cc  +  {\snr} \right) \right )
\Big\}
\nonumber\
\\&\leq 
\min\Big\{
     \log\left ( 1+(\sqrt{{\snr}}+\sqrt{{\inr}})^2\right ), 
   \log\left ( 1 +  \Cc + {\snr} \right ) 
\Big\}
=:r^{(\rm{RC})}({\snr},{\snr},\Cc).
\label{eq:app rCS}
\end{align}
The behavior of the rate $r^{(\rm{RC})}({\snr},{\inr},\Cc)$ in~\eqref{eq:app rCS} at high SNR, with ${\inr}={\snr}^\alpha, \Cc={\snr}^\beta$, is given by~\eqref{eq:Dof upper asym:CS gen}.

For an IC with cooperative sources, the rate of a given source cannot be larger than the rate that this source can achieve when the other source acts as a pure relay. Therefore we have 
%The cut-set upper bound for the general interference channel with unilateral source cooperation with channel gains as in~\eqref{eq:paramet} is equivalent to
\begin{align}
\Rp &\leq r^{(\rm{RC})}(\Sp,\Ic,\Cc) \label{eq:cusetupper general rp}\\
\Rc &\leq r^{(\rm{RC})}(\Sc,\Ip,0  ) \label{eq:cusetupper general rc}
\end{align}
which are the upper bounds on the individual rates in~\eqref{eq:CutSetRc} and~\eqref{eq:CutSetRp}, which imply the sum-rate upper bound in~\eqref{eq:CutSet}.

%\item
\subsection{Sum-rate bounds from~\cite{TuninettiITA10}}
From~\cite{TuninettiITA10} %(from which the DT upper bound in~\eqref{eq:Tuninetti} was derived)
 we have
\begin{align*}
\Rp+ \Rc
  & \leq  \max_{|\rho|\leq1}
    \log \left( \frac{1+ \left ( 1-|\rho|^2 \right )\left ( \Cc + \max\{\Ip,\Sp \} \right )}{1+ \left ( 1-|\rho|^2 \right) \Ip} \right)
 +\log \left ( 1+ \Ip + \Sc + 2 |\rho|\sqrt{\Sc \Ip }  \right )
\\& \leq    \log \left( \frac{1+ \Cc + \max\{\Ip,\Sp \} }{1+ \Ip} \right)
 +\log \left( 1+ (\sqrt{\Ip} + \sqrt{\Sc})^2 \right ).
% =: r^{(\rm{DT})}(S,I_9,C)
\end{align*}
%By swapping the role of the cognitive and primary pairs we have
%\begin{align*}
%&\Rp + \Rc 
%\leq  r^{(\rm{DT})}(S,I_p,0).
%\end{align*}
By swapping the role of the users, we obtain a similar sum-rate upper bound, and
the combination of the two gives the sum-rate upper bound in~\eqref{eq:Tuninetti}. 

The function
\[
r^{(\rm{DT})}({\snr},{\inr},\Cc) := 
 \log \left( \frac{1+ \Cc + \max\{{\snr},{\inr}\} }{1+ {\inr}} \right)
+\log \left( 1+ (\sqrt{{\inr}} + \sqrt{{\snr}})^2 \right )
\]
with ${\inr}={\snr}^\alpha, \Cc={\snr}^\beta$, has the high SNR behavior given by~\eqref{eq:Dof upper asym:DT gen}.
%Therefore, for the case where the direct links are the same and with the parameterization in~\eqref{eq:paramet}, the bounds from~\cite{TuninettiITA10} imply~\eqref{eq:Dof upper asym:DT}.

%\item
\subsection{Sum-rate bound from~\cite{PVIT11}}
From~\cite{PVIT11} we have the sum-rate upper bound reported in~\eqref{eq:PV},
%\begin{align*}
%&\Rp+\Rc \leq
%    \log \left( 1+C \right)
%\\&+\log \left( 1+ \left( \frac{\sqrt{S}}{\max \{ 1,\sqrt{I_9} \}} + \sqrt{I_p} \right)^2 \right) 
%\\&+\log \left( 1+ \left( \frac{\sqrt{S}}{\max \{ 1,\sqrt{I_p} \}} + \frac{\sqrt{I_9}}{\max \{1,\sqrt{C} \}} \right)^2 \right)=: r^{(\rm{PV})}
%\end{align*}
whose behavior at high SNR, with the parameterization in \eqref{eq:paramet}, gives~\eqref{eq:Dof upper asym:PV gen}.
%Hence we get the gDoF upper bound in~\eqref{eq:Dof upper asym:PV} are the same. 

%\end{itemize}

\section{Achievable Schemes Based on Superposition Coding only}
\label{sec:allachschemsappSUP}

\subsection{Superposition-only Achievable Scheme}
\label{sec:scheme sup}

\begin{figure}[!h]
\centering
\includegraphics[width=0.7\textwidth]{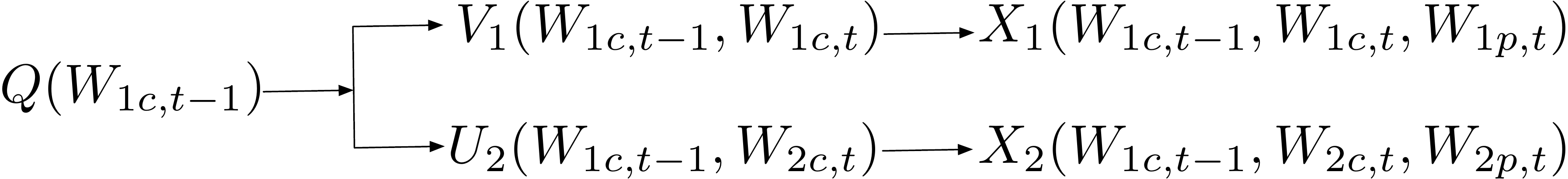}
\caption{Achievable scheme based on superposition only.}
\label{fig:achsuponly}
\end{figure}

We specialize the `superposition only' achievable scheme in~\cite[Thereom IV.1]{YANG-TUNINETTI} to the case of unilateral cooperation.
In~\cite[Thereom IV.1]{YANG-TUNINETTI}, the network comprises four nodes numbered from 1 to 4; nodes 1 and 2 are sources and nodes 3 and 4 destinations; source node $j\in[1:2]$, with input to the channel $X_j$ and output from the channel $Y_j$, has a message $W_j$ for node $j+2$; destination node $j\in[3:4]$ has channel output $Y_{j}$ from which it decodes the message $W_{j-2}$. 

Both users do rate splitting, where only the common message of user 1 is cooperative, while all other messages are noncooperative.
We set $Q=V_2, \ Y_1=\emptyset, \ T_2=X_2, \ U_1=\emptyset, \ T_1=X_1$ in~\cite[Thereom IV.1]{YANG-TUNINETTI},
i.e., then $R_1=R_{11n}+R_{10c},  \ R_2=R_{22n}+R_{20n}$, to obtain a scheme that comprises:
a    cooperative common message  (carried by the pair $(Q  ,V_1)$ at rate $R_{10c}$) for user 1,
a noncooperative private message (carried by $X_1$ at rate $R_{11n}$) for user 1,
a noncooperative common message  (carried by $U_2$ at rate $R_{20n}$) for user 2 and
a noncooperative private message (carried by $X_2$ at rate $R_{22n}$) for user 2.
Here $Q$ carries the `past cooperative common message', and $V_1$ the `new cooperative common message' in a block Markov encoding scheme. 

The set of possible input distributions is
\begin{align}
 &P_{Q, V_1,X_1, U_2,X_2}
= P_{Q} P_{V_1,X_1|Q} P_{U_2,X_2|Q}.
\label{eq:input pdf sup}
\end{align}

A schematic representation of the achievable scheme is given in Fig.~\ref{fig:achsuponly}, where an arrow indicates superposition coding.

Regarding encoding. 
Source~2 cooperates with source~1 by using decode-and-forward in a block Markov coding scheme.
In a given slot 
         the old cooperative common message of source~1 is carried by $Q$,
to which the new cooperative common message of source~1 is superposed and carried by $V_1$,
to which the     noncooperative private message of source~1 is superposed and carried by $X_1$.
After source~2 decodes the new cooperative common message of source~1 carried by $V_1$, 
with knowledge of $Q$ and by treating the noncooperative private message of source~1 in $X_1$ as noise,
it superposes its noncooperative common message carried by $U_2$ to the old cooperative common message of source~1 carried by $Q$,
and then 
it superposes its noncooperative private message carried by $X_2$.
In this scheme the common messages are jointly (backward) decoded at all destinations while treating the non-intended private massage as noise.

Regarding decoding.
There are three decoding nodes in the network and therefore three groups of rate constraints.
These are:
\begin{subequations}
\begin{itemize}
\item
Node~2/CTx decodes $V_1$ from its channel output with knowledge of $(Q,U_2,X_2)$.  Successful decoding is possible if (see \cite[eq(6a)]{YANG-TUNINETTI} 
\begin{align}
R_{10c} \leq I(Y_2;V_1|U_2,X_2,Q). \label{eq:suponly 6a} %6a
\end{align}

\item
Node~3/PRx jointly decodes $(Q,V_1,X_1,U_2)$ from its channel output, with knowledge of some message indices in $V_1$, by treating $X_2$ as noise. Successful decoding is possible if (see \cite[eq(6b)-(6f)]{YANG-TUNINETTI} 
\begin{align}
R_{10c}+R_{20n}+R_{11n} &\leq I(Y_3;Q,V_1,X_1,U_2) \label{eq:suponly 6f}\\ %6f
        R_{20n}+R_{11n} &\leq I(Y_3;X_1,U_2|Q,V_1) \label{eq:suponly 6c}\\ %6c=6e
                R_{11n} &\leq I(Y_3;X_1|Q,V_1,U_2).\label{eq:suponly 6b}   %6b=6d
\end{align}

\item
Node~4/CRx jointly decodes $(Q,V_1,U_2,X_2)$ from its channel output, with knowledge of some message index in $V_1$, by treating $X_1$ as noise. 
Successful decoding is possible if (see \cite[eq(7b)-(7f)]{YANG-TUNINETTI} 
\begin{align}
R_{10c}+R_{20n}+R_{22n} &\leq I(Y_4;Q,V_1,X_2,U_2) \label{eq:suponly 7f}\\ %7f
        R_{20n}+R_{22n} &\leq I(Y_4;X_2,U_2|Q,V_1) \label{eq:suponly 7d}\\ %7d=7e
                R_{22n} &\leq I(Y_4;X_2|Q,V_1,U_2).\label{eq:suponly 7b}   %7b=7c
\end{align}
\end{itemize}
\label{eq:suponly all constrints}
\end{subequations}

The achievable region, after Fourier-Motzkin elimination, is given by~\cite[Thereom IV.1]{YANG-TUNINETTI}
\begin{subequations}
\begin{align}
  R_1     &\leq {\rm eq\eqref{eq:suponly 6f}} \label{eq:ach reg 2 not in a very insightful form 8a}
\\R_1     &\leq {\rm eq\eqref{eq:suponly 6a}} 
              + {\rm eq\eqref{eq:suponly 6b}} \label{eq:ach reg 2 not in a very insightful form 8b}
\\R_2     &\leq {\rm eq\eqref{eq:suponly 7d}} \label{eq:ach reg 2 not in a very insightful form 8d}
\\R_1+R_2 &\leq {\rm eq\eqref{eq:suponly 6f}} 
              + {\rm eq\eqref{eq:suponly 7b}} \label{eq:ach reg 2 not in a very insightful form 8e}
\\R_1+R_2 &\leq {\rm eq\eqref{eq:suponly 7f}} 
              + {\rm eq\eqref{eq:suponly 6b}} \label{eq:ach reg 2 not in a very insightful form 8f}
\\R_1+R_2 &\leq {\rm eq\eqref{eq:suponly 6a}} 
              + {\rm eq\eqref{eq:suponly 6c}} 
              + {\rm eq\eqref{eq:suponly 7b}} \label{eq:ach reg 2 not in a very insightful form 8g}
\\R_1+2R_2&\leq {\rm eq\eqref{eq:suponly 6c}} 
              + {\rm eq\eqref{eq:suponly 7b}} 
              + {\rm eq\eqref{eq:suponly 7f}} \label{eq:ach reg 2 not in a very insightful form 8l}
\end{align}
\label{eq:ach sup FME}
\end{subequations}
for all distributions that factor as~\eqref{eq:input pdf sup}.

\bigskip
Remark~1. 
The rate bound in~\eqref{eq:ach reg 2 not in a very insightful form 8l} is redundant if
\[
\min\{{\rm eq\eqref{eq:ach reg 2 not in a very insightful form 8e}},
       {\rm eq\eqref{eq:ach reg 2 not in a very insightful form 8f}}, 
       {\rm eq\eqref{eq:ach reg 2 not in a very insightful form 8g}} \}       
+ {\rm eq\eqref{eq:ach reg 2 not in a very insightful form 8d}}
\leq 
 {\rm eq\eqref{eq:ach reg 2 not in a very insightful form 8l}} 
\]
that is, if for the considered input distribution we have
\begin{subequations}
\begin{align}
&\text{either} \
 {\rm eq\eqref{eq:suponly 6f}} + {\rm eq\eqref{eq:suponly 7d}}% + {\rm eq\eqref{eq:suponly 7b}}
\leq
 {\rm eq\eqref{eq:suponly 6c}} + {\rm eq\eqref{eq:suponly 7f}}% + {\rm eq\eqref{eq:suponly 7b}} 
\Longleftrightarrow
I(Y_3; Q,V_1)\leq I(Y_4; Q,V_1), %common 1 better decoded at 2
%\text{i.e., Rx2 better decodes the common message of Tx1},
\\&\text{or} \
 {\rm eq\eqref{eq:suponly 6b}} + {\rm eq\eqref{eq:suponly 7d}}%+ {\rm eq\eqref{eq:suponly 7f}} 
\leq
 {\rm eq\eqref{eq:suponly 6c}} + {\rm eq\eqref{eq:suponly 7b}}%+ {\rm eq\eqref{eq:suponly 7f}}
\Longleftrightarrow
I(Y_4; U_2|Q,V_1)\leq I(Y_3; U_2|Q,V_1), %common 2 better decoded at 1
%\text{i.e., Rx1 better decodes the common message of Tx2},
\\&\text{or} \
 {\rm eq\eqref{eq:suponly 6a}} + {\rm eq\eqref{eq:suponly 7d}}%+ {\rm eq\eqref{eq:suponly 6c}}  + {\rm eq\eqref{eq:suponly 7b}}
\leq 
 {\rm eq\eqref{eq:suponly 7f}}%{\rm eq\eqref{eq:suponly 6c}} + {\rm eq\eqref{eq:suponly 7b}} + 
\Longleftrightarrow
I(Y_2;V_1|U_2,X_2,Q) \leq I(Y_4; Q,V_1),
\end{align}
\label{eq:ach reg 2 not in a very insightful form:when is r1+2r2 redundant}
\end{subequations}
%We will impose one of the conditions in~\eqref{eq:ach reg 2 not in a very insightful form:when is r1+2r2 redundant} to claim that the bound in~\eqref{eq:ach reg 2 not in a very insightful form 8l} is redundant when evaluating the region in~\eqref{eq:ach sup FME} in Gaussian noise. 

\bigskip
Remark~2. If the private message of user~1 carried by $X_1$ is also decoded at Node~2 (a strategy that could be leading to a larger region than the one in~\eqref{eq:ach sup FME} when the link between PTx and CTX is very large), then successful decoding at the cooperating source is possible if
\begin{subequations}
\begin{align}
R_1=R_{10c} +R_{11n} &\leq I(Y_2;V_1,X_1|U_2,X_2,Q), \label{eq:suponly 6a new sum}\\
             R_{11n} &\leq I(Y_2;X_1|V_1,U_2,X_2,Q). \label{eq:suponly 6a new private}
\end{align}
\label{eq:suponly 6a new}
\end{subequations}
If we now do Fourier-Motzkin elimination of the region in~\eqref{eq:suponly all constrints}, by replacing the constraint in~\eqref{eq:suponly 6a} with those in~\eqref{eq:suponly 6a new}, we obtain a new achievable region where the bounds that depend on~\eqref{eq:suponly 6a} in~\eqref{eq:ach sup FME} change as follows:
the bound in~\eqref{eq:ach reg 2 not in a very insightful form 8b} is replaced by~\eqref{eq:suponly 6a new sum}, and 
the one in~\eqref{eq:ach reg 2 not in a very insightful form 8g} by $R_1+R_2\leq{\rm eq\eqref{eq:suponly 7f}}+{\rm eq\eqref{eq:suponly 6a new private}}$. In Appendix~\ref{sec:allachschemsappDPC} we shall further improve on this scheme by using DCP to cancel the `known interference' due to the private message decoded at the cooperating source.

\subsection{Achievable Scheme 1: message~1 is common, and message~2 is split}
\label{sec:scheme 1}
By identifying Node1 with the PTx (i.e., $\Xp=X_1$), Node2 with the CTx  (i.e., $\Xc=X_2, \Yf=Y_2$), Node3 with the PRx  (i.e., $\Yp=Y_3$) and Node4 with the CRx (i.e., $\Yc=Y_4$), by further setting $Q=\emptyset, V_1=X_1$ (i.e., $R_{11n}=0, \ R_1=R_{10c}$) in the scheme in~\eqref{eq:ach sup FME} in Appendix~\ref{sec:scheme sup},  the following region is achievable
\begin{subequations}
\begin{align}
   \Rp     &\leq I(\Yf;\Xp|U_2,\Xc)
\\ \Rc     &\leq I(\Yc;U_2,\Xc|\Xp)
\\ \Rp+\Rc &\leq I(\Yp;U_2,\Xp)+I(\Yc;\Xc|U_2,\Xp)
\\ \Rp+\Rc &\leq I(\Yc;\Xp,U_2,\Xc)
\end{align}
\label{eq:sup.only ach.region with common messages only in unilateral cooperation V1=X1 and U2}
\end{subequations}
for all input distributions that factor as $P_{\Xp,U_2,\Xc} = P_{\Xp}P_{\Xc,U_2}$.

\bigskip
In Gaussian noise, we choose $\Xp,U_2,L_2$  to be i.i.d. $\mathcal{N}(0,1)$, and $\Xc=\gc U_2 +\sqrt{1-|\gc|^2} L_2$ for $|\gc| \leq 1$. With this choice of inputs, the channel outputs are
\begin{align*}
  &\Yf = \sqrt{\Cc}\Xp+\Zf
\\&\Yp = \sqrt{\Sp}\Xp
       + \sqrt{\Ic} \eac \left(\gc U_2+ \sqrt{1-|\gc |^2} L_2\right)+\Zp
\\&\Yc = \sqrt{\Sc}\left(\gc U_2+ \sqrt{1-|\gc |^2} L_2\right)
       + \sqrt{\Ip} \eap \Xp+\Zc
\end{align*}
and the achievable region in~\eqref{eq:sup.only ach.region with common messages only in unilateral cooperation V1=X1 and U2} reduces to
%\begin{subequations}
\begin{align*}
   \Rp &\leq \log(1+\Cc)
\\ \Rc &\leq \log(1+\Sc)
\\ \Rp+\Rc &\leq \log(1+\Sp+\Ic)-\log(1+(1-|\gc |^2)\Ic)+\log(1+(1-|\gc |^2)\Sc)
\\ \Rp+\Rc &\leq \log(1+\Sc+\Ip)
\end{align*}
%\label{eq:sup.only ach.region with common messages only in unilateral cooperation V1=X1 gaussian}
%\end{subequations}
for all $|\gc| \leq 1$. 
If $\Sc \leq \Ic$ we choose $|\gc |=1$ otherwise $|\gc |=0$ to obtain
\begin{subequations}
\begin{align}
   \Rp &\leq \log(1+\Cc)
\\ \Rc &\leq \log(1+\Sc)
\\ \Rp+\Rc &\leq \log(1+\Sp+\Ic)+\log^+\left(\frac{1+\Sc}{1+\Ic}\right)
\\ \Rp+\Rc &\leq \log(1+\Sc+\Ip).
\end{align}
\label{eq:sup.only ach.region with common messages only in unilateral cooperation V1=X1 gaussian}
\end{subequations}

\subsection{Achievable Scheme 2: both messages are split}
\label{sec:scheme 2}

For the GCCIC we identifying Node1 with the PTx (i.e., $\Xp=X_1$), Node2 with the CTx  (i.e., $\Xc=X_2, \Yf=Y_2$), Node3 with the PRx  (i.e., $\Yp=Y_3$) and Node4 with the CRx (i.e., $\Yc=Y_4$) in the scheme in~\eqref{eq:ach sup FME} in Appendix~\ref{sec:scheme sup}.

In Gaussian noise, in order to comply with~\eqref{eq:input pdf sup}, we choose $Q=\emptyset$, $V_1,L_1,U_2,L_2$ i.i.d. $\mathcal{N}(0,1)$ and we let
\begin{align*}
  & \Xc = \gc U_2+ \sqrt{1-|\gc |^2} L_2 : |\gc |^2 \leq 1
\\& \Xp = \gp V_1+ \sqrt{1-|\gp |^2} L_1 : |\gp |^2 \leq 1.
\end{align*}
With this choice of inputs the channel outputs are given by 
\begin{align*}
  &\Yf = \sqrt{\Cc} \left (\gp V_1+ \sqrt{1-|\gp |^2} L_1 \right )+\Zf
\\&\Yp = \sqrt{\Sp} \left (\gp V_1+ \sqrt{1-|\gp |^2} L_1 \right )
       + \sqrt{\Ic} \eac \left (\gc U_2+ \sqrt{1-|\gc |^2} L_2 \right )+\Zp
\\&\Yc = \sqrt{\Sc}\left (\gc U_2+ \sqrt{1-|\gc |^2} L_2 \right)
       + \sqrt{\Ip} \eap \left (\gp V_1+ \sqrt{1-|\gp |^2} L_1 \right )+\Zc.
\end{align*}
Inspired by~\cite{etw} for the noncooperative IC in weak interference, we set $( 1-|\gc|^2 )\Ip = ( 1-|\gp|^2 )\Ic = 1$ (here we assume $1\leq \min\{\Ip,\Ic\}$) so that the scheme in~\eqref{eq:ach sup FME} in Appendix~\ref{sec:scheme sup} results in the following achievable region 
%\begin{subequations}
%\begin{align}
%   \cite[\rm eq(6a)]{YANG-TUNINETTI}&
%= \log \left( \frac{1+\Cc}{1+ \Cc/\Ip} \right)
%%
%\\ \cite[\rm eq(6b)]{YANG-TUNINETTI}
%=  \cite[\rm eq(6d)]{YANG-TUNINETTI}&
%= \log \left( 1+\frac{\Sp/\Ip}{2} \right)
%%
%\\ \cite[\rm eq(6c)]{YANG-TUNINETTI}
%=  \cite[\rm eq(6e)]{YANG-TUNINETTI}&
%= \log \left( \frac{1+\Ic+\Sp/\Ip}{2} \right)
%%
%\\ \cite[\rm eq(6f)]{YANG-TUNINETTI}&
%= \log \left( \frac{1+\Sp + \Ic}{2} \right) 
%%
%%
%\\ \cite[\rm eq(7b)]{YANG-TUNINETTI}
%=  \cite[\rm eq(7c)]{YANG-TUNINETTI}&
% = \log \left( 1+ \frac{\Sc/\Ic}{2} \right)
%%
%\\ \cite[\rm eq(7d)]{YANG-TUNINETTI}
%=  \cite[\rm eq(7e)]{YANG-TUNINETTI}&
%= \log \left( 1+\frac{\Sc}{2} \right)
%%
%\\ \cite[\rm eq(7f)]{YANG-TUNINETTI}&
%= \log \left( \frac{1+\Sc+\Ip}{2} \right).
%\end{align}
%\label{eq:lots of mutual info}
%\end{subequations}
%With these choices, the achievable region in~\eqref{eq:ach reg 2 not in a very insightful form} reduces to
\begin{subequations}
\begin{align}
   \Rp     &\leq \log \left( \frac{1+\Sp + \Ic}{2} \right)   \label{eq:ach reg sup only with etw power split rp1}%\cite[\rm eq(6f)]{YANG-TUNINETTI}
\\ \Rp     &\leq \log \left( \frac{1+\Cc}{1+ \Cc/\Ip} \right)%\cite[\rm eq(6a)]{YANG-TUNINETTI} 
               + \log \left( 1+\frac{\Sp/\Ip}{2} \right)\label{eq:ach reg sup only with etw power split rp2}%\cite[\rm eq(6d)]{YANG-TUNINETTI} 
\\ \Rc     &\leq \log \left( 1+\frac{\Sc}{2} \right)\label{eq:ach reg sup only with etw power split rc}%\cite[\rm eq(7d)]{YANG-TUNINETTI}
\\ \Rp+ \Rc&\leq \log \left( \frac{1+\Sp + \Ic}{2} \right)%\cite[\rm eq(6f)]{YANG-TUNINETTI}
               + \log \left( 1+ \frac{\Sc/\Ic}{2} \right)\label{eq:ach reg sup only with etw power split rsum1}%\cite[\rm eq(7b)]{YANG-TUNINETTI}
\\ \Rp+ \Rc&\leq \log \left( \frac{1+\Sc+\Ip}{2} \right)%\cite[\rm eq(7f)]{YANG-TUNINETTI} 
               + \log \left( 1+\frac{\Sp/\Ip}{2} \right)\label{eq:ach reg sup only with etw power split rsum2}%\cite[\rm eq(6b)]{YANG-TUNINETTI}
\\ \Rp+ \Rc&\leq \log \left( \frac{1+\Cc}{1+ \Cc/\Ip} \right)%\cite[\rm eq(6a)]{YANG-TUNINETTI} 
               + \log \left( \frac{1+\Ic+\Sp/\Ip}{2} \right)%\cite[\rm eq(6e)]{YANG-TUNINETTI} 
               + \log \left( 1+ \frac{\Sc/\Ic}{2} \right)\label{eq:ach reg sup only with etw power split rsum3}%\cite[\rm eq(7b)]{YANG-TUNINETTI}
\\ \Rp+2\Rc&\leq \log \left( \frac{1+\Ic+\Sp/\Ip}{2} \right)%\cite[\rm eq(6c)]{YANG-TUNINETTI} 
               + \log \left( 1+ \frac{\Sc/\Ic}{2} \right)%\cite[\rm eq(7b)]{YANG-TUNINETTI} 
               + \log \left( \frac{1+\Sc+\Ip}{2} \right)\label{eq:ach reg sup only with etw power split else}%\cite[\rm eq(7f)]{YANG-TUNINETTI}
\end{align}
\label{eq:ach reg sup only with etw power split}
\end{subequations}
{Note that 
the sum-rate in~\eqref{eq:ach reg sup only with etw power split rsum1} and the first upper bound in~\eqref{eq:Tuninetti} differ by at most 3~bits, and
and the sum-rate in~\eqref{eq:ach reg sup only with etw power split rsum2} and the second upper bound in~\eqref{eq:Tuninetti}  by at most 4~bits when $\Cc \leq  \max\{\Sp,\Ip\}$.
}

\bigskip
For the symmetric case, i.e., $\Sc=\Sp={\snr}, \Ic=\Ip={\inr}$, 
%inspired by~\cite{etw} for the noncooperative IC in weak interference, we set $( 1-|\gc|^2 ) = ( 1-|\gp|^2 ) = \frac{1}{{\inr}}$ (here we assume ${\inr}\geq 1$; if ${\inr}\leq 1$ then the interference is at the level of the noise and can be treated as noise) %, or alternatively set $( 1-|\gc|^2 ) = ( 1-|\gp|^2 ) = \frac{1}{1+{\inr}}$),  to get from~\eqref{eq:ach sup FME} 
the following  sum-rate is achievable from~\eqref{eq:ach reg sup only with etw power split}
\begin{subequations}
\begin{align}
&\Rp+\Rc
\leq \max \min\{
\nonumber\\&  \min\{
{\rm eq\eqref{eq:ach reg 2 not in a very insightful form 8a}},
{\rm eq\eqref{eq:ach reg 2 not in a very insightful form 8b}}\}+
{\rm eq\eqref{eq:ach reg 2 not in a very insightful form 8d}}
\label{eq:sumrate weak 2 first}
\\&{\rm eq\eqref{eq:ach reg 2 not in a very insightful form 8e}},
   {\rm eq\eqref{eq:ach reg 2 not in a very insightful form 8f}},
   {\rm eq\eqref{eq:ach reg 2 not in a very insightful form 8g}},
\label{eq:sumrate weak 2 second}
\\&\frac{\min\{
{\rm eq\eqref{eq:ach reg 2 not in a very insightful form 8a}},
{\rm eq\eqref{eq:ach reg 2 not in a very insightful form 8b}}\}+
{\rm eq\eqref{eq:ach reg 2 not in a very insightful form 8l}}}{2} 
\Big \}
\label{eq:sumrate weak 2 third}
\end{align}
\label{eq:sumrate weak 2}
\end{subequations}
with
\begin{align*}
  &{\rm eq\eqref{eq:ach reg 2 not in a very insightful form 8a}}
 = {\rm eq\eqref{eq:ach reg sup only with etw power split rp1}}
 = \log \left( \frac{{\snr} + {\inr}+1}{2} \right)
\\ &{\rm eq\eqref{eq:ach reg 2 not in a very insightful form 8b}}
 = {\rm eq\eqref{eq:ach reg sup only with etw power split rp2}}
= \log \left( \frac{1+\Cc}{1+\frac{\Cc}{{\inr}}} \right)
+ \log \left( 1+\frac{{\snr}}{2{\inr}} \right)
\\ &{\rm eq\eqref{eq:ach reg 2 not in a very insightful form 8d}}
 = {\rm eq\eqref{eq:ach reg sup only with etw power split rc}}
 = \log \left( 1+\frac{{\snr}}{2} \right)
\\ &{\rm eq\eqref{eq:ach reg 2 not in a very insightful form 8e}}
 = {\rm eq\eqref{eq:ach reg sup only with etw power split rsum1}}
 = \log \left( \frac{{\snr} + {\inr}+1}{2} \right)
 + \log \left( 1+ \frac{{\snr}}{2{\inr}} \right)
\\ &{\rm eq\eqref{eq:ach reg 2 not in a very insightful form 8f}}
 = {\rm eq\eqref{eq:ach reg sup only with etw power split rsum2}}
 = \log \left( 1+\frac{{\snr}}{2{\inr}} \right)
 + \log \left(\frac{{\snr}+{\inr}+1}{2} \right)
\\ &{\rm eq\eqref{eq:ach reg 2 not in a very insightful form 8g}}
 = {\rm eq\eqref{eq:ach reg sup only with etw power split rsum3}}
 = \log \left( \frac{1+\Cc}{1+\frac{\Cc}{{\inr}}} \right)
 + \log \left( \frac{\frac{{\snr}}{{\inr}}+{\inr}+1}{2} \right)
 + \log \left( 1+ \frac{{\snr}}{2{\inr}} \right)
\\ &{\rm eq\eqref{eq:ach reg 2 not in a very insightful form 8l}}
 = {\rm eq\eqref{eq:ach reg sup only with etw power split else}}
= \log \left( \frac{\frac{{\snr}}{{\inr}}+{\inr}+1}{2} \right)
+\log \left( 1+ \frac{{\snr}}{2{\inr}} \right)
+\log \left( \frac{{\snr}+{\inr}+1}{2} \right).
\end{align*}

We next show that the sum-rate in~\eqref{eq:sumrate weak 2} is equal to the term in~\eqref{eq:sumrate weak 2 second}.
In order to show that the term in~\eqref{eq:sumrate weak 2 first} is redundant, consider the following facts:
\begin{itemize}

\item
${\rm eq\eqref{eq:ach reg 2 not in a very insightful form 8a}} 
+{\rm eq\eqref{eq:ach reg 2 not in a very insightful form 8d}}$ is always greater than 
${\rm eq\eqref{eq:ach reg 2 not in a very insightful form 8e}}$ because
${\snr} \geq \frac{{\snr}}{{\inr}}$, since we assume ${\inr} \geq 1$. 

\item
${\rm eq\eqref{eq:ach reg 2 not in a very insightful form 8b}}
+{\rm eq\eqref{eq:ach reg 2 not in a very insightful form 8d}}$ is always greater than 
${\rm eq\eqref{eq:ach reg 2 not in a very insightful form 8g}}$ since
$2{\inr}+{\snr} {\inr} \geq {\snr}+{\inr}^2+{\inr}
\Longleftrightarrow
{\snr} \geq {\inr}$,
which is always satisfied since we are in the weak interference regime.

\end{itemize}

In order to show that the term in~\eqref{eq:sumrate weak 2 third} is redundant, consider the following facts:
\begin{itemize}
\item
the bound 
$\frac{
 {\rm eq\eqref{eq:ach reg 2 not in a very insightful form 8a}}
+{\rm eq\eqref{eq:ach reg 2 not in a very insightful form 8l}}
}{2}$ is always bigger than 
${\rm eq\eqref{eq:ach reg 2 not in a very insightful form 8e}}$ and it is therefore redundant.

\item
the bound 
$\frac{
 {\rm eq\eqref{eq:ach reg 2 not in a very insightful form 8b}}
+{\rm eq\eqref{eq:ach reg 2 not in a very insightful form 8l}}
}{2}$ is equal to 
$\frac{
 {\rm eq\eqref{eq:ach reg 2 not in a very insightful form 8f}}
+{\rm eq\eqref{eq:ach reg 2 not in a very insightful form 8g}}
}{2}$ and it is therefore redundant.

\end{itemize}
Therefore we conclude that in the weak interference regime $1\leq{\inr}\leq{\snr}$ the sum-rate in~\eqref{eq:sumrate weak 2} is equal to~\eqref{eq:sumrate weak 2 second}  and, since ${\rm eq\eqref{eq:ach reg 2 not in a very insightful form 8f}}$ is equal to ${\rm eq\eqref{eq:ach reg 2 not in a very insightful form 8e}}$, is given by
\begin{subequations}
\begin{align}
\Rp + \Rc 
%  = \min \left \{\cite[\rm eq(8e)]{YANG-TUNINETTI},\cite[\rm eq(8g)]{YANG-TUNINETTI} \right \}  \nonumber
&\leq \min 
\left \{ \log \left( 1+\frac{{\snr}}{2{\inr}} \right) + \log \left( \frac{{\snr}+{\inr}+1}{2} \right),
\label{eq:after long computations scheme 2 sym 1}
 \right.\\& \left.
 \log \left( 1+\frac{{\snr}}{2{\inr}} \right) 
+\log \left( \frac{1+\Cc}{{\inr}+\Cc} \right)
+\log \left( \frac{{\snr}+{\inr}^2+{\inr}}{2} \right)
\label{eq:after long computations scheme 2 sym 2}
\right \}.
\end{align}
\label{eq:after long computations scheme 2 sym}
\end{subequations}
%Now comparing the last term of the first argument of the $\max \min$ with the last two terms of the second argument we get
For future use, the term in~\eqref{eq:after long computations scheme 2 sym 2} is the smallest term if
\begin{align*}
& 
({\snr}+{\inr}+1)({\inr}+\Cc) \geq {\snr}+{\inr}^2+{\inr}+{\snr} \Cc+\Cc {\inr}^2+\Cc{\inr}
%\\& SI + SC + I^2 + IC+ I + C \geq S+I^2+I+SC+CI^2+CI
%\\& SI + C \geq S+CI^2
%\\& S(I-1) \geq C(I-1)(I+1)
%\\& 
%\\& 
\Longleftrightarrow
{\snr} \geq \Cc({\inr}+1).
\end{align*}

\subsection{Constant gap result for the sum-capacity of the symmetric GCCIC in Regime 6 of Fig.~\ref{fig:fig3}}
\label{app:gap sym regime 6}
We analyze the regime $\Ip = \Ic = {\inr} < \Sp = \Sc = {\snr}$.

\textit{Parameter Range}: ${\snr}({\snr}+{\inr})> {\inr}^2({\inr}+1)$ and $\Cc \geq \frac{{\inr}^2}{{\snr}}$. In order to find the tightest upper bound we need to split this region in different subregions, namely: %{\red  MAKE SURE THE FIG HAS THE SAME NUMBERS AND LETTERS}
\begin{itemize}

\item Regime 6a) \label{item: weak 2 a}
$\mathsf{S < C \left( I+1 \right)}$: here the tightest gDoF upper bound gives
\[
{\gdof}(\alpha,\beta) \leq 1-\frac{\alpha}{2};
\]

\item Regime 6b) \label{item: weak 2 b}
$\mathsf{S \geq C \left( I+1 \right)}$ and $\mathsf{C \geq I}$: here the tightest gDoF upper bound gives
\[
{\gdof}(\alpha,\beta) \leq 1-\frac{\alpha}{2};
\]

\item Regime 6c) \label{item: weak 2 c} 
$\mathsf{S \geq C \left( I+1 \right)}$, $\mathsf{I^2 \leq S}$ and $\mathsf{C< I}$: here the tightest gDoF upper bound gives
\[
{\gdof}(\alpha,\beta) \leq 1-\alpha +\frac{\beta}{2};
\]

\item Regime 6d) \label{item: weak 2 d} 
$\mathsf{S \geq C \left( I+1 \right)}$, $\mathsf{I^2 > S}$, $\mathsf{C<I}$ and $\mathsf{S(S+I)>I^2(I+1)}$: here the tightest gDoF upper bound gives
\[
{\gdof}(\alpha,\beta) \leq \frac{1+\beta}{2}.
\]

\end{itemize}

\textit{Inner Bound}:
We use the achievable scheme in~\eqref{eq:after long computations scheme 2 sym} developed in Appendix~\ref{sec:scheme 2}.
%% to show that the following sum-rate is achievable with superposition coding with common and private messages
%%\begin{align*}
%%\left(\Rp+\Rc \right)^{\rm(IB)} 
%%  &\geq \min \left \{ \log \left( 1+\frac{{\snr}}{2{\inr}} \right) + \log \left( \frac{{\snr}+{\inr}+1}{2} \right), \right.
%%\\&            \left. \log \left( 1+\frac{{\snr}}{2{\inr}} \right) + \log \left( \frac{1+\Cc}{{\inr}+\Cc} \right)
%%                     +\log \left( \frac{{\snr}+{\inr}^2+{\inr}}{2} \right) \right \}.
%%\end{align*}
%%Note that, for $1\leq {\inr}$ the sum-rate $\log \left( 1+\frac{{\snr}}{2{\inr}} \right) + \log \left( \frac{{\snr}+{\inr}+1}{2} \right)$ is the tightest for ${\snr} < \Cc(1+{\inr})$.
%%This achievable rate 
which in the weak interference regime (i.e., $\alpha\leq 1$) implies that the following gDoF is achievable
%\eqref{eqdofnew} given at the top of the page holds.
%\begin{figure*}
\begin{align}
 {\gdof}(\alpha,\beta)  &\geq 
 %\lim_{{\snr}\to\infty}\frac{\left(\Rp+\Rc \right)^{\rm(IB)}}{2\log(1+{\snr})} \nonumber \\&= 
 \frac{1}{2}{\min\{
[1-\alpha]^+ +\max\{1,\alpha\},
[1-\alpha]^+ +\beta-\max\{\alpha,\beta\}+\max\{1,2\alpha\}
\}}
\nonumber
%\\& = \frac{\min\{
%[1-\alpha]^+ +\max\{1,\alpha\},
%[1-\alpha]^+ +\min\{\alpha,\beta\}+\max\{1-\alpha,\alpha\}
%\}}{2} \nonumber
\\&=
\left\{\begin{array}{ll}
1-\alpha/2          & \text{for} \ \beta \geq \min\{\alpha, 1-\alpha\} \\
1-\alpha+\beta/2    & \text{for} \ \beta<   \alpha, \ \alpha\in[0,1/2] \\
(1+\beta)/2         & \text{for} \ \beta< 1-\alpha, \ \alpha\in[1/2,1]. \\
%1-\alpha/2          & \text{for} \ 1<\alpha+\beta \\
%1-\alpha/2          & \text{for} \ 1\geq \alpha+\beta, \ \beta\geq \alpha\\
%1-\alpha+\beta/2 & \text{for} \ 1\geq \alpha+\beta, \ \beta< \alpha, \ \alpha< 1/2 \\
%(1+\beta)/2        & \text{for} \ 1\geq \alpha+\beta, \ \beta< \alpha, \ \alpha\in[1/2,1]. \\
\end{array}\right.
\label{eqdofnew}
\end{align}
%\end{figure*}
This shows the achievability of the gDoF upper bound in Regime~6 of Fig.~\ref{fig:fig3}.
Actually, the proposed scheme is gDoF optimal in the whole weak interference regime $\alpha \leq 1$ except for $\beta\leq \min\{1-\alpha,[2\alpha-1]^+\}$, where a noncooperative scheme is gDoF optimal.

\textit{Outer Bound}: 
For the regime $\beta \geq \min\{\alpha, 1-\alpha\}$, where ${\gdof}(\alpha,\beta)\leq 1-\alpha/2$ (regimes 6a and 6b), we use the upper bound in~\eqref{eq:Tuninetti};
%under the  weak interference condition $S \geq I$, that is
%\begin{align*}
%&\left(\Rp + \Rc \right)^{\rm(OB)}  \leq  \mathsf{\log  \left( \frac{1 + S}{1 + I} \right)} + \mathsf{\log \left( 1 +  S + I  + 2 \sqrt{SI}  \right)}
%\end{align*}
otherwise (regimes 6c and 6d) we use the upper bound in~\eqref{eq:PV}. 
%\begin{align*}
%&\left(\Rp+\Rc \right)^{\rm(OB)} 
%%  & \leq \log_2 \left( 1+ \left(\frac{\sqrt{S}}{\max \left(1,\sqrt{I} \right)}+\sqrt{I} \right)^2 \right) + \log_2 \left( 1+C \right)+
%%\\ &+\log_2 \left( 1+ \left(\frac{\sqrt{S}}{\max \left(1,\sqrt{I} \right)}+\frac{\sqrt{I}}{\max \left( 1,\sqrt{C} \right)} \right)^2 \right)
%\leq \log \left( 1+ \left(\mathsf{ \frac{\sqrt{S}}{\sqrt{I}}} + \sqrt{{\inr}}\right)^2 \right)+
%\\& +\log \left( 1+\Cc \right) +   \log \left( 1+ \left( \frac{\sqrt{{\snr}}}{\sqrt{{\inr}}} + \frac{\sqrt{{\inr}}}{\sqrt{\Cc}}\right)^2 \right).
%\end{align*}

\textit{Gap}: We analyze separately the different sub regimes:
\begin{itemize}

%\item[\underline{gap~6a)}] 
\item Regime 6a)\label{item: weak 2 a gap}
For the regime ${\snr} < \Cc(1+{\inr})$ within ${\inr} \leq {\snr}$
\begin{align*}
  & \mathsf{GAP} %=\left(\Rp+\Rc \right)^{\rm(OB)} - \left(\Rp+\Rc \right)^{\rm(IB)}
  \leq {\rm eq\eqref{eq:Tuninetti}}-{\rm eq\eqref{eq:after long computations scheme 2 sym 1}}
\\& \leq   \log \left( \frac{1+{\snr}}{1+{\inr}} \right)
+\log \left( 1+ (\sqrt{\snr}+\sqrt{\inr})^2  \right)+
%+\log \left( 1+ {\snr}+{\inr} +2 \sqrt{SI}  \right)+
%\\& \quad 
-\log \left( 1+\frac{{\snr}}{2{\inr}} \right) 
- \log \left( \frac{{\snr}+{\inr}+1}{2} \right)
%\\& \leq 2+\log_2 \left( \frac{1+S}{1+I} \right)+\log_2 \left( 1+ S+I +2 \sqrt{SI}  \right)-\log_2 \left( 1+\frac{S}{I} \right)- \log_2 \left( S+I+1 \right)
\\&\leq 2\log(2)+\max_{1\leq {\inr} \leq {\snr}}\log \left( \frac{1}{1+{\inr}} \ \frac{1+{\snr}}{1+\frac{{\snr}}{2{\inr}}}\right)
\\&=    2\log(2)+\max_{1\leq {\inr}            }\log \left( \frac{2{\inr}}{1+{\inr}}\right) = 3\log(2).
%+\log \left( \frac{1+{\snr}}{2\mathsf{I+S}} \cdot \frac{{\inr}}{1+{\inr}} \right)+
%\\& \quad +\log \left( \frac{1+ {\snr}+{\inr} +2 \sqrt{{\snr} {\inr}}}{{\snr}+{\inr}+1} \right)\leq  3 \ \rm{bits}
\end{align*}
%since $\frac{{\inr}}{1+{\inr}}\leq 1$, $\frac{1+{\snr}}{2\mathsf{I+S}}\leq \max\{1,\frac{1}{2I}\} = 1$ because ${\inr}>1$, and $2 \sqrt{SI}\leq 1+{\snr}+{\inr} \Longleftrightarrow 0\leq 1+(\sqrt{{\snr}}-\sqrt{{\inr}})^2$ and $\mathsf{I < S}$.

%\item[\underline{gap~6b)}] 
\item Regime 6b)\label{item: weak 2 b gap}
For the regime ${\snr} \geq \Cc({\inr}+1)$ and $\Cc\geq {\inr}$
\begin{align*}
&\mathsf{GAP} %=\left(\Rp+\Rc \right)^{\rm(OB)} - \left(\Rp+\Rc \right)^{\rm(IB)}
  \leq {\rm eq\eqref{eq:Tuninetti}}-{\rm eq\eqref{eq:after long computations scheme 2 sym 2}}
\\&\leq
    \log \left( \frac{1+{\snr}}{1+{\inr}} \right)
   +\log \left( 1+ (\sqrt{\snr}+\sqrt{\inr})^2  \right)+
\\&-\log \left( 1+\frac{{\snr}}{2{\inr}}   \right) 
   -\log \left( \frac{1+\Cc}{{\inr}+\Cc}  \right)
   -\log \left( \frac{{\snr}+{\inr}^2+{\inr}}{2}\right)
\\&\leq
    \log \left( \frac{1+{\snr}}{1+{\inr}} \right)
   +\log \left( \frac{1+ {\snr}+{\inr} }{ {\snr}+{\inr}^2+{\inr}} \right)+2\log(2)
%\\&
+\log \left( \frac{2{\inr}}{2{\inr}+{\snr}}   \right) 
   +\log \left( \frac{2{\inr}}{1+{\inr}}  \right)
%\\& = 2 + \log \left( \frac{1+{\snr}}{1+{\inr}} \right) + \log \left( 1 +  {\snr} + {\inr}  + 2 \sqrt{{\snr} {\inr}}  \right)+
%\\& - \log \left( 2\mathsf{I+S} \right) + \log \left( {\inr} \right) -\log \left( \frac{1+\Cc}{{\inr}+\Cc} \right)+
%\\&-\log \left( {\snr}+{\inr}^2+{\inr}\right)
%\\& \leq 2 + \log \left( \frac{1+{\snr}}{1+{\inr}} \right)+\log \left( 1+ {\snr}+{\inr} +2 \sqrt{{\snr} {\inr}}  \right)
%\\& - \log \left( 2{\inr}+{\snr} \right) +\log \left( {\inr} \right) -\log \left( \frac{1+{\inr}}{2{\inr}} \right)+
%\\&-\log \left( {\snr}+{\inr}^2+{\inr}\right)
%\\& = 3 + \log_2 \left( 1+S \right)- \log_2 \left( 1+I \right)+\log_2 \left( 1+ S+I +2 \sqrt{SI}  \right)- \log_2 \left( 2I+S \right) + \log_2 \left( I^2 \right) +
%\\& -\log_2 \left( 1+I\right)-\log_2 \left( S+I^2+I\right)
\\& = 4\log(2)
+ \log \left( \frac{1+{\snr}}{2{\inr}+{\snr}} \right)
+2\log \left( \frac{{\inr}}{1+{\inr}} \right) 
+ \log \left( \frac{1+ {\snr}+{\inr} }{ {\snr}+{\inr}^2+{\inr}} \right)
\leq 4\log(2)
\end{align*}
since $1+ {\snr}+{\inr} <{\snr}+{\inr}^2+{\inr}$, $1\leq {\inr}$, and where we upper bounded the gap by evaluating it for $\Cc={\inr}$, i.e., minimum possible value for $\Cc$, since the function is decreasing in $\Cc$.

%\item[\underline{gap~6c)}] 
\item Regime 6c)\label{item: weak 2 c gap}
For the regime ${\snr} \geq \Cc({\inr}+1)$, $\Cc<{\inr}$ and ${\inr}^2\leq {\snr}$
\begin{align*}
&\mathsf{GAP}%=\left(\Rp+\Rc \right)^{\rm(OB)} - \left(\Rp+\Rc \right)^{\rm(IB)}
  \leq {\rm eq\eqref{eq:PV}}-{\rm eq\eqref{eq:after long computations scheme 2 sym 2}}
\\&\leq \log \left( 1+ \left( \frac{\sqrt{{\snr}}}{\sqrt{{\inr}}} + \sqrt{{\inr}}\right)^2 \right)
       +\log \left( 1+\Cc \right) 
       +\log \left( 1+ \left( \frac{\sqrt{{\snr}}}{\sqrt{{\inr}}} +  \frac{\sqrt{{\inr}}}{\sqrt{\Cc}}\right)^2 \right)
\\&    -\log \left( 1+\frac{{\snr}}{2{\inr}} \right)
       -\log \left( \frac{1+\Cc}{{\inr}+\Cc} \right)
       -\log \left( \frac{{\snr}+{\inr}^2+{\inr}}{2} \right)
\\&\leq \log \left( 1+ \frac{{\snr}}{{\inr}} + {\inr} \right)
       +\log \left( 2{\inr} \right) 
       +\log \left( 2+\frac{{\snr}}{{\inr}} \right)
\\&    -\log \left( 1+\frac{{\snr}}{2{\inr}} \right)
       -\log \left( {\snr}+{\inr}^2+{\inr} \right) + 3\log(2) \leq 5\log(2),
%\\& = 2+\log \left({\inr}+{\snr}+{\inr}^2+2{\inr}\sqrt{{\snr}}  \right)-\log\left({\inr} \right)+
%\\&+  \log \left( 1 + \Cc \right)+\log \left( {\inr} \Cc + {\snr} \Cc + {\inr}^2+2{\inr}\sqrt{{\snr} \Cc} \right) +
%\\& - \log \left( {\inr} \Cc \right)-\log \left(2{\inr}+{\snr} \right)+\log\left({\inr} \right)+
%\\&-\log \left(1+\Cc \right)+ \log \left({\inr}+\Cc \right) - \log \left( {\snr}+{\inr}^2+{\inr} \right)
%\\& = 2+\log \left({\inr}+{\snr}+{\inr}^2+2{\inr}\sqrt{{\snr}}  \right)+
%\\&+\log \left( {\inr} \Cc+{\snr} \Cc+{\inr}^2+2{\inr}\sqrt{{\snr} \Cc} \right)+
%\\&+\log \left({\inr}+\Cc \right) - \log \left( {\inr} \Cc \right)+
%\\&-\log \left(2{\inr}+{\snr} \right)-\log \left( {\snr}+{\inr}^2+{\inr} \right)
%\\& \leq 3 + \log \left({\inr}+{\snr}+{\inr}^2+2{\inr}\sqrt{{\snr}}  \right)+
%\\& + \log \left( 2{\inr}^2+{\snr} {\inr}+2{\inr}\sqrt{{\snr} {\inr}}\right)+
%\\& -\log \left( {\inr} \right)-\log \left(2{\inr}+{\snr} \right)-\log \left( {\snr}+{\inr}^2+{\inr} \right)
%\\& = 3+\log \left( \frac{{\inr}+{\snr}+{\inr}^2+2{\inr}\sqrt{{\snr}}}{{\snr}+{\inr}^2+{\inr}}\right)+
%\\&+\log \left( \frac{2{\inr}^2+{\snr} {\inr}+2{\inr}\sqrt{{\snr} {\inr}}}{2{\inr}^2+{\snr} {\inr}} \right)
%\\& = 3+ \log \left(1+\frac{2{\inr}\sqrt{{\snr}}}{{\snr}+{\inr}^2+{\inr}} \right)+
%\\&+\log \left( 1+\frac{2{\inr}\sqrt{{\snr} {\inr}}}{2{\inr}^2+{\snr} {\inr}} \right)
%\\& \leq 3+ \log (2)+\log(2)=5 \ \rm{bits}
\end{align*}
%since $2{\inr}\sqrt{{\snr}} < {\snr}+{\inr}^2+{\inr}$ and $2{\inr} < 2{\inr}^2+{\snr} {\inr}$. Here 
where we upper bounded the gap by evaluating it for $\Cc={\inr}$, i.e., the maximum possible value for $\Cc$, since the function is increasing in $\Cc$.

%\item[\underline{gap~6d)}] 
\item Regime 6d)\label{item: weak 2 d gap}
For the regime ${\snr} \geq \Cc({\inr}+1)$, $\Cc<{\inr}$, ${\inr}^2> {\snr}$ and ${\snr}({\snr}+{\inr})\geq {\inr}^2({\inr}+1)$
\begin{align*}
&\mathsf{GAP} %=\left(\Rp+\Rc \right)^{\rm(OB)} - \left(\Rp+\Rc \right)^{\rm(IB)}
  \leq {\rm eq\eqref{eq:PV}}-{\rm eq\eqref{eq:after long computations scheme 2 sym 2}}
%\\&\leq \log \left( 1+ \left( \frac{\sqrt{{\snr}}}{\sqrt{{\inr}}} + \sqrt{{\inr}}\right)^2 \right)
%       +\log \left( 1+\Cc \right) 
%       +\log \left( 1+ \left( \frac{\sqrt{{\snr}}}{\sqrt{{\inr}}} +  \frac{\sqrt{{\inr}}}{\sqrt{\Cc}}\right)^2 \right)
%\\&    -\log \left( 1+\frac{{\snr}}{2{\inr}} \right) 
%       -\log \left( \frac{1+\Cc}{{\inr}+\Cc} \right)
%       -\log \left( \frac{{\snr}+{\inr}^2+{\inr}}{2} \right)
\leq5\log(2),
\end{align*}
by following exactly the same steps as done for Regime 6c) above.
\end{itemize}

%\end{enumerate}
This shows the achievability of the sum-capacity upper bound to within a constant gap of 2.5~bits (per user) in this regime.

\section{Achievable Schemes Based on Superposition Coding and DPC}
\label{sec:allachschemsappDPC}

\subsection{DPC-based Achievable Scheme}
\label{sec:scheme dpc}

\begin{figure}[!h]
\centering
\includegraphics[width=0.7\textwidth]{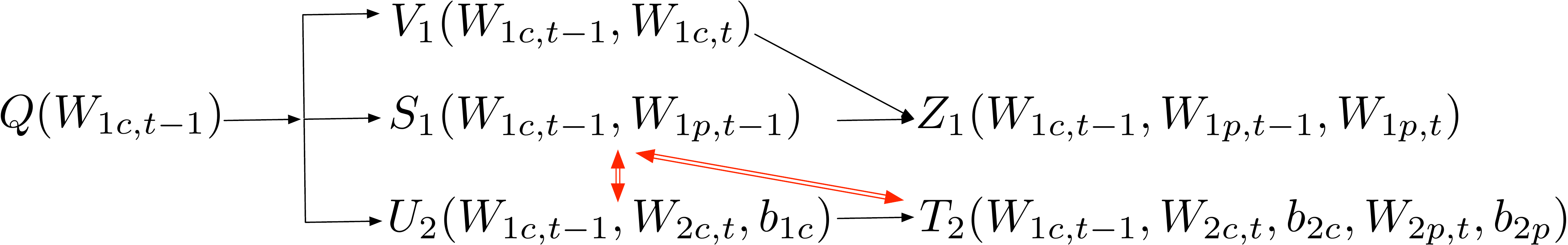}
\caption{Achievable scheme based on binning and superposition.}
\label{fig:achdpc}
\end{figure}

We specialize the `binning+superposition' achievable scheme in~\cite[Section V]{YANG-TUNINETTI}.
In~\cite[Thereom IV.1]{YANG-TUNINETTI}, the network comprises four nodes numbered from 1 to 4; nodes 1 and 2 are sources and nodes 3 and 4 destinations; source node $j\in[1:2]$, with input to the channel $X_j$ and output from the channel $Y_j$, has a message $W_j$ for node $j+2$; destination node $j\in[3:4]$ has channel output $Y_{j}$ from which it decodes message $W_{j-2}$. 

Both users do rate splitting, where the messages of user 1 are cooperative while the messages of user 2 are noncooperative.
In~\cite[Section V]{YANG-TUNINETTI}, we set $Y_1=U_1=T_1=S_2=V_2=Z_2=\emptyset$, 
i.e., then $R_1=R_{11c}+R_{10c},  \ R_2=R_{22n}+R_{20n}$, to obtain a scheme that comprises:
a cooperative common message  (carried by the pair $(Q  ,V_1)$ at rate $R_{10c}$) for user 1,
a cooperative private message (carried by the pair $(S_1,Z_1)$ at rate $R_{11c}$) for user 1,
a noncooperative common message  (carried by $U_2$ at rate $R_{20n}$) for user 2 and
a noncooperative private message (carried by $T_2$ at rate $R_{22n}$) for user 2.
Here the pair $(Q,S_1)$ carries the `past cooperative messages', and the pair $(V_1,Z_1)$ the `new cooperative messages' in a block Markov encoding scheme.  
The channel inputs are functions of the auxiliary random variables, where $X_1$ is a function of $(Q,S_1,V_1,Z_1)$ and $X_2$ a function of $(Q,S_1,U_2,T_2)$.

The set of possible input distributions is
\begin{align}
 &P_{Q,S_1, V_1,Z_1,X_1, U_2,T_2,X_2}
= P_{Q} P_{V_1|Q}
  P_{S_1|Q}P_{Z_1|Q,S_1,V_1} % was P_{S_1,Z_1|Q}  
  P_{U_2,T_2|S_1,Q}
  P_{X_1|Q,S_1,V_1,Z_1}
  P_{X_2|Q,S_1,U_2,T_2}.
\label{eq:input pdf dpc}
\end{align}

A schematic representation of the achievable scheme is given in Fig.~\ref{fig:achdpc}, where an black arrow indicates superposition coding and a red arrow indicates binning.

Regarding encoding. The codebooks are generated as follows:
first the codebook $Q$ is generated; 
then the codebook $V_1$ is superposed to $Q$; 
independently of $V_1$,           the codebook $S_1$ is superposed to $Q$ and then the codebook $Z_1$ is superposed to $(Q,S_1,V_1)$;
independently of $(V_1,S_1,Z_1)$, the codebook $U_2$ is superposed to $Q$ and then the codebook $T_2$ is superposed to $(Q,U_2)$.
With this random coding codebook generation, the pair $(U_2,T_2)$ is independent of $S_1$ conditioned on $Q$.
\cite[Theorem V.1]{YANG-TUNINETTI} involves several binning steps to allow for a large set of input distributions. Here, in order to simplify the scheme, we do not bin $V_1$ against $S_1$; the only binning steps are for $(U_2,T_2)$ against $S_1$.
\begin{subequations}
We use a block Markov coding scheme to convey the message of user~1 to user~2.
In particular, at the end of any given time slot in a block Markov coding scheme, encoder~2 knows $(Q,S_1,U_2,T_2)$ and decodes $(V_1,Z_1)$ from its channel output; the decoded pair $(V_1,Z_1)$ becomes the pair $(Q,S_1)$ of the next time slot; then, at the beginning of each time slot, encoder~2, by binning, finds the new pair $(U_2,T_2)$ that is jointly typical with $(Q,S_1)$; for this to be possible, we must generate several $(U_2,T_2)$ sequences for each message of user 2 so as to be able to find one pair to send with the correct joint distribution with $(Q,S_1)$; 
this entails the rate penalties in \cite[eq(20)]{YANG-TUNINETTI} for user~1 and then again \cite[eq(20)]{YANG-TUNINETTI} for user~2 by swapping the role of the subscripts~1 and~2, with $S_2=Z_2=V_2=U_1=T_1=\emptyset$ and with $V_1$ independent of $S_1$, i.e.,
\begin{align}
   R_{20n}^{'} + R_{22n}^{'} &\geq I(S_1;U_2,T_2|Q).\label{eq:const for dpc:node2 u2 t2}
\\ R_{20n}^{'}               &\geq I(U_2;S_1|Q)     \label{eq:const for dpc:node2 u2}   
\end{align}

Regarding decoding.
There are three decoding nodes in the network and therefore three groups of rate constraints.
These are:
\begin{itemize}
\item

Node~2/CTx jointly decodes $(V_1,Z_1)$ from its channel output with knowledge of the indices in $(Q,S_1,U_2,T_2,X_2)$. 
Successful decoding is possible if (i.e., use \cite[eq(21)]{YANG-TUNINETTI} by swapping the role of the subscripts~1 and~2, with $S_2=Z_2=V_2=U_1=T_1=\emptyset$ and with $V_1$ independent of $S_1$)
\begin{align}
   R_{10c} + R_{11c} &\leq I(Y_2;Z_1,V_1|U_2,T_2,X_2,S_1,Q) \label{like eq:suponly 6a new sum}    %\label{eq:const for dpc:node2 z1 v1}
\\           R_{11c} &\leq I(Y_2;Z_1|U_2,T_2,X_2,S_1,Q,V_1).\label{like eq:suponly 6a new private}%\label{eq:const for dpc:node2 z1}
\end{align}

\item
Node~3/PRx jointly decodes $(Q,S_1,U_2)$ from its channel output, with knowledge of some message indices in $(V_1,Z_1)$, by treating $T_2$ as noise. Successful decoding is possible if (see \cite[eq(22)]{YANG-TUNINETTI} where only the bounds in \cite[eq(22a)]{YANG-TUNINETTI}, \cite[eq(22f)]{YANG-TUNINETTI}, and \cite[eq(22g)]{YANG-TUNINETTI} remain after setting several auxiliary random variables to zero and removing the redundant constraints)
\begin{align}
   R_{10c} + R_{20n} + R_{11c} &\leq I(Y_3;Q,V_1,S_1,Z_1,U_2)-(R_{20n}^{'}-I(U_2;S_1|Q))      \label{like eq:suponly 6f}%\label{eq:const for dpc:node3 u2 s1 q}
\\           R_{20n} + R_{11c} &\leq I(Y_3;      S_1,Z_1,U_2|V_1,Q)-(R_{20n}^{'}-I(U_2;S_1|Q))\label{like eq:suponly 6c}%\label{eq:const for dpc:node3 u2 s1}
\\                     R_{11c} &\leq I(Y_3;      S_1,Z_1|V_1,Q,U_2).                          \label{like eq:suponly 6b}%\label{eq:const for dpc:node3 s1}
\end{align}

\item
Node~4/CRx jointly decodes $(Q,U_2,T_2)$ from its channel output, with knowledge of some message index in $V_1$, by treating $Z_1$ as noise (recall that the pair $(U_2,T_2)$ has been precoded/binned against $S_1$).  Successful decoding is possible if (see \cite[eq(22)]{YANG-TUNINETTI}, with the role of the users swapped, where only the bounds in \cite[eq(22a)]{YANG-TUNINETTI}, \cite[eq(22i)]{YANG-TUNINETTI}, and \cite[eq(22k)]{YANG-TUNINETTI} remain after setting several auxiliary random variables to zero and removing the redundant constraints)
\begin{align}
   R_{10c}+R_{20n}+ R_{22n} &\leq I(Y_4;U_2,T_2,V_1,Q)-(R_{20n}^{'} + R_{22n}^{'}) \label{like eq:suponly 7f}%\label{eq:const for dpc:node4 u2 t2 q}
\\         R_{20n}+ R_{22n} &\leq I(Y_4;U_2,T_2|V_1,Q)-(R_{20n}^{'} + R_{22n}^{'}) \label{like eq:suponly 7d}%\label{eq:const for dpc:node4 u2 t2}
\\                  R_{22n} &\leq I(Y_4;T_2|V_1,Q,U_2)-R_{22n}^{'}.                \label{like eq:suponly 7b}%\label{eq:const for dpc:node4 t2}
\end{align}

\end{itemize}
\label{eq:DPC unified}
\end{subequations}

From Remark~2 in Appendix~\ref{sec:scheme sup}, after Fourier-Motzkin elimination of the achievable region in~\eqref{eq:DPC unified} where we take the the constraints in~\eqref{eq:const for dpc:node2 u2 t2} and~\eqref{eq:const for dpc:node2 u2} to hold with equality (i.e., $R_{20n}^{'} = I(U_2;S_1|Q), \ R_{22n}^{'} = I(S_1;T_2|Q,U_2)$),
we get
%THERE CAN BE A PROBLEM SINCE J IS NOT THE SMALLEST IN PRINCIPLE; not the case with the costa choice we make in awgn
\begin{subequations}
\begin{align}
  R_1     &\leq {\rm eq\eqref{like eq:suponly 6f}}         \label{like eq:ach reg 2 not in a very insightful form 8a}
\\R_1     &\leq {\rm eq\eqref{like eq:suponly 6a new sum}} \label{like eq:ach reg 2 not in a very insightful form 8b}
%COMMON          {\rm eq\eqref{like eq:suponly 6a}} 
%ONLY          + {\rm eq\eqref{like eq:suponly 6b}} 
\\R_2     &\leq {\rm eq\eqref{like eq:suponly 7d}} \label{like eq:ach reg 2 not in a very insightful form 8d}
\\R_1+R_2 &\leq {\rm eq\eqref{like eq:suponly 6f}} 
              + {\rm eq\eqref{like eq:suponly 7b}} \label{like eq:ach reg 2 not in a very insightful form 8e}
\\R_1+R_2 &\leq {\rm eq\eqref{like eq:suponly 7f}} 
              + {\rm eq\eqref{like eq:suponly 6b}} \label{like eq:ach reg 2 not in a very insightful form 8f}
\\R_1+R_2 &\leq {\rm eq\eqref{like eq:suponly 7f}}
               +{\rm eq\eqref{like eq:suponly 6a new private}} \label{like eq:ach reg 2 not in a very insightful form 8g}
%COMMON           {\rm eq\eqref{like eq:suponly 6a}} 
%ONLY           + {\rm eq\eqref{like eq:suponly 6c}} 
%               + {\rm eq\eqref{like eq:suponly 7b}} 
\\R_1+2R_2&\leq {\rm eq\eqref{like eq:suponly 6c}} 
              + {\rm eq\eqref{like eq:suponly 7b}} 
              + {\rm eq\eqref{like eq:suponly 7f}} \label{like eq:ach reg 2 not in a very insightful form 8l}
\end{align}
\label{eq:ach dpc FME}
\end{subequations}
for all distributions that factor as~\eqref{eq:input pdf dpc}.
%full region
%\begin{align*}
%  \Rp     &\leq I(\Yf;Z_1,V_1|U_2,T_2,\Xc,S_1,Q) 
%\\\Rp     &\leq I(\Yp;Q,V_1,S_1,Z_1,U_2)         
%\\\Rc     &\leq I(\Yc;U_2,T_2|V_1,Q)-I(S_1;U_2,T_2|Q)
%\\\Rp+\Rc &\leq I(\Yc;T_2|V_1,Q,U_2)-I(S_1;T_2|Q,U_2)+I(\Yp;Q,V_1,S_1,Z_1,U_2) 
%\\\Rp+\Rc &\leq I(\Yc;U_2,T_2,V_1,Q)-I(S_1;U_2,T_2|Q)+I(\Yp;      S_1,Z_1|V_1,Q,U_2)
%\\\Rp+\Rc &\leq I(\Yc;U_2,T_2,V_1,Q)-I(S_1;U_2,T_2|Q)+ I(\Yf;Z_1|U_2,T_2,\Xc,S_1,Q,V_1)
%\\\Rp+2\Rc&\leq I(\Yc;T_2|V_1,Q,U_2)-I(S_1;T_2|Q,U_2)+I(\Yc;U_2,T_2,V_1,Q)-I(S_1;U_2,T_2|Q)+I(\Yp;      S_1,Z_1,U_2|V_1,Q) 
%\end{align*}

\bigskip
Remark~3.
As done in Remark~1 in Appendix~\ref{sec:scheme sup}, 
the rate bound in~\eqref{like eq:ach reg 2 not in a very insightful form 8l} is redundant if
\[
\min\{{\rm eq\eqref{like eq:ach reg 2 not in a very insightful form 8e}},
      {\rm eq\eqref{like eq:ach reg 2 not in a very insightful form 8f}}  \}
      %{\rm eq\eqref{like eq:ach reg 2 not in a very insightful form 8g}}  \}
 + {\rm eq\eqref{like eq:ach reg 2 not in a very insightful form 8d}}
\leq 
 {\rm eq\eqref{like eq:ach reg 2 not in a very insightful form 8l}} 
\]
that is, if for the considered input distribution we have
\begin{subequations}
\begin{align}
&\text{either} \
 {\rm eq\eqref{like eq:suponly 6f}} + {\rm eq\eqref{like eq:suponly 7d}}% + {\rm eq\eqref{like eq:suponly 7b}}
\leq
 {\rm eq\eqref{like eq:suponly 6c}} + {\rm eq\eqref{like eq:suponly 7f}}% + {\rm eq\eqref{like eq:suponly 7b}} 
\Longleftrightarrow
I(Y_3; Q,V_1)\leq I(Y_4; Q,V_1),
%\text{i.e., Rx2 better decodes the common message of Tx1},
\\
&\text{or} \
 {\rm eq\eqref{like eq:suponly 6b}} + {\rm eq\eqref{like eq:suponly 7d}}%+ {\rm eq\eqref{like eq:suponly 7f}} 
\leq
 {\rm eq\eqref{like eq:suponly 6c}} + {\rm eq\eqref{like eq:suponly 7b}}%+ {\rm eq\eqref{like eq:suponly 7f}}
\Longleftrightarrow
I(Y_4; U_2|Q,V_1)-I(U_2; S_1|Q)\leq I(Y_3; U_2|Q,V_1).
%\text{i.e., Rx1 better decodes the common message of Tx2},
%%%\\
%%%&\text{or} \
%%% {\rm eq\eqref{like eq:suponly 6a new private}} + {\rm eq\eqref{like eq:suponly 7d}}%+ {\rm eq\eqref{like eq:suponly 6c}}  + {\rm eq\eqref{like eq:suponly 7b}}
%%%\leq 
%%% {\rm eq\eqref{like eq:suponly 6c}} + {\rm eq\eqref{like eq:suponly 7b}} %{\rm eq\eqref{like eq:suponly 7f}}
%%%\nonumber\\& \Longleftrightarrow
%%%I(Y_2;Z_1|U_2,T_2,X_2,S_1,Q,V_1) + I(Y_4; U_2|Q,V_1)-I(U_2; S_1|Q) \leq I(Y_3;      S_1,Z_1,U_2|V_1,Q).
\end{align}
\label{like eq:ach reg 2 not in a very insightful form:when is r1+2r2 redundant}
\end{subequations}

\subsection{DPC region for the Gaussian noise channel}\label{subsect:monster dpc region in awgn}
We identify Node1 with the PTx (i.e., $\Xp=X_1$), Node2 with the CTx  (i.e., $\Xc=X_2, \Yf=Y_2$), Node3 with the PRx  (i.e., $\Yp=Y_3$) and Node4 with the CRx (i.e., $\Yc=Y_4$). 
For the Gaussian noise channel, in the achievable region in~\eqref{eq:ach dpc FME}, we choose $Q=\emptyset$, we let $S_1,V_1,Z_1,U_2,T_2^{\prime}$ to be i.i.d. $\mathcal{N}(0,1)$, and
\begin{align*}
  &\Xp = |a_1|{\eac} S_1 + b_1 V_1 + c_1 Z_1          &&: |a_1|^2+|b_1|^2+|c_1|^2= 1,
\\&\Xc = |a_2|       S_1 + b_2 U_2 + c_2 T_2^{\prime} &&: |a_2|^2+|b_2|^2+|c_2|^2= 1,
\\&T_2 = T_2^{\prime}+\lambda S_1 &&: \lambda = \frac{\Sc|c_2|^2}{\Sc|c_2|^2+1+\Ip|c_1|^2} \ \frac{\sqrt{\Ip}{\eap}{\eac} |a_1|+\sqrt{\Sc}|a_2|}{\sqrt{\Sc} c_2},
\end{align*}
where the choice of $\lambda$ is so as to ``pre-cancel'' $S_1$ from $\Yc$ in decoding $T_2$,
i.e., so as to have $I(\Yc;T_2|V_1,Q,U_2)-I(S_1;T_2|Q,U_2)=I(\Yc;T_2|V_1,Q,U_2,S_1)$.
With these choices, the channel outputs are
\begin{align*}
  & \Yf = \sqrt{\Cc}  \left(|a_1|{\eac} S_1 + b_1 V_1          + c_1 Z_1 \right)+\Zf,
\\& \Yp = (\sqrt{\Sp} |a_1| +\sqrt{\Ic}|a_2|) {\eac}S_1
  +\sqrt{\Sp}      \left( b_1 V_1          + c_1 Z_1 \right)+\sqrt{\Ic}{\eac}\left( b_2 U_2 + c_2 T_2^{\prime} \right) +\Zp,
\\& \Yc = (\sqrt{\Ip}{\eap}{\eac} |a_1|+\sqrt{\Sc}|a_2|)S_1
  +\sqrt{\Ip}{\eap}\left( b_1 V_1          + c_1 Z_1 \right)+\sqrt{\Sc}      \left( b_2 U_2 + c_2 T_2^{\prime} \right) +\Zc,
\end{align*}
and the achievable region in~\eqref{eq:ach dpc FME} (notice that we have $I(S_1;U_2|Q)=0$ since $U_2$ is not precoded against $S_1$) becomes
\begin{align*}
  \Rp     &\leq I(\Yf;Z_1,V_1|U_2,T_2,\Xc,S_1,Q)
   \\&=\log\left( 1+\Cc(|b_1|^2+|c_1|^2) \right),
\\\Rp     &\leq I(\Yp;Q,V_1,S_1,Z_1,U_2)  
   \\&=\log\left( \frac{1+\Sp+\Ic+2\sqrt{\Sp \Ic |a_1|^2|a_2|^2}}{1+\Ic|c_2|^2}  \right),       
\\\Rc     &\leq I(\Yc;U_2,T_2|V_1,Q)-I(S_1;T_2|Q,U_2)
   \\&= I(\Yc;U_2|V_1,Q)+I(\Yc;T_2|V_1,Q,U_2,S_1)
   \\&=\log\left( 1+\frac{\Sc|b_2|^2}{1+\Ip|c_1|^2+\Sc|c_2|^2+|\sqrt{\Ip}{\eap}{\eac} |a_1|+\sqrt{\Sc}|a_2||^2} \right)
              \\& \quad+ \log\left( 1+\frac{\Sc|c_2|^2}{1+\Ip|c_1|^2} \right),
\\\Rp+\Rc &\leq I(\Yc;T_2|V_1,Q,U_2)-I(S_1;T_2|Q,U_2)+I(\Yp;Q,V_1,S_1,Z_1,U_2) 
       \\&=\log\left( \frac{1+\Sp+\Ic+2\sqrt{\Sp \Ic |a_1|^2|a_2|^2}}{1+\Ic|c_2|^2}  \right)
              + \log\left( 1+\frac{\Sc|c_2|^2}{1+\Ip|c_1|^2} \right),
   %\\ & \stackrel{!!!}{=}I(\Yc;T_2|V_1,Q,U_2,S_1)+I(\Yp;Q,V_1,S_1,Z_1,U_2) 
\end{align*}
and
\begin{align*}
\Rp+\Rc &\leq I(\Yc;U_2,T_2,V_1,Q)-I(S_1;T_2|Q,U_2)+I(\Yp;      S_1,Z_1|V_1,Q,U_2)
       \\&=\log\left( 1+\frac{\Sc|b_2|^2+\Ip|b_1|^2}{1+\Ip|c_1|^2+\Sc|c_2|^2+|\sqrt{\Ip}{\eap}{\eac} |a_1|+\sqrt{\Sc}|a_2||^2} \right)
              \\&+ \log\left( 1+\frac{\Sc|c_2|^2}{1+\Ip|c_1|^2} \right)
         + \log\left( 1+\frac{|\sqrt{\Sp} |a_1| +\sqrt{\Ic}|a_2||^2+\Sp|c_1|^2}{1+\Ic|c_2|^2} \right)
   %\\ & \stackrel{!!!}{=} I(\Yc;U_2,V_1,Q)+I(\Yc;T_2|V_1,Q,U_2,S_1)+I(\Yp;      S_1,Z_1|V_1,Q,U_2)
\\\Rp+\Rc &\leq I(\Yc;U_2,T_2,V_1,Q)-I(S_1;T_2|Q,U_2)+ I(\Yf;Z_1|U_2,T_2,\Xc,S_1,Q,V_1)
       \\&=\log\left( 1+\frac{\Sc|b_2|^2+\Ip|b_1|^2}{1+\Ip|c_1|^2+\Sc|c_2|^2+|\sqrt{\Ip}{\eap}{\eac} |a_1|+\sqrt{\Sc}|a_2||^2} \right)
              \\&+ \log\left( 1+\frac{\Sc|c_2|^2}{1+\Ip|c_1|^2} \right)
           + \log\left( 1+\Cc |c_1|^2 \right)
   %\\ & \stackrel{!!!}{=} I(\Yc;U_2,V_1,Q)+I(\Yc;T_2|V_1,Q,U_2,S_1)+I(\Yf;Z_1|U_2,T_2,\Xc,S_1,Q,V_1)
\\
\Rp+2\Rc&\leq I(\Yc;T_2|V_1,Q,U_2)-I(S_1;T_2|Q,U_2)+I(\Yc;U_2,T_2,V_1,Q)-I(S_1;T_2|Q,U_2)
\\&\quad+I(\Yp;      S_1,Z_1,U_2|V_1,Q) 
       \\&=2\log\left( 1+\frac{\Sc|c_2|^2}{1+\Ip|c_1|^2} \right)
          \\&+\log\left( 1+\frac{\Sc|b_2|^2+\Ip|b_1|^2}{1+\Ip|c_1|^2+\Sc|c_2|^2+|\sqrt{\Ip}{\eap}{\eac} |a_1|+\sqrt{\Sc}|a_2||^2} \right)
       \\&+\log\left( 1+\frac{|\sqrt{\Sp} |a_1| +\sqrt{\Ic}|a_2||^2+\Sp|c_1|^2+\Ip|b_2|^2}{1+\Ic|c_2|^2} \right)
   %\\ &\stackrel{!!!}{=} 2I(\Yc;T_2|V_1,Q,U_2,S_1)+I(\Yc;U_2,V_1,Q)+I(\Yp;      S_1,Z_1,U_2|V_1,Q)
\end{align*}
%GENERAL REGION
%\begin{subequations}
%\begin{align*}
%  \Rp     &\leq \log\left( 1+\Cc(|b_1|^2+|c_1|^2) \right)
%\\\Rp     &\leq \log\left( \frac{1+\Sp+\Ic+2\sqrt{\Sp \Ic |a_1|^2|a_2|^2}}{1+\Ic|c_2|^2}  \right)
%\\\Rc     &\leq \log\left( 1+\frac{\Sc|b_2|^2}{1+\Ip|c_1|^2+\Sc|c_2|^2+|\sqrt{\Ip}{\eap}{\eac} |a_1|+\sqrt{\Sc}|a_2||^2} \right)
%              + \log\left( 1+\frac{\Sc|c_2|^2}{1+\Ip|c_1|^2} \right)
%\\\Rp+\Rc &\leq \log\left( \frac{1+\Sp+\Ic+2\sqrt{\Sp \Ic |a_1|^2|a_2|^2}}{1+\Ic|c_2|^2}  \right)
%              + \log\left( 1+\frac{\Sc|c_2|^2}{1+\Ip|c_1|^2} \right)
%\\\Rp+\Rc &\leq \log\left( 1+\frac{\Sc|b_2|^2+\Ip|b_1|^2}{1+\Ip|c_1|^2+\Sc|c_2|^2+|\sqrt{\Ip}{\eap}{\eac} |a_1|+\sqrt{\Sc}|a_2||^2} \right)
%              + \log\left( 1+\frac{\Sc|c_2|^2}{1+\Ip|c_1|^2} \right)
%         \\&  + \log\left( 1+\frac{|\sqrt{\Sp} |a_1| +\sqrt{\Ic}|a_2||^2+\Sp|c_1|^2}{1+\Ic|c_2|^2} \right)
%\\\Rp+\Rc &\leq\log\left( 1+\frac{\Sc|b_2|^2+\Ip|b_1|^2}{1+\Ip|c_1|^2+\Sc|c_2|^2+|\sqrt{\Ip}{\eap}{\eac} |a_1|+\sqrt{\Sc}|a_2||^2} \right)
%              + \log\left( 1+\frac{\Sc|c_2|^2}{1+\Ip|c_1|^2} \right)
%         \\&  + \log\left( 1+\Cc |c_1|^2 \right)
%\\\Rp+2\Rc& \ \text{to be made redundant}
%\end{align*}
%\end{subequations}

%Table of conversions
%\[
%\begin{array}{|l|| l l l ||l l l |}
%\hline
%  & S_1 & V_1 & Z_1   & S_1 & U_2 & T_2^\prime \\ 
%  & a_1 & b_1 & c_1   & a_2 & b_2 & c_2        \\
%\hline
%3 & \gp & 0 & \sqrt{1-|\gp|^2} & \gc & 0 & \sqrt{1-|\gc|^2} \\  
%4 & 0 & \sqrt{1-|\gp|^2} & \gp & \gc & \sqrt{1-|\gc|^2} & 0 \\  
%5 & a & c & b & 0 & 0 & 1 \\     
%\hline
%\end{array}
%\]

Remark~4. 
Motivated by the observation in~\cite{etw} that all terms that appears as noise should be at most at the level of the noise, we set 
\begin{align*}
& |a_1|=0, \ |b_1|^2 = \frac{\Ip}{1+\Ip}, \ |c_1|^2 = \frac{1}{1+\Ip}, 
\\ 
& |a_2|^2= \frac{\Ic}{1+\Ic}\frac{1}{1+\Sc}, \ |b_2|^2=\frac{\Ic}{1+\Ic}\frac{\Sc}{1+\Sc}, \ |c_2|^2=\frac{1}{1+\Ic},
\end{align*}
so that the achievable region derived in this section is included into
\begin{subequations}
\begin{align}
  \Rp     &\leq \log\left( 1+\Cc \right)
\\\Rp     &\leq \log\left( {1+\Sp+\Ic}  \right)-\log\left(2\right)
\\\Rc     &\leq \log\left( {1+\Sc}\right)
              - 2\log\left(2\right)
\\\Rp+\Rc &\leq \log\left( {1+\Sp+\Ic} \right)
              + \log\left( {1+\frac{\Sc}{1+\Ic}} \right)
              - 2\log\left(2\right)
\\\Rp+\Rc &\leq \log\left( {1+\Ip+\Sc}\right)
              + \log\left( {1+\frac{\Ic}{1+\Ic}\frac{\Ic}{1+\Sc}+\frac{\Sp}{1+\Ip}} \right)
              - 3\log\left(2\right) \label{eq:dt dpc surprisingly simple: sumrate tough to match}
\\\Rp+\Rc &\leq \log\left( {1+\Ip+\Sc}\right)
              + \log\left( 1+\frac{\Cc}{1+\Ip} \right)
              - 2\log\left(2\right)
%\\\Rp+2\Rc& \ \text{to be made redundant}
\end{align}
for either
\begin{align}
&I(\Yp; V_1)\leq I(\Yc; V_1)%\nonumber\\& 
\Longleftrightarrow
\frac{\Sp|b_1|^2}{1+\Sp|c_1|^2+\Ic}\leq \frac{\Ip|b_1|^2}{1+\Ip|c_1|^2+\Sc}
\nonumber\\&\Longleftrightarrow \Sp(1+\Sc) \leq \Ip(1+\Ic),
\end{align}
or
\begin{align}
&I(\Yc; U_2|V_1)\leq I(\Yp; U_2|V_1)
\Longleftrightarrow
\frac{\Sc|b_2|^2}{1+\Sc(1-|b_2|^2)+\Ip|c_1|^2}\leq \frac{\Ic|b_2|^2}{1+\Ic(1-|b_2|^2)+\Sp|c_1|^2}
\nonumber\\&\Longleftrightarrow %b_2=0, \ or \ 
\Sc\frac{1+\Ip+\Sp}{1+2\Ip} \leq \Ic,
\end{align}
\label{eq:dt dpc surprisingly simple}
\end{subequations}
so that the bound on $\Rp+2\Rc$ is redundant (see conditions in~\eqref{like eq:ach reg 2 not in a very insightful form:when is r1+2r2 redundant}).
%In the regime $\Sp < \Cc < \Ip, \ \Ic < \Sc$ (see Fig.~\ref{fig:gen case plot of where we have a gap} on the right, for which the superposition-based scheme in Section~\ref{subsec:cap regime Sp<Cc and Ip<C and Sc<Ic} did not give a gap)
In the regime $\Cc > \max\{\Sp,\Ip\}$ (see Fig.~\ref{fig:gen case plot of where we have a gap} on the right) the gap would be 2~bits if one could neglect the sum-rate bound in~\eqref{eq:dt dpc surprisingly simple: sumrate tough to match}.

%%%{\red
%%%GENERAL REGION
%%%\[
%%%|a_1|^2 = \frac{\Ip}{1+\Ip} ?, 
%%%|b_1|^2 = \frac{\Ip}{1+\Ip} ?, 
%%%|c_1|^2 = \frac{1}{1+\Ip}, 
%%%|a_2|^2 = 0
%%%|b_2|^2 = \frac{\Ic}{1+\Ic}, 
%%%|c_2|^2 = \frac{1}{1+\Ic}, 
%%%\]
%%%\begin{subequations}
%%%\begin{align*}
%%%  \Rp     &\leq \log\left( 1+\Cc(|b_1|^2+\frac{1}{1+\Ip}) \right)
%%%\\\Rp     &\leq \log\left( \frac{1+\Sp+\Ic}{1+\frac{\Ic}{1+\Ic}}  \right)
%%%\\\Rc     &\leq \log\left( 1+\frac{\Sc\frac{\Ic}{1+\Ic}}{1+\frac{\Ip}{1+\Ip}+\Sc\frac{1}{1+\Ic}+\Ip |a_1|^2} \right)
%%%              + \log\left( 1+\frac{\Sc\frac{1}{1+\Ic}}{1+\frac{\Ip}{1+\Ip}} \right)
%%%\\\Rp+\Rc &\leq \log\left( \frac{1+\Sp+\Ic}{1+\frac{\Ic}{1+\Ic}}  \right)
%%%              + \log\left( 1+\frac{\Sc\frac{1}{1+\Ic}}{1+\frac{\Ip}{1+\Ip}} \right)
%%%\\\Rp+\Rc &\leq \log\left( 1+\frac{\Sc\frac{\Ic}{1+\Ic}+\Ip|b_1|^2}{1+\frac{\Ip}{1+\Ip}+\Sc\frac{1}{1+\Ic}+\Ip |a_1|^2} \right)
%%%              + \log\left( 1+\frac{\Sc\frac{1}{1+\Ic}}{1+\frac{\Ip}{1+\Ip}} \right)
%%%         \\&  + \log\left( 1+\frac{\Sp |a_1|^2+\Sp\frac{1}{1+\Ip}}{1+\frac{\Ic}{1+\Ic}} \right)
%%%\\\Rp+\Rc &\leq \log\left( 1+\frac{\Sc\frac{\Ic}{1+\Ic}+\Ip|b_1|^2}{1+\frac{\Ip}{1+\Ip}+\Sc\frac{1}{1+\Ic}+\Ip |a_1|^2} \right)
%%%              + \log\left( 1+\frac{\Sc\frac{1}{1+\Ic}}{1+\frac{\Ip}{1+\Ip}} \right)
%%%         \\&  + \log\left( 1+\Cc \frac{1}{1+\Ip} \right)
%%%\\\Rp+2\Rc& \ \text{to be made redundant}
%%%\end{align*}
%%%\end{subequations}
%%%}

\subsection{Achievable Scheme 3: both messages are private}
\label{sec:scheme 3}

From the general region in Section~\ref{subsect:monster dpc region in awgn}, we set 
\begin{align*}
  &a_1=\gp, \ b_1=0, \ c_1=\sqrt{1-|\gp|^2}, \ |\gp|\leq 1, \quad
\\&a_2=\gc, \ b_2=0, \ c_2=\sqrt{1-|\gc|^2}, \ |\gc|\leq 1, 
\end{align*}
to obtain
\begin{subequations}
\begin{align}
   \Rp &\leq \log \left( 1+\Cc(1-|\gp|^2) \right)
\\ \Rp &\leq \log \left( \frac{1+\Sp +\Ic + 2|\gc||\gp| \sqrt{\Sp \Ic}}{1+ (1-|\gc|^2)\Ic} \right)
\\ \Rc &\leq \log \left( 1 +  \frac{(1-|\gc|^2)\Sc}{1+(1-|\gp|^2)\Ip} \right)
\end{align}
\label{eq:the simple DPC region}
\end{subequations}
for all $(|\gc |,|\gp |)\in[0,1]^2$.

From~\eqref{eq:the simple DPC region} the following sum-rate is achievable
\begin{align*}
\Rp + \Rc  
&\leq  \max_{(|\gc|,|\gp|)\in[0,1]^2} 
  \log \left( 1 +  \frac{(1-|\gc|^2)\Sc}{1+(1-|\gp|^2)\Ip} \right) +
\\+\min 
  &\left\{ 
  \log \left( 1+\Cc(1-|\gp|^2) \right), \
  \log \left( 1+\frac{\Sp +|\gc|^2 \Ic + 2|\gc||\gp| \sqrt{\Sp \Ic}}{1+ (1-|\gc|^2)\Ic} \right) \right\}.
\end{align*}
For the symmetric case, i.e., $\Sc=\Sp={\snr}, \Ic=\Ip=\inr$, instead of solving analytically the optimization involved in the sum-rate maximization, which does not seem to lead to a closed-form expression, we choose to set $|\gc|=0$ and $(1-|\gp|^2)=1$ if $\Cc < \frac{{\snr}}{1+{\inr}}$ and $(1-|\gp|^2)=\frac{{\snr}}{\Cc(1+{\inr})}$ otherwise (i.e., these values are not necessarily optimal). With these choices the following sum-rate is achievable
\begin{subequations}
\begin{align}
\Rp + \Rc 
\leq 
& 
 \log\left( 1+\frac{{\snr}}{1+{\inr}} \right)
+\log\left( \frac{1+{\snr}}{1+\frac{{\snr}/\Cc}{1+{\inr}} {\inr}} \right)
&&\text{for $\Cc \geq \frac{{\snr}}{1+{\inr}}$}
\label{eq:sumrate weak2 1}
\\
\Rp + \Rc 
\leq 
& \log\left( 1+\frac{{\snr}}{1+{\inr}} \right)+\log\left( 1+\Cc \right) 
&&\text{for $\Cc < \frac{{\snr}}{1+{\inr}}$.}
\label{eq:sumrate weak2 2}
\end{align}
\label{eq:sumrate weak2}
\end{subequations}

\subsection{Constant gap result for the sum-capacity of the symmetric GCCIC in Regimes 4 and 6 of Fig.~\ref{fig:fig3} for $\alpha < 1\leq \beta$}
%\subsection{Gap for the regime $\alpha < 1$ and $\beta > 1$ for the symmetric Gaussian CCIC}
\label{sec:scheme 3gap}

With the DPC-based achievable scheme in Appendix~\ref{sec:scheme 3} an achievable sum-rate is given by~\eqref{eq:sumrate weak2 1}, which we now use to derive a smaller gap than those in Section \ref{subsec:sum regime 4} and Appendix \ref{app:gap sym regime 6} in the regime ${\inr}<{\snr}$ and $\Cc > {\snr}$ (that corresponds to parts of regimes 4 and 6 of Fig.~\ref{fig:fig3}). The achievable sum-rate in~\eqref{eq:sumrate weak2 1} implies
\begin{align*}
& {\gdof}(\alpha,\beta)  \geq 
%\lim_{{\snr}\to\infty}\frac{\left(\Rp+\Rc \right)^{\rm(IB)}}{2\log(1+{\snr})}  \\& = 
\lim_{{\snr}\to\infty}\frac{ \log\left( 1+\frac{{\snr}}{1+{\inr}} \right)
+\log\left( \frac{1+{\snr}}{1+\frac{{\snr}/\Cc}{1+{\inr}} {\inr}} \right)
}{2\log(1+{\snr})} 
\\& = \frac{1}{2}\left( [1-\alpha]^+ + 1 -[1-\beta]^+ \right)
\stackrel{\text{$\alpha< 1 \leq \beta$}}{=} \frac{2-\alpha}{2}.
\end{align*}
This shows the achievability of the gDoF upper bound by means of~\eqref{eq:sumrate weak2 1}.

By using the sum-capacity upper bound in~\eqref{eq:Tuninetti} under the condition ${\snr} \geq {\inr}$ and the achievable sum-rate in~\eqref{eq:sumrate weak2 1} we obtain the following gap
\begin{align*}
  \mathsf{GAP} %&=\left(\Rp+\Rc \right)^{\rm(OB)} - \left(\Rp+\Rc \right)^{\rm(IB)} 
  &\leq \log \left( \frac{1+{\snr}}{1+{\inr}} \right)+\log \left( 1+ {\snr}+{\inr} \right)+\log(2)
\\&    - \log\left( 1+\frac{{\snr}}{1+{\inr}} \right)
        -\log\left( \frac{1+{\snr}}{1+\frac{{\snr}/\Cc}{1+{\inr}} {\inr}} \right)
\\&\leq  \log\left(1+ \frac{{\snr}}{\Cc} \right)+\log(2)
\\&\leq  2\log(2),
\end{align*}
using ${\snr}\leq \Cc$. %$0\leq {\inr}$ and 
This example shows that an achievable scheme more complex than simple superposition coding, like the DPC-based one, can achieve a smaller gap.

\subsection{Achievable Scheme 4: message~1 is split, and message~2 is common but not precoded}
\label{sec:scheme 4}
From the general region in Section~\ref{subsect:monster dpc region in awgn}, we set 
\begin{align*}
  &a_1=0, \ b_1=\sqrt{1-|\gp|^2}, \ c_1=\gp, \ |\gp|\leq 1, \quad
\\&a_2=\gc, \ b_2=\sqrt{1-|\gc|^2}, \ c_2=0, \ |\gc|\leq 1, 
\end{align*}
to obtain
\begin{subequations}
\begin{align}
   \Rp &\leq \log \left( 1+\Cc \right)
\\ \Rc & \leq \log \left( 1+\frac{\Sc \left( 1-|\gc|^2\right)}{1+|\gp|^2 \Ip + |\gc|^2 \Sc} \right)
\\ \Rp+\Rc & \leq \log \left( 1+ \Sp + \Ic  \right)
\\ \Rp + \Rc &  \leq  \log \left( 1 + |\gp|^2 \Cc \right )  +  \log \left( 1 + \frac{\Sc \left( 1-|\gc|^2\right) + \Ip \left( 1-|\gp|^2\right)}{1 + |\gp|^2 \Ip  +  |\gc|^2 \Sc } \right)
\\ \Rp + \Rc & \leq \log \left(  1 + |\gp|^2 \Sp + |\gc|^2 \Ic  \right) +  \log \left( 1 + \frac{\Sc \left( 1-|\gc|^2\right) + \Ip \left( 1-|\gp|^2\right)}{1 + |\gp|^2 \Ip  +  |\gc|^2 \Sc } \right)
\\ \Rp +2 \Rc & \leq \log \left(  1+ |\gp|^2 \Sp + \Ic   \right)
 +  \log \left( 1 + \frac{\Sc \left( 1-|\gc|^2\right) + \Ip \left( 1-|\gp|^2\right)}{1 + |\gp|^2 \Ip  +  |\gc|^2 \Sc } \right).
\label{eq:scheme fully connected strong interference Gaussian r1+2r2}
\end{align}
\label{eq:scheme fully connected strong interference Gaussian}
\end{subequations}

In the rate region in~\eqref{eq:scheme fully connected strong interference Gaussian}, the constraint on $\Rp +2\Rc$ becomes redundant if one of the conditions in~\eqref{like eq:ach reg 2 not in a very insightful form:when is r1+2r2 redundant} holds; in particular, if 
\begin{align}
  &I(\Yp;V_1) \leq I(\Yc;V_1) 
  \quad \Longleftrightarrow
\nonumber\\&\frac{(1- |\gp|^2)\Sp}{1+ |\gp|^2\Sp+\Ic} \leq \frac{(1- |\gp|^2)\Ip}{1+ |\gp|^2\Ip+\Sc}
\quad \Longleftrightarrow
\nonumber\\&\text{either $|\gp|=1$, or $\Sp \left(1+\Sc \right) \leq \Ip \left(1+\Ic \right)$},
\label{eq:scheme fully connected strong interference Gaussian drop r1+2rc condition 1}
\end{align}
or if
\begin{align}
  &I(\Yc;U_2|V_1) \leq I(\Yc;U_2|V_1)
  \quad \Longleftrightarrow
\nonumber\\&\frac{(1- |\gc|^2)\Sc}{1+ |\gp|^2\Ip+|\gc|^2\Sc} \leq \frac{(1- |\gc|^2)\Ic}{1+ |\gp|^2\Sp+|\gc|^2\Ic}
  \quad \Longleftrightarrow
\nonumber\\&\text{either $|\gc|=1$, or $\Sc \frac{1+ |\gp|^2\Sp}{1+ |\gp|^2\Ip} \leq \Ic$}.
\label{eq:scheme fully connected strong interference Gaussian drop r1+2rc condition 2}
\end{align}

\subsection{Achievable Scheme 5: message~1 is split, and message~2 is private; gap for the S-channel}
%\label{sec:scheme 5}
%\subsection{Gap for the S-channel}
\label{app:gap for the general S-channel}

From the general region in Section~\ref{subsect:monster dpc region in awgn}, we set $c_2=1$ to obtain
\begin{subequations}
\begin{align}
   \Rp &\leq \log(1+\Cc(|c_1|^2+|b_1|^2))\label{eq:final dpc ach region Q=0 rp 1} %=I(Z_1,V_1;\Yf|T_2,X_c,S_1)
\\ \Rp &\leq \log\left(1+\frac{\Sp}{1+\Ic}\right)\label{eq:final dpc ach region Q=0 rp 2}%=I(\Yp;V_1,S_1,Z_1)
\\ \Rc &\leq \log\left(1+\frac{\Sc}{1+\Ip|c_1|^2}\right)%=I(\Yc;T_2|V_1)-I(T_2;S_1)= I(\Yc;L_2|V_1,S_1)
\\ \Rp+\Rc &\leq  
  \log\left(\frac{1+\Sc+\Ip}{1+\Ip(|a_1|^2+|c_1|^2)+\Sc}\right)
 +\log\left(1+\frac{\Sc}{1+\Ip|c_1|^2}\right)
 +\log(1+\Cc |c_1|^2)\label{eq:final dpc ach region Q=0 rp+rc 1}
 %I(\Yc;V_1)+[I(\Yc;T_2|V_1)-I(T_2;S_1)]+I(\Yf;     Z_1|T_2,X_c,S_1,V_1)
\\ \Rp+\Rc &\leq 
  \log\left(\frac{1+\Sc+\Ip}{1+\Ip(|a_1|^2+|c_1|^2)+\Sc}\right)
 +\log\left(1+\frac{\Sc}{1+\Ip|c_1|^2}\right) \nonumber
 \\& +\quad \log\left(1+\frac{\Sp(|a_1|^2+|c_1|^2)}{1+\Ic}\right).\label{eq:final dpc ach region Q=0 rp+rc 2}
%I(\Yc;V_1)+[I(\Yc;T_2|V_1)-I(T_2;S_1)]+I(\Yp; S_1,Z_1|V_1)
% I(\Yc;V_1) = \log\left(1+\frac{I_p|b_1|^2}{1+I_p(1-|b_1|^2)+\Sc}\right)
% I(\Yc;T_2|V_1)-I(T_2;S_1) = \log\left(1+\frac{\Sc}{1+I_p|c_1|^2}\right)
% I(\Yp; S_1,Z_1|V_1) = \log\left(1+\frac{\Sp(1-|b_1|^2)}{1+\Ic}\right)
\end{align}
\label{eq:final dpc ach region Q=0}
\end{subequations}

An achievable region for the S-channel is obtained by setting $\Ic=0$ in~\eqref{eq:final dpc ach region Q=0}. 
Here we concentrate on the regime $\Sp \leq \Cc \leq (1+\Ip)\Sp$
and evaluate the region in~\eqref{eq:final dpc ach region Q=0} for
\[
\Ic=0, \
|a_1|^2 = \frac{\Cc-\Sp}{(1+\Ip)\Sp}, \
|b_1|^2 = \frac{(1+\Ip)\Sp-\Cc}{(1+\Ip)\Sp}, \
|c_1|^2 = \frac{1}{1+\Ip}.
\]
With these choices the region in~\eqref{eq:final dpc ach region Q=0} reduces to
\begin{subequations}
\begin{align}
   \Rp &\leq \log(1+\Sp)
\\ \Rc &\leq \log\left(1+\frac{\Sc}{1+\frac{\Ip}{1+\Ip}}\right)
\\ \Rp+\Rc &\leq  
  \log\left(\frac{1+\Sc+\Ip}{1+\Sc+\frac{\Ip}{1+\Ip} \ \frac{\Cc}{\Sp}}\right)
 +\log\left(1+\frac{\Sc}{1+\frac{\Ip}{1+\Ip}}\right)
 +\log\left(1+\frac{\Cc}{1+\Ip}\right)\label{eq:final dpc ach region Q=0 Ic=0 2 rp+pc}
\end{align}
\label{eq:final dpc ach region Q=0 Ic=0 2}
\end{subequations}
since the bound on $\Rp$ in~\eqref{eq:final dpc ach region Q=0 rp 1} would give $\Rp \leq \log\left(1+\Cc\frac{2+\Ip-\Cc/\Sp}{1+\Ip}\right)$
which is redundant because 
\[
\Sp \leq \Cc\frac{2+\Ip-\frac{\Cc}{\Sp}}{1+\Ip} \Longleftrightarrow
1-2\frac{\Cc}{\Sp}+\left(\frac{\Cc}{\Sp}\right)^2 \leq \Ip\left(\frac{\Cc}{\Sp}-1\right)\Longleftrightarrow
\Sp\leq \Cc \leq (1+\Ip)\Sp;
\]
and the  sum-rate bound in~\eqref{eq:final dpc ach region Q=0 rp+rc 2} would give
$\Rp+\Rc \leq 
  \log\left(\frac{1+\Sc+\Ip}{1+\Sc+\frac{\Ip}{1+\Ip} \ \frac{\Cc}{\Sp}}\right)
 +\log\left(1+\frac{\Sc}{1+\frac{\Ip}{1+\Ip}}\right)
 +\log\left(1+\Cc\right)$,
which is clearly redundant because of~\eqref{eq:final dpc ach region Q=0 Ic=0 2 rp+pc}.

%reduces to 
%\begin{subequations}
%\begin{align}
%   \Rp &\leq \log(1+\Cc(1-|a|^2))%=I(Z_1,V_1;\Yf|T_2,X_c,S_1)
%\\ \Rp &\leq \log\left(1+\Sp\right)%=I(\Yp;V_1,S_1,Z_1)
%\\ \Rc &\leq \log\left(1+\frac{\Sc}{1+\Ip|b|^2}\right)%=I(\Yc;T_2|V_1)-I(T_2;S_1)= I(\Yc;L_2|V_1,S_1)
%\\ \Rp+\Rc &\leq  
%  \log\left(\frac{1+\Sc+\Ip}{1+\Ip(|a|^2+|b|^2)+\Sc}\right)
% +\log\left(1+\frac{\Sc}{1+\Ip|b|^2}\right)
% +\log(1+\Cc |b|^2)
%\\ \Rp+\Rc &\leq 
%  \log\left(\frac{1+\Sc+\Ip}{1+\Ip(|a|^2+|b|^2)+\Sc}\right)
% +\log\left(1+\frac{\Sc}{1+\Ip|b|^2}\right)
% +\log\left(1+\Sp(|a|^2+|b|^2)\right).
%\end{align}
%\label{eq:final dpc ach region Q=0 Ic=0}
%\end{subequations}
%
%\[
%\Cc(1-|a|^2) = (\Cc-\Sp)\left(1- \frac{\Cc}{(1+\Ip)\Sp}\right) + \Sp \geq \Sp
%\]
%we simplify the region in~\eqref{eq:final dpc ach region Q=0 Ic=0} to

We next match the achievable region in~\eqref{eq:final dpc ach region Q=0 Ic=0 2} to the outer bound
\begin{subequations}
\begin{align}
%O^{\rm(S)}=
%\left\{\begin{array}{rl}
  \Rp &\leq \log\left ( 1 + \Sp\right )
\\\Rc &\leq \log\left ( 1 + \Sc\right )
\\\Rp+\Rc &\leq \log\left ( 1 + (\sqrt{\Sc} + \sqrt{\Ip})^2\right ) 
              + \log \left( \frac{1+\Cc+\max\{\Sp,\Ip\}}{1+\Ip} \right).
%\end{array}\right.
\end{align}
\label{eq:outerBoundS appendix}
\end{subequations}
from~\eqref{eq:ourRate} with $\Ic=0$.
The bounds on $\Rp$ in~\eqref{eq:final dpc ach region Q=0 Ic=0 2} and~\eqref{eq:outerBoundS appendix} are the same, and
the bounds on $\Rc$ in~\eqref{eq:final dpc ach region Q=0 Ic=0 2} and~\eqref{eq:outerBoundS appendix} are are at most 1~bit apart.
For the sum-rate, if $\Cc/\Sp \leq \Sc$ (and recall that we focus on $\Sp \leq \Cc$) then  
\begin{align*}
\mathsf{GAP}& 
\leq \log\left ( 1 + (\sqrt{\Sc} + \sqrt{\Ip})^2\right ) 
   + \log \left( \frac{1+\Cc+\max\{\Sp,\Ip\}}{1+\Ip} \right)+
\\& -\log\left(\frac{1+\Sc+\Ip}{1+\Sc+\frac{\Ip}{1+\Ip} \ \frac{\Cc}{\Sp}}\right)
    -\log\left(1+\frac{\Sc}{1+\frac{\Ip}{1+\Ip}}\right)
    -\log\left(1+\frac{\Cc}{1+\Ip}\right)
\\& \leq \log(2)
+\log\left ( \frac{ {1+\Sc+\frac{\Ip}{1+\Ip} \ \frac{\Cc}{\Sp}}}{ 1+\frac{\Sc}{1+\frac{\Ip}{1+\Ip}}}\right )
+\log\left ( \frac{1+\Cc+\max\{\Sp,\Ip\}}{  1+\Cc+\Ip}\right )
\\& \leq \log(2)
+\log\left ( \frac{ {1+\Sc\left(1+\frac{\Ip}{1+\Ip}\right)}}{ 1+\frac{\Sc}{1+\frac{\Ip}{1+\Ip}}}\right )
+\log\left ( \frac{1+2\max\{\Cc,\Ip\}}{  1+\Cc+\Ip}\right )
\\& \leq \log(2)+2\log(2)+ \log(2) = 4 \log(2);
\end{align*}
while if $\Cc/\Sp > \Sc$ then
\begin{align*}
\mathsf{GAP}& \leq \log(1+\Sp)
+\log(1+\Sc)+
\\& -\log\left(\frac{1+\Sc+\Ip}{1+\Sc+\frac{\Ip}{1+\Ip} \ \frac{\Cc}{\Sp}}\right)
 -\log\left(1+\frac{\Sc}{1+\frac{\Ip}{1+\Ip}}\right)
 -\log\left(1+\frac{\Cc}{1+\Ip}\right)
\\& \leq
 \log\left( \frac{(1+\Sp)(1+2 \Cc/\Sp)}{1+\Ip+\Cc}) \right)
+\log\left( \frac{1+\Sc}{1+\frac{\Sc}{1+\frac{\Ip}{1+\Ip}}} \right)
+\log\left( \frac{1+\Ip}{1+\Sc+\Ip} \right)
\\& \stackrel{1\leq \Sp \leq \Cc}{\leq}
\log\left(\max\left\{\frac{2(1+2\Cc)}{1+\Cc}, 3 \right\}\right)
 +\log(2)+\log(1) = 3\log(2).
\end{align*}

\bibliographystyle{IEEEtran}
\bibliography{AllertonBib}

\begin{thebibliography}{10}
\providecommand{\url}[1]{#1}
\csname url@samestyle\endcsname
\providecommand{\newblock}{\relax}
\providecommand{\bibinfo}[2]{#2}
\providecommand{\BIBentrySTDinterwordspacing}{\spaceskip=0pt\relax}
\providecommand{\BIBentryALTinterwordstretchfactor}{4}
\providecommand{\BIBentryALTinterwordspacing}{\spaceskip=\fontdimen2\font plus
\BIBentryALTinterwordstretchfactor\fontdimen3\font minus
  \fontdimen4\font\relax}
\providecommand{\BIBforeignlanguage}[2]{{%
\expandafter\ifx\csname l@#1\endcsname\relax
\typeout{** WARNING: IEEEtran.bst: No hyphenation pattern has been}%
\typeout{** loaded for the language `#1'. Using the pattern for}%
\typeout{** the default language instead.}%
\else
\language=\csname l@#1\endcsname
\fi
#2}}
\providecommand{\BIBdecl}{\relax}
\BIBdecl

\bibitem{goldsmith:spectrymgridlock}
A.~Goldsmith, S.~A. Jafar, I.~Maric, and S.~Srinivasa, ``Breaking spectrum
  gridlock with cognitive radios: An information theoretic perspective,''
  \emph{Proocedings of the IEEE}, vol.~97, no.~5, pp. 894 --914, May 2009.

\bibitem{Devroye}
N.~Devroye, P.~Mitran, and V.~Tarokh, ``Achievable rates in cognitive radio
  channels,'' \emph{IEEE Trans. on Info. Theory}, vol.~52, no.~5, pp. 1813 --
  1827, May 2006.

\bibitem{3GPPRel10doc}
LTE-A, \emph{3rd Generation Partnership Project; Technical Specification Group
  Radio Access Network; Evolved Universal Terrestrial Radio Access
  ({EUTRA})}.\hskip 1em plus 0.5em minus 0.4em\relax 3GPP TR 36.806 V9.0.0,
  2010.

\bibitem{RICEpaper}
M.~Duarte, C.~Dick, and A.~Sabharwal, ``Experiment-driven characterization of
  full-duplex wireless systems,'' \emph{IEEE Trans. on Wireless
  Communications}, vol.~11, no.~12, pp. 4296--4307, 2012.

\bibitem{HK}
T.~Han and K.~Kobayashi, ``A new achievable rate region for the interference
  channel,'' \emph{IEEE Trans. on Info. Theory}, vol.~27, no.~1, pp. 49 -- 60,
  Jan. 1981.

\bibitem{costaDPC}
M.~Costa, ``Writing on dirty paper (corresp.),'' \emph{IEEE Trans. on Info.
  Theory}, vol.~29, no.~3, pp. 439 -- 441, May 1983.

\bibitem{HostMadsenIT06}
A.~Host-Madsen, ``Capacity bounds for cooperative diversity,'' \emph{IEEE
  Trans. on Info. Theory}, vol.~52, no.~4, pp. 1522 --1544, April 2006.

\bibitem{YANG-TUNINETTI}
S.~Yang and D.~Tuninetti, ``Interference channel with generalized feedback
  (a.k.a. with source cooperation): Part i: Achievable region,'' \emph{IEEE
  Trans. on Info. Theory}, vol.~57, no.~5, pp. 2686 --2710, May 2011.

\bibitem{PVIT11}
V.~Prabhakaran and P.~Viswanath, ``Interference channels with source
  cooperation,'' \emph{IEEE Trans. on Info. Theory}, vol.~57, no.~1, pp. 156
  --186, Jan. 2011.

\bibitem{TuninettiITA10}
D.~Tuninetti, ``An outer bound region for interference channels with
  generalized feedback,'' in \emph{Information Theory and Applications Workshop
  (ITA), 2010}, Feb. 2010, pp. 1--5.

\bibitem{TndonUlukusIT11}
R.~Tandon and S.~Ulukus, ``Dependence balance based outer bounds for gaussian
  networks with cooperation and feedback,'' \emph{IEEE Trans. on Info. Theory},
  vol.~57, no.~7, pp. 4063 --4086, July 2011.

\bibitem{Kramer}
G.~Kramer, ``Outer bounds on the capacity of gaussian interference channels,''
  \emph{IEEE Transactions on Information Theory}, vol.~20, no.~3, pp. 581--586,
  2004.

\bibitem{TuninettiITW12}
D.~Tuninetti, ``An outer bound for the memoryless two-user interference channel
  with general cooperation,'' in \emph{IEEE Information Theory Workshop (ITW),
  2012}, 2012, pp. 217--221.

\bibitem{etw}
R.~Etkin, D.~Tse, and H.~Wang, ``Gaussian interference channel capacity to
  within one bit,'' \emph{IEEE Trans. on Info. Theory}, vol.~54, no.~12, pp.
  5534 --5562, Dec. 2008.

\bibitem{TelatarTse}
E.~Telatar and D.~Tse, ``Bounds on the capacity region of a class of
  interference channels,'' in \emph{IEEE ISIT 2007}, June 2007, pp. 2871
  --2874.

\bibitem{kekstrawillems}
A.~Hekstra and F.~Willems, ``Dependence balance bounds for single-output
  two-way channels,'' \emph{IEEE Trans. on Info. Theory}, vol.~35, no.~1, pp.
  44 --53, Jan. 1989.

\bibitem{gelfandpinsker}
S.~Gelfand and M.~Pinsker, ``Coding for channel with random parameters,''
  \emph{Problem of Contr. Inf. Theory}, vol.~9, no.~1, pp. 19 -- 31, 1980.

\bibitem{coveElGamal}
T.~Cover and A.~Gamal, ``Capacity theorems for the relay channel,'' \emph{IEEE
  Trans. on Info. Theory}, vol.~25, no.~5, pp. 572 -- 584, Sep. 1979.

\bibitem{YangHighCoop}
S.~Yang and D.~Tuninetti, ``Interference channels with source cooperation in
  the strong cooperation regime: Symmetric capacity to within 2 bits/s/hz with
  dirty paper coding,'' in \emph{Asilomar 2011}, Nov. 2011, pp. 2140 --2144.

\bibitem{DBLP:journals/corr/abs-1111-3966}
Z.~Wu and M.~Vu, ``Partial decode-forward binning schemes for the causal
  cognitive relay channels,'' \emph{CoRR}, vol. abs/1111.3966, 2011.

\bibitem{MirmohseniIT2012}
M.~Mirmohseni, B.~Akhbari, and M.~Aref, ``On the capacity of interference
  channel with causal and noncausal generalized feedback at the cognitive
  transmitter,'' \emph{IEEE Trans. on Info. Theory}, vol.~58, no.~5, pp.
  2813--2837, May 2012.

\bibitem{riniJ1}
S.~Rini, D.~Tuninetti, and N.~Devroye, ``Inner and outer bounds for the
  gaussian cognitive interference channel and new capacity results,''
  \emph{IEEE Trans. on Info. Theory}, vol.~58, no.~2, pp. 820 --848, Feb. 2012.

\bibitem{ElGamalKimBook}
A.~E. Gamal and Y.-H. Kim, \emph{Network Information Theory}.\hskip 1em plus
  0.5em minus 0.4em\relax Cambridge U.K.: Cambridge Univ. Press,, 2011.

\bibitem{suhtse:ICwithfeedback}
C.~Suh and D.~Tse, ``Feedback capacity of the gaussian interference channel to
  within 2 bits,'' \emph{IEEE Trans. on Info. Theory}, vol.~57, no.~5, pp.
  2667--2685, 2011.

\bibitem{sriranCICweak}
W.~Wu, S.~Vishwanath, and A.~Arapostathis, ``Capacity of a class of cognitive
  radio channels: Interference channels with degraded message sets,''
  \emph{IEEE Trans. on Info. Theory}, vol.~53, no.~11, pp. 4391--4399, 2007.

\bibitem{viswanathCICweak}
A.~Jovicic and P.~Viswanath, ``Cognitive radio: An information-theoretic
  perspective,'' \emph{IEEE Trans. on Info. Theory}, vol.~55, no.~9, pp.
  3945--3958, 2009.

\bibitem{SASON}
I.~Sason, ``On achievable rate regions for the gaussian interference channel,''
  \emph{IEEE Trans. on Info. Theory}, vol.~50, no.~6, pp. 1345 -- 1356, June
  2004.

\end{thebibliography}
%\end{thebibliography}

\begin{figure*}
\centering
\includegraphics[width=0.8\textwidth]{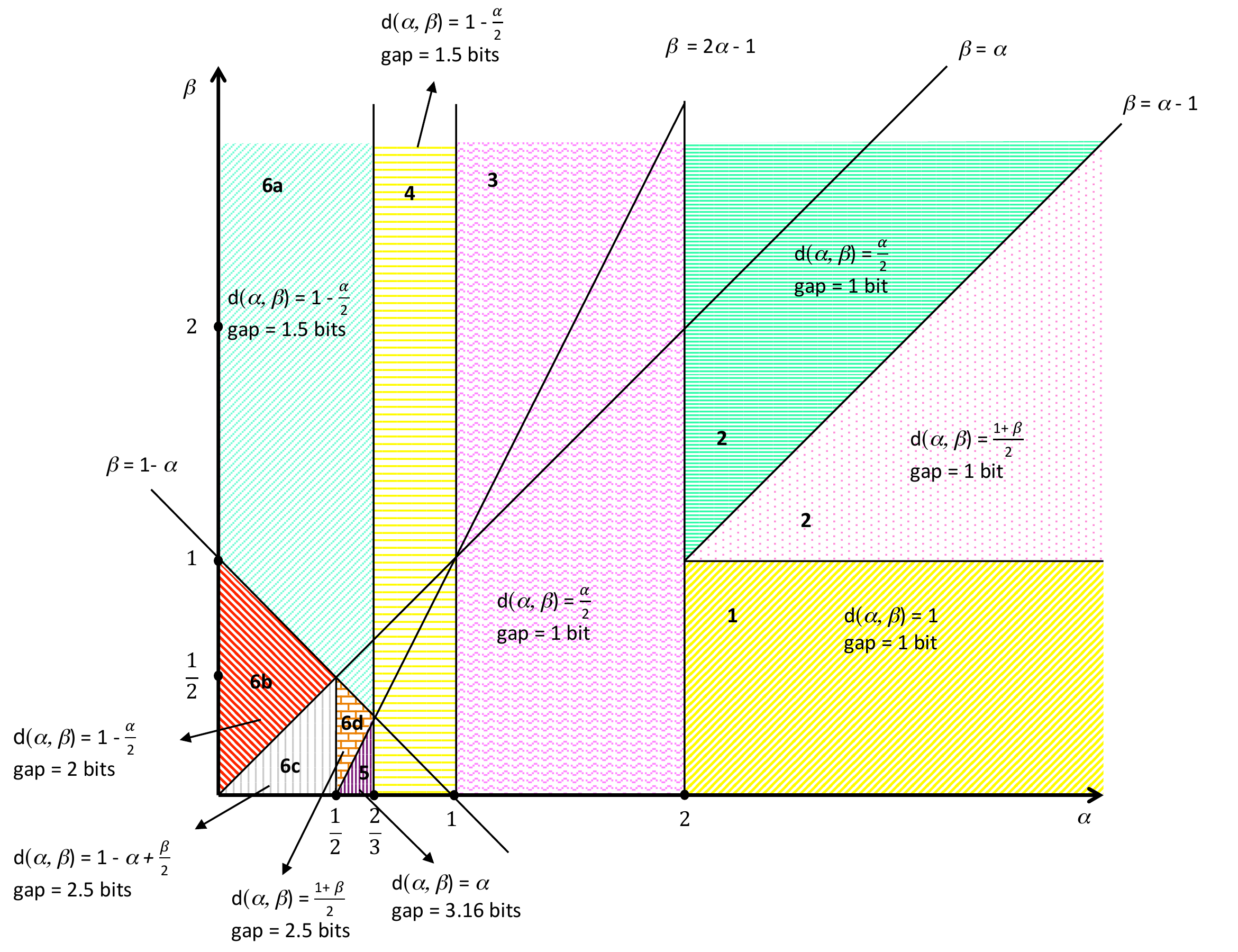}
\vspace{-0.5cm}
\caption{Optimal gDoF and constant gap for the symmetric channel in the different regimes of $(\alpha,\beta)$.}
\label{fig:fig3}
\end{figure*}
\begin{figure*}
\centering
%\vspace{6mm}
\includegraphics[width=0.8\textwidth]{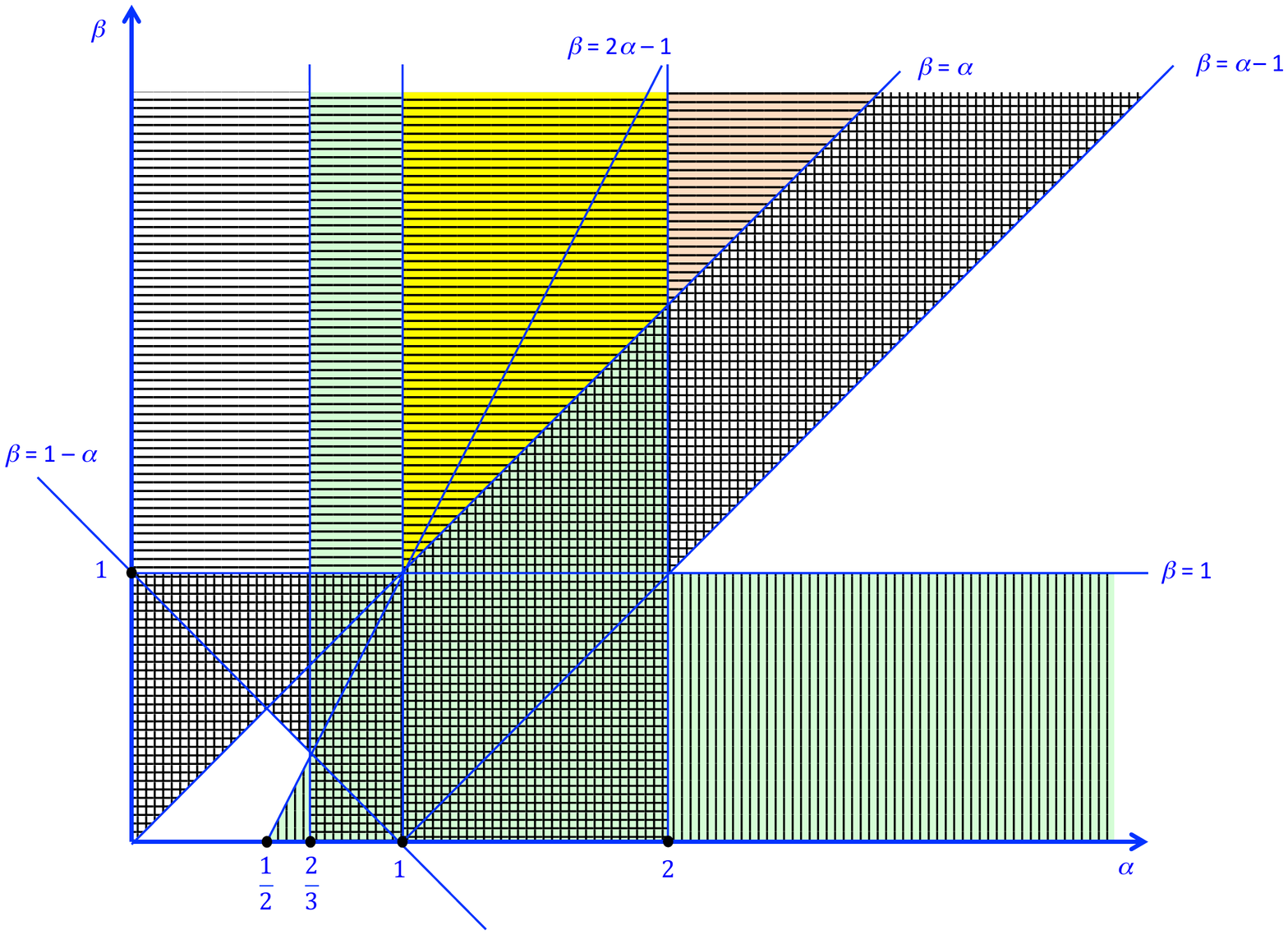}
%\vspace{2cm}
\caption{Regions in which the gDoF of the symmetric channel is equal to that 
of the noncooperative IC (green and yellow regions), 
of the RC (red and yellow regions), 
of the non-causal cognitive IC (region with horizontal lines), and
of bilateral source cooperation (region with vertical lines). 
Note that the different regions can overlap.}
\label{fig:fig6}
\end{figure*}

\begin{figure*}
\centering
\includegraphics[width=0.75\textwidth]{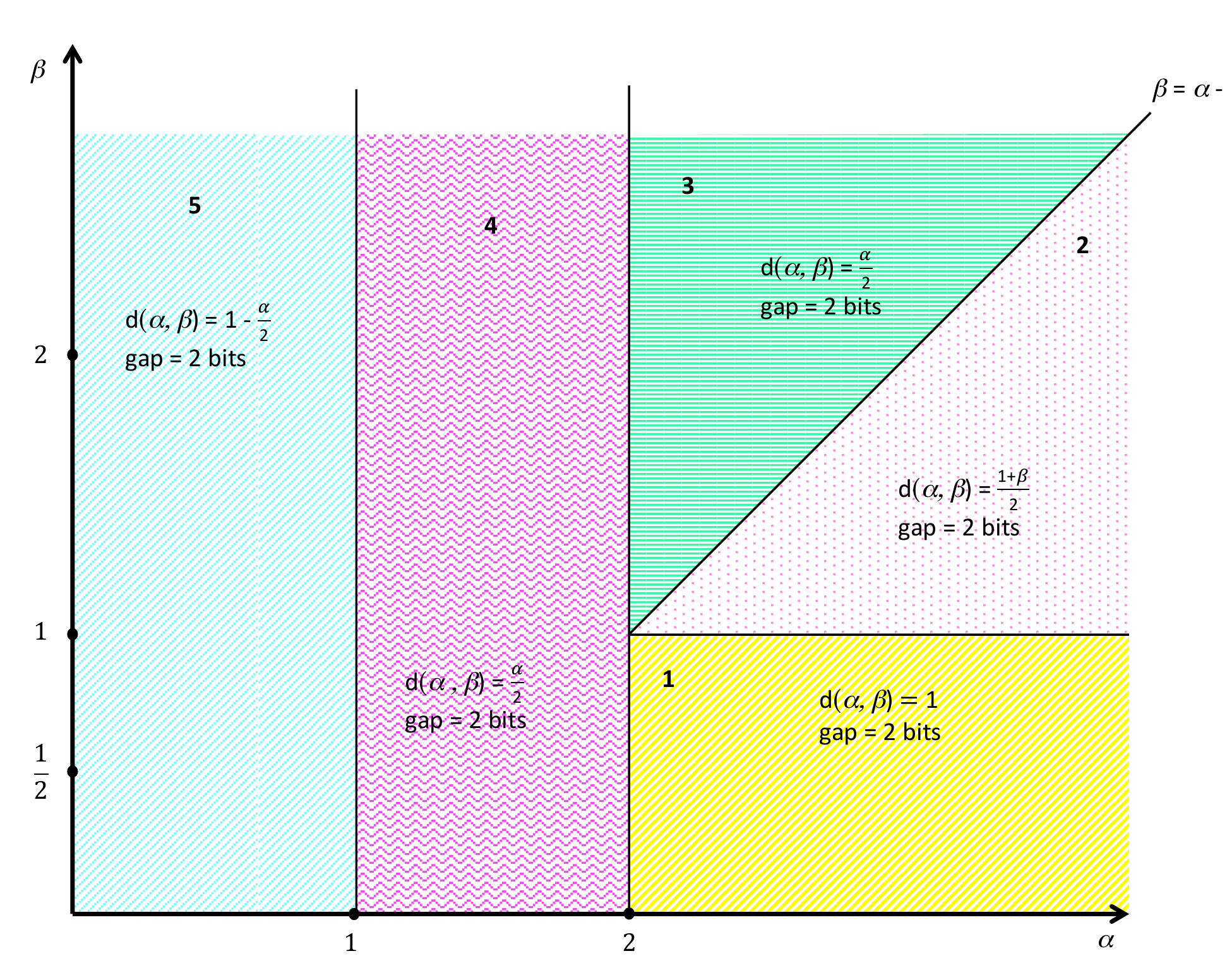}
%\vspace{2cm}
\vspace{-0.5cm}
\caption{Optimal gDoF and constant gap for the Z-channel in the different regimes of $(\alpha,\beta)$.}
\label{fig:fig4}
\end{figure*}
\begin{figure*}
\centering
%\vspace{8mm}
\includegraphics[width=0.7\textwidth]{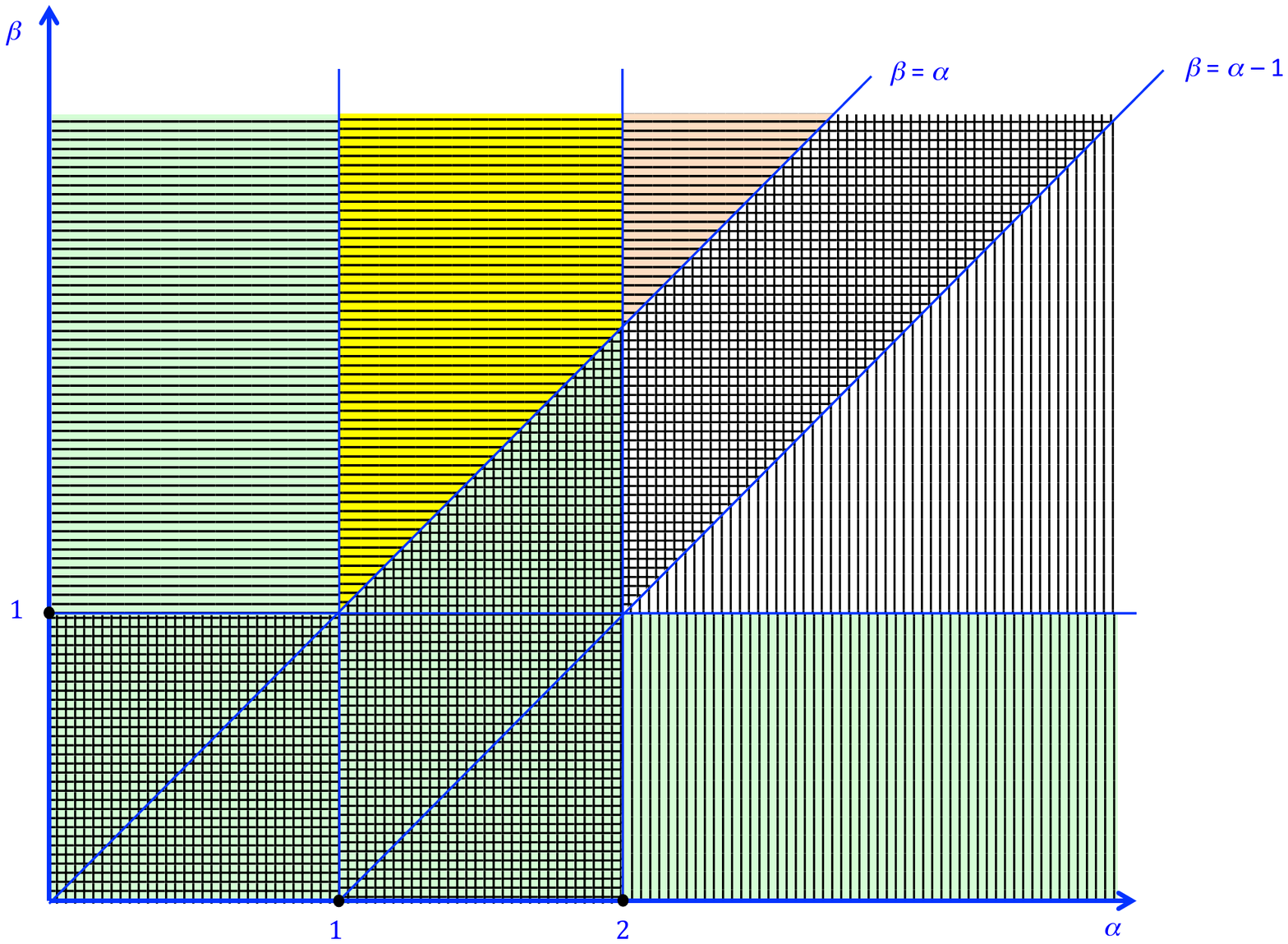}
%\vspace{2cm}
\caption{Regions in which the gDoF of the Z-channel is equal to that 
of the noncooperative IC (green and yellow regions), 
of the RC (red and yellow regions), 
of the non-causal cognitive IC (region with horizontal lines), and
of bilateral source cooperation (region with vertical lines). 
Note that the different regions can overlap.}
\label{fig:fig7}
\end{figure*}

\begin{figure*}
\centering
\includegraphics[width=0.75\textwidth]{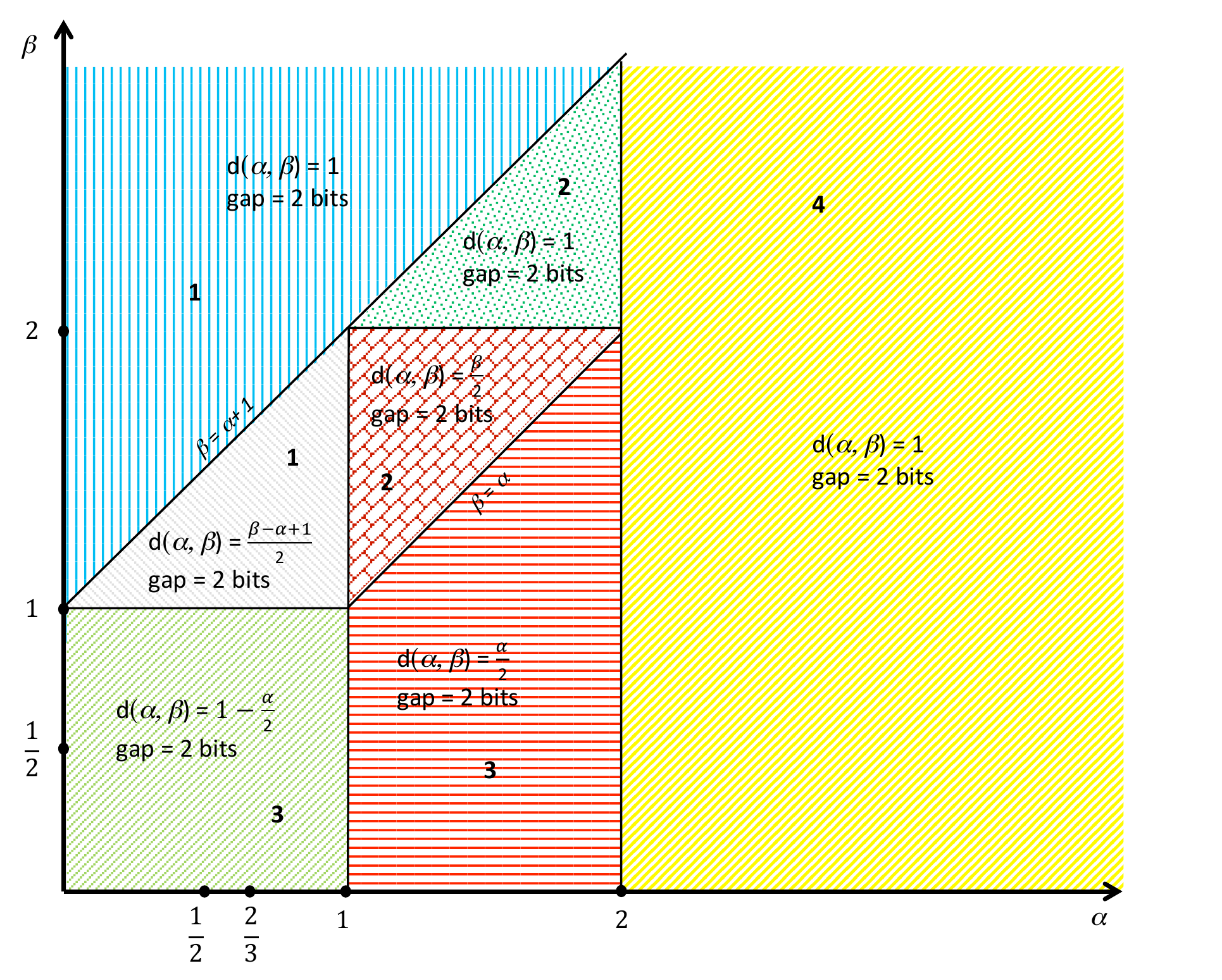}
%\vspace{2cm}
\vspace{-0.5cm}
\caption{Optimal gDoF and constant gap for the S-channel in the different regimes of $(\alpha,\beta)$.}
\label{fig:fig5}
\end{figure*}

\begin{figure*}
\centering
%\vspace{8mm}
\includegraphics[width=0.7\textwidth]{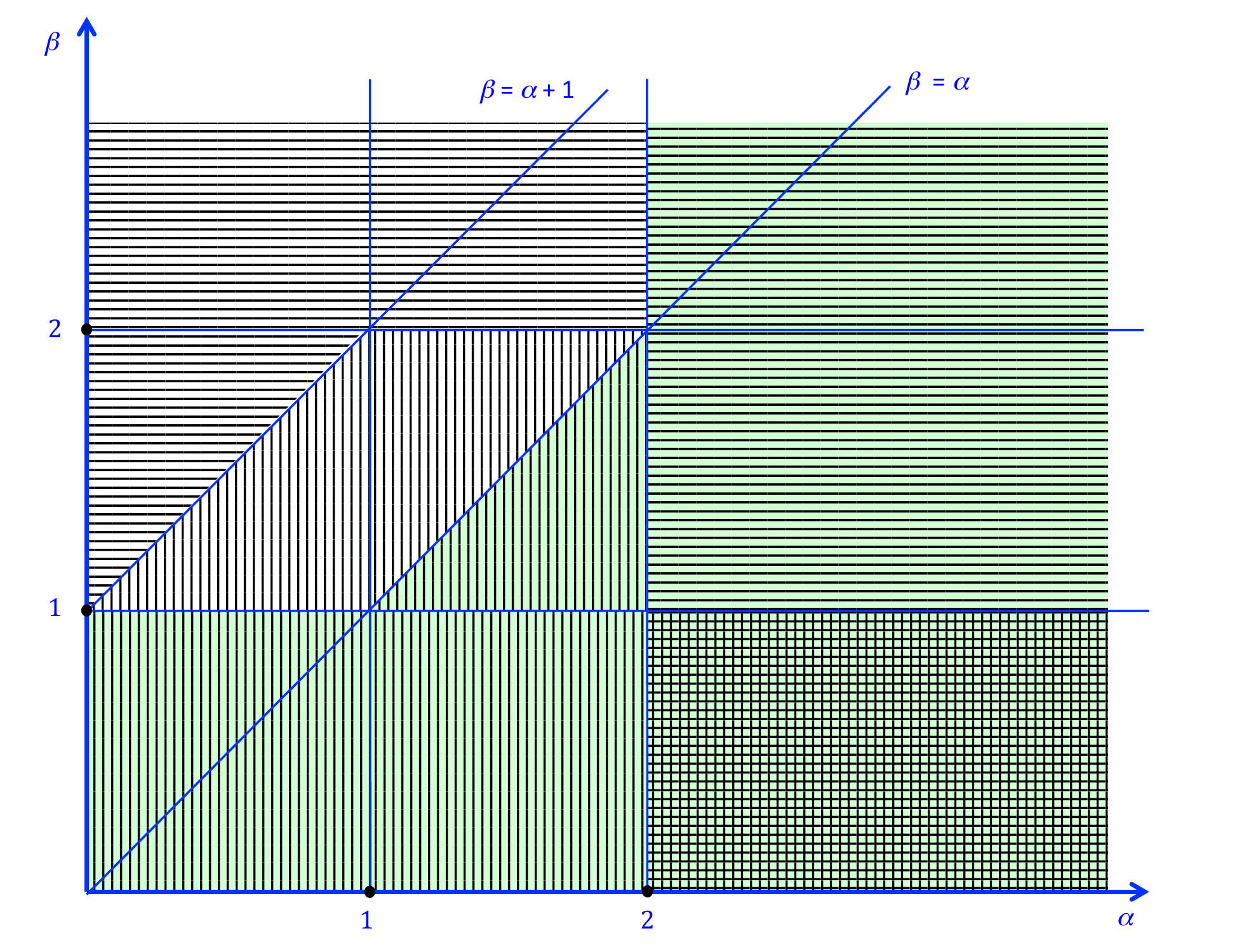}
%\vspace{2cm}
\caption{Regions in which the gDoF of the S-channel is equal to that 
of the noncooperative IC (green region), 
of the non-causal cognitive IC (region with horizontal lines), and
of bilateral source cooperation (region with vertical lines). 
Note that the different regions can overlap.}
\label{fig:fig8}
\end{figure*}

\begin{figure*}
\centering
\includegraphics[width=0.7\textwidth]{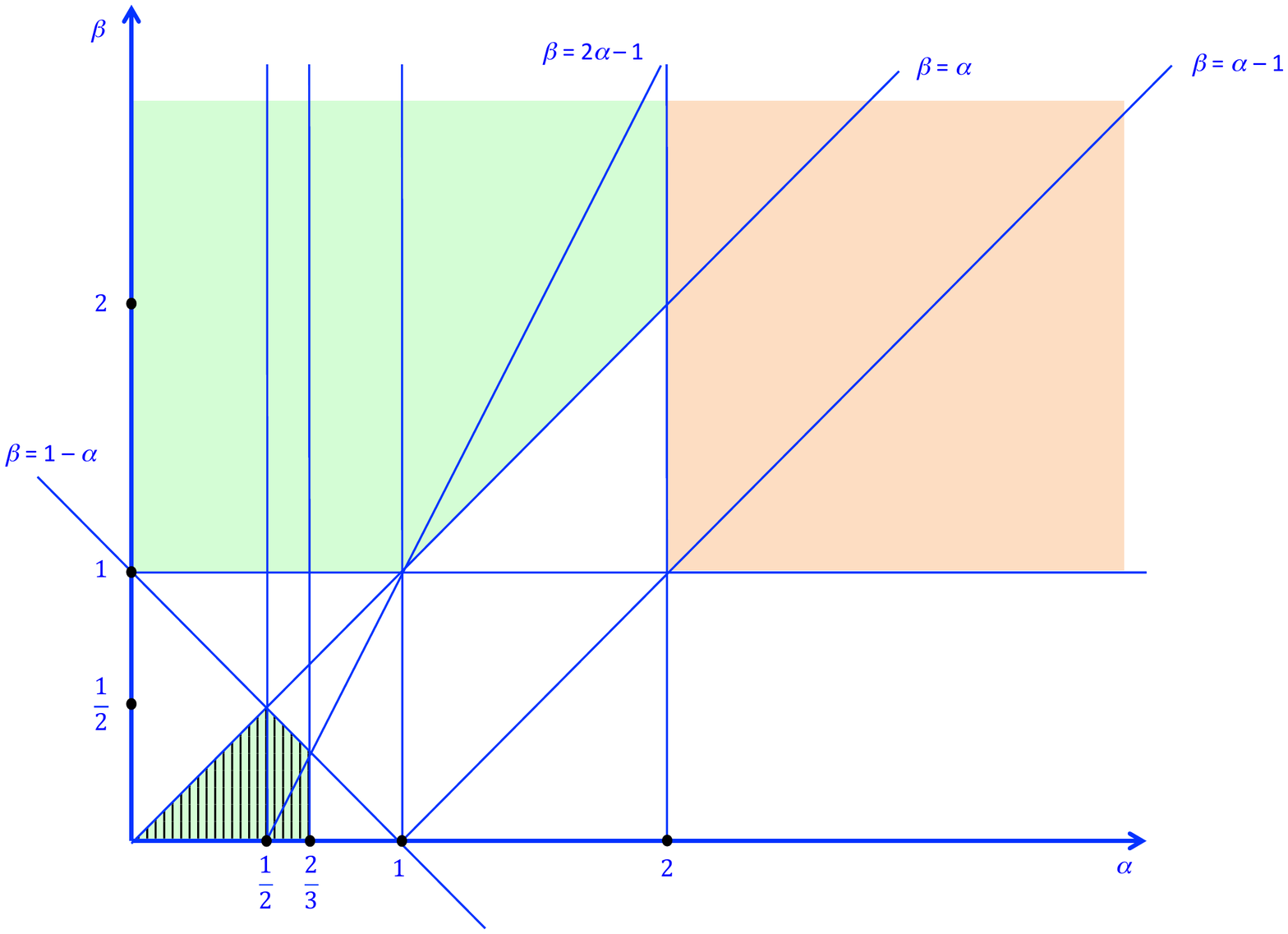}
%\vspace{2cm}
\caption{Regions in which
the S-channel outperforms the symmetric channel (green region), 
the symmetric channel outperforms the S-channel (red region), 
the Z-channel outperforms the symmetric channel (region with vertical lines).
Note that the different regions can overlap.}
\label{fig:fig9}
\end{figure*}

\end{document}